\documentclass{aa}

\usepackage{graphicx}
\usepackage{txfonts}
\usepackage{hyperref}
\usepackage{multirow}
\usepackage{xcolor}
\usepackage{tikz}
\usepackage{placeins}
\usetikzlibrary{automata,positioning}
\usepackage{pgfplots}
\usepackage{subcaption}

\newcommand{\orcit}[1]{\protect\href{https://orcid.org/#1}{\protect\includegraphics[width=8pt]{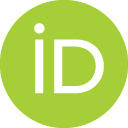}}}

\defcitealias{Giacobbo2020}{GM20}
\defcitealias{Hobbs2005}{H05}
\defcitealias{Speagle2014}{S14}
\defcitealias{Boogaard2018}{B18}
\defcitealias{Popesso2023}{P23}

\begin{document} 

   \title{The more accurately the metal-dependent star formation rate is modeled, the larger the predicted excess of binary black hole mergers}

    \titlerunning{Merger rate density of binary black holes}

   \subtitle{}

   \author{Cecilia Sgalletta\orcit{0009-0003-7951-4820} \inst{1,2,3}\thanks{\href{mailto:cecilia.sgalletta@sissa.it}{cecilia.sgalletta@sissa.it}},
   Michela Mapelli \orcit{0000-0001-8799-2548}\inst{4,5,6}\thanks{\href{mailto:mapelli@uni-heidelberg.de}{mapelli@uni-heidelberg.de}},
   Lumen Boco\orcit{0000-0003-3127-922X} \inst{1,4},
   Filippo Santoliquido\orcit{0000-0003-3752-1400} \inst{7, 8},
   M. Celeste Artale\orcit{0000-0003-0570-785X} \inst{9},
   Giuliano Iorio\orcit{0000-0003-0293-503X} \inst{5,6,10},
   Andrea Lapi\orcit{0000-0002-4882-1735} \inst{1,11,12},
   Mario Spera\orcit{0000-0003-0930-6930}\inst{1,2,3}
          }

    \institute{
    SISSA, via Bonomea 365, I--34136 Trieste, Italy
    \and
    National Institute for Nuclear Physics – INFN, Sezione di Trieste, I--34127 Trieste, Italy
    \and
    Istituto Nazionale di Astrofisica – Osservatorio Astronomico di Roma, Via Frascati 33, I--00040, Monteporzio Catone, Italy
    \and
    Institut f{\"u}r Theoretische Astrophysik, ZAH, Universit{\"a}t Heidelberg, Albert-Ueberle-Stra{\ss}e 2, D--69120, Heidelberg, Germany
    \and
   Physics and Astronomy Department Galileo Galilei, University of Padova, Vicolo dell'Osservatorio 3, I--35122, Padova, Italy 
   \and
   INFN - Padova, Via Marzolo 8, I--35131 Padova, Italy
   \and
    Gran Sasso Science Institute, Via F. Crispi 7, I-67100 L'Aquila, Italy
      \and
      INFN, Laboratori Nazionali del Gran Sasso, I--67100 Assergi, Italy
    \and
    Universidad Andres Bello, Facultad de Ciencias Exactas, Departamento de Fisica y Astronomia, Instituto de Astrofisica, Fernandez Concha 700, Las Condes, Santiago RM, Chile 
    \and
      Departament de F\'isica Qu\`antica i Astrof\'isica, Institut de Ci\`encies del Cosmos, Universitat de Barcelona, Mart\'i i Franqu\`es 1, E-08028 Barcelona, Spain 
    \and
    Institute for Fundamental Physics of the Universe – IFPU, Via Beirut 2, I--34014 Trieste, Italy
    \and
    Istituto di Radioastronomia – INAF/IRA, Via Piero Gobetti 101, I--40129 Bologna, Italy\\}

  \abstract
   {
   As the number of gravitational wave detections grows, the merger rate of binary black holes (BBHs) can help us to constrain their formation, the properties of their progenitors, and their birth environment.
   Here, we aim to address the impact of the metal-dependent star formation rate (SFR) on the BBH merger rate. 
   To this end, we have developed a fully data-driven approach to model the metal-dependent SFR and coupled it to BBH evolution. We have adopted the most up-to-date  scaling relations, based on recent observational  results, and we have studied how the BBH merger rate density varies over a wide grid of galaxy and binary evolution parameters. 
   Our results show that including a realistic metal-dependent SFR evolution yields a value of the merger rate density that is too high compared to the one inferred from gravitational wave data. Moreover, variations in the SFR in low-mass galaxies ($M_\ast \lesssim 10^8 \mathrm{M}_{\odot}$) do not contribute more than a factor $\sim 2$ to the overall merger rate density at redshift $z=0$. 
   These results suggest that the discrepancy between the BBH merger rate density inferred from data and theoretical models is not caused by approximations in the treatment of the metal-dependent SFR, but rather stems from stellar evolution models and/or BBH formation channels.
   }


   \authorrunning{C. Sgalletta et al.}
    
   \maketitle

\section{Introduction}

The advent of gravitational wave (GW) detections has significantly expanded our understanding of stellar-mass black holes (BHs) and neutron stars (NSs,  \citealt{Abbott2016b, Abbott2016a, Abbott2017b, Abbott2017a, Abbott2021gwtc21,  Abbott2023gwtc3, Abbott2024gwtc21}). LIGO--Virgo--KAGRA (LVK) data offers valuable insights into the  mass function of binary BHs (BBHs), challenging existing stellar evolutionary models with observations of BHs within the lower and upper mass gaps \citep[e.g.,][]{LVK2020_gap_1, LVK2020_PIgap, LVK2020_gap_2, LVK2024_gap}.
Moreover, as the dataset grows, a possible evolution of BBH properties with redshift $z$ is beginning to surface, even with current observations limited to  $ z \lesssim 1.5$ \citep{Biscoveanu2022, Abbott2023population, Nitz2023, Ray2023, Callister2024, Rinaldi2024}. 
Next-generation GW detectors will further advance our ability to explore the Universe, allowing us to probe BBH mergers at redshifts up to $ z \lesssim 100$ \citep{Hall2019, Kalogera2021, Branchesi2023}.  
Specifically, the study of the BBH merger rates as a function of redshift can help us to test and distinguish among different formation channels of such objects \citep[e.g.,][]{Belczynski2002, Dominik2013,Neijssel2019, Baibhav2019, Mapelli2020, Mapelli2021, Zevin2021, Broekgaarden2021, Mandel2022, Belczynski2022, VanSon2022, Broekgaarden2022}.
The merger rate density is the result of the interplay between binary star evolution and the history of cosmic star formation \citep{Lamberts2018, Santoliquido2020, Broekgaarden2021, Santoliquido2022, Boesky2024MRD, Boesky2024Delay, Chruslinska2024}.

Most previous work relies on averaged star formation rate (SFR)  density and metallicity distribution relations \citep[e.g.,][]{VanSon2022, Santoliquido2020, Neijssel2019}. This assumption makes it extremely straightforward to calculate the merger rate density evolution, but contains several approximations in the description of the star formation process and metallicity evolution in the Universe. Such assumptions affect both the shape and absolute values of the estimated merger rates, as shown by several authors  \citep{Boco2019,Boco2021, Santoliquido2022, Briel2022}. 
Current predictions for BBH merger rates span over three orders of magnitude \citep[see e.g.,][]{Mandel2022}. However, emerging trends indicate that state-of-the-art binary population synthesis models tend to predict higher rates than those inferred by LVK, particularly when incorporating the latest observational data to model the Universe \citep[e.g.,][]{Neijssel2019, Broekgaarden2021, Broekgaarden2022, Santoliquido2022, Srinivasan2023, Boesky2024MRD}.

This discrepancy is largely driven by the interplay of merger efficiency (i.e., the number of BHs that merge per total stellar mass) and metal-poor star formation \citep{Belczynski2010, Mapelli2019, Chruslinska2019, Neijssel2019, Tang2020, Santoliquido2021, Broekgaarden2021, Chruslinska2024}. In fact, the merger efficiency of BBHs is 1--4 orders of magnitude higher in metal-poor than in metal-rich progenitor systems (\citealt{Klencki2018,Giacobbo2018, Spera2019, Broekgaarden2021, Iorio2023, Vanson2025}). Hence, the merger rate density evolution should be largely boosted by star formation in metal-poor regions, often associated with low-mass galaxies. As discussed by \citet{Chruslinska2021, Boco2021, Chruslinska2024} the large uncertainties concerning the metallicity evolution in the Universe and the role of metal-poor galaxies might explain a fraction of the overall discrepancy.

Additionally, there is growing evidence that current models  provide a poor description of the common envelope (CE) phase and binary evolution via stable mass transfer episodes might be more important than previously assumed for the formation of BBHs (\citealt{Neijssel2019, Bavera2021, Marchant2021, Gallegos2021, Olejak2021}). Specifically, the stability criteria commonly adopted in population synthesis codes appear to produce unstable mass transfer for a wider range of parameters compared to detailed stellar modeling \citep{Ge2010, Ge2015, Ge2020, Marchant2021}. This in turn may lead to an overestimate of the BBH merger rates \citep{Gallegos2021}. 

This work is motivated by the considerations above. Firstly, we aim to assess the impact of the metal-dependent SFR on the BBH merger rates. To this end, we adopt and upgrade a detailed Universe model built from observational scaling relations \citep{Boco2019, Boco2021} with \textsc{galaxy$\mathcal{R}$ate} \citep{Santoliquido2022}. We test several SFR -- galaxy mass relations, fundamental metallicity relations, and we explore the impact of low-mass galaxies. Secondly, we investigate the relative contribution of different formation channels to the BBH merger rates.
We show that our state-of-the-art data-driven approach produces BBH merger rates consistently above the LVK inferred range. Such a discrepancy is robust against variations in the metal-dependent SFR model. In particular, low-mass galaxies ($M_\ast<10^8 \mathrm{M}_{\odot}$) do not contribute more than a factor of $\sim 2$ to the overall BBH merger rate density and their relative contribution strongly depends on the assumed SFR -- galaxy mass relation. 

The paper is organized as follows: Section \ref{sec:method} describes the methodology and the main codes used in this work. Section \ref{sec:universecomparison} provides an overview and a comparison of the adopted Universe models. Sections \ref{sec:smallgalaxies} and \ref{sec:mrd} present the results of the cosmic merger rate density as a function of the cosmic model. Section \ref{sec:channels} details the merger rate density results for the different BBH formation channels. Finally, we discuss the implications of our results in Section~\ref{sec:discussion}. We summarize our findings in Section~\ref{sec:summary}. 

\section{Method}\label{sec:method}
We investigated the BBH merger rate density varying five model parameters, for a total of 360 parameter combinations.
We used the code \textsc{galaxy$\mathcal{R}$ate} \citep{Santoliquido2022, Santoliquido2023} to couple binary compact object populations to a synthetic Universe built from observational scaling relations \citep{Mapelli2017, Artale2020, Santoliquido2021, Santoliquido2022, Santoliquido2023}. The methodology adopted is described below.

\subsection{Population-synthesis code \textsc{sevn}}\label{sec:sevn}

The stellar evolution for N-body code (\textsc{sevn}) \footnote{  In this work, we used the \textsc{sevn} version V 2.10.1 (commit \href{https://gitlab.com/sevncodes/sevn/-/tree/22c923637bd6a7fe7546bf4585455f9a1d97b71c}{22c9236}). \textsc{sevn} is publicly available at the gitlab repository \url{https://gitlab.com/sevncodes/sevn}} evolves stellar properties through on-the-fly interpolation of pre-computed stellar tracks \citep{Spera2017, Spera2019, Mapelli2020,Iorio2023}. In this work, we used  the stellar tracks evolved with \textsc{parsec} \citep{Bressan2012, Costa2019b, Costa2019a, nguyen2022}. 
\textsc{sevn} models binary evolution with semi-analytical prescriptions. We refer to \citet{Iorio2023} for a detailed description of the code.

For the purpose of this work, we adopted the default \textsc{sevn} settings as described by \cite{Iorio2023}. We assumed the \textit{DelayedGauNS} supernova (SN) prescription, with  BH masses  determined according to the \textit{delayed} model by \cite{Fryer2012}. 
In our fiducial model, natal kicks are sampled as  \citet[][hereafter, GM20]{Giacobbo2020}. Specifically, kick magnitudes are drawn from a Maxwellian distribution with one-dimensional root mean square $\sigma = 265 \mathrm{km s}^{-1}$ \citep{Hobbs2005} and then re-scaled by a factor $\propto M_{\rm ej} / M_{\rm rem}$, where $M_{\rm ej}$ and $M_{\rm rem}$ are the mass of the ejecta and the compact remnant, respectively. We varied the kick prescription testing also a model
in which BHs get the same natal kicks as derived for Galactic neutron stars \citep[H05, ][]{Hobbs2005}. 

One of the biggest uncertainties in population-synthesis codes are the mass transfer stability criteria. The most common approach relies on the evaluation of the mass ratio $q= M_{\rm d}/M_{\rm a}$ and its comparison to a critical value $q_{\rm crit}$, depending on the evolutionary  phases of the stars involved. Here, $M_{\rm d}$ is the mass of the donor star and $M_{\rm a}$ of the accretor. If $q>q_{\rm crit}$ the mass transfer episode is considered unstable on a dynamical timescale, triggering a CE \citep{Ivanova2013}. Our fiducial model assumes the \textsc{sevn} default $q_{\rm crit}$ prescription (QCRS), that is the same as described by \cite{Hurley2002}, but for one difference: we assume that mass transfer with donor stars in the main sequence and Hertzsprung gap is always stable \citep{Iorio2023, Sgalletta2023}.

\textsc{sevn} handles  CE evolution with the standard $( \alpha \,{}\lambda )-$formalism \citep{Webbink1984, Hurley2002}. We adopted the $\lambda$ parameters specified by \cite{Claeys2014}. Our fiducial model assumes a CE efficiency parameter  $\alpha=1$ (this corresponds to assuming that the change in the orbital energy of the core contributes to unbind the envelope with an efficiency of 100\%). However, we also explored different $\alpha$ parameters within our models, as detailed in section \ref{sec:parameters}. 

\subsection{Initial conditions} \label{sec:initialconditions}

We sampled the masses of the primary star ($M_1$) from a Kroupa initial mass function \citep{kroupa2001}, in a range between $5$ and $150\,{} \text{M}_\odot$:
\begin{equation}
\mathcal{F}(M_1) \propto M_{1}^{-2.3}.
\end{equation}
We drew the secondary star mass ($M_2$) within the range $[2.2, 150]\,{} {\rm M}_{\odot}$ from the observational distribution on $q=M_2/M_1$ \citep{Sana2012}:
\begin{equation}
\mathcal{F}(q) \propto q^{-0.1},
\end{equation}
with $\max \left( \frac{2.2\,{} \text{M}_\odot}{M_{1}}, 0.1 \right)\leq q \leq 1 $.
The orbital periods and eccentricities were also drawn from the distributions derived by \cite{Sana2012}:
\begin{equation}
\mathcal{F}(P_{\rm orb}) \propto (\log P_{\rm orb})^{-0.55},
\end{equation}
with $0.15 \leq \log \left(P_{\rm orb}/{\rm d}\right) \leq 5.5$,  
and
\begin{equation}
\mathcal{F}(e) \propto e^{-0.42},
\end{equation}
with $0\leq e \leq 1-\left(P/2 \ \mathrm{days}\right)^{-2/3}$, following the correction by \cite{Moe2017}. These distributions are fits to observations of a sample of massive binary stars in open clusters. 

We sampled $10^{7}$ binaries and used them as initial conditions for 12 metallicities: $Z=0.0002$, 0.0004, 0.0008, 0.0012, 0.0016, 0.002, 0.004, 0.006, 0.008, 0.012, 0.016, and 0.02. The total simulated mass for each metallicity is thus $M_{\rm sim} = 2.2\times 10^8 \mathrm{M}_{\odot}$. 

\subsection{Formation channels} \label{sec:formationchannels}
We distinguish four main formation channels of BBHs, following the classification by \cite{Broekgaarden2022} and \cite{Iorio2023}. 
\begin{itemize}
    \item Channel I: The systems experience a stable mass transfer episode before the first compact object formation and only after they evolve through at least one CE phase.
    \item Channel II: The systems interact only via stable mass transfer episodes. 
    \item Channel III: At the time of the first compact remnant formation the system has already experienced at least one CE phase and the system is composed of one H-rich star and one without the H envelope. 
    \item Channel IV: Similarly to Channel III, the systems experience at least one CE before the first compact remnant formation, but in this case, at the time of compact remnant formation both stars have lost their H envelopes. 
\end{itemize}

\subsection{Observational scaling relations} \label{sec:osr}

\begin{figure}
    \centering
    \includegraphics[width=.5\textwidth]{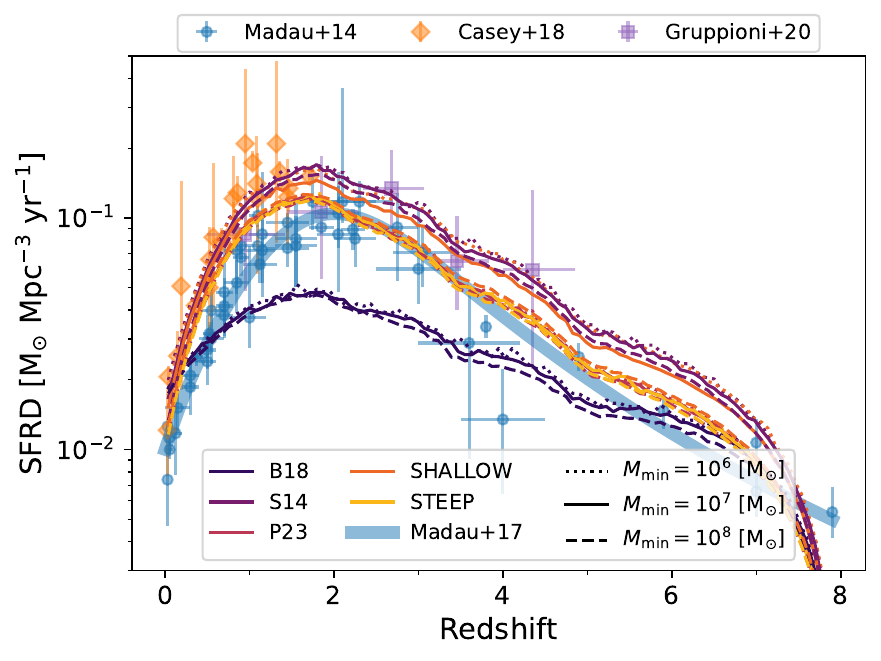}
    \caption{SFR density as a function of redshift for different SFR-$M_\ast$ relations (different colors). The different linestyles show the results by cutting the galaxy main sequence at different minimum galaxy masses, $M_{\rm min} = 10^6 \mathrm{M}_{\odot}$ (dotted), $10^7 \mathrm{M}_{\odot}$ (solid) and $10^8 \mathrm{M}_{\odot}$ (dashed). The wide blue line shows the fit by \citet{Madau2017}. The markers show observed data points: the blue dots are are taken from \citet{Madau2014}, the orange diamonds are from \citet{Casey2018}, and the purple squares are taken from \citet{Gruppioni2020}.}
    \label{fig:SFRDs}
\end{figure}

In order to model the Universe and populate it with  BBHs, we adopted the code \textsc{galaxy$\mathcal{R}$ate} \footnote{ The code \textsc{galaxy$\mathcal{R}$ate} can be found at \url{https://gitlab.com/Filippo.santoliquido/galaxy_rate_open}} \citep{Santoliquido2022}. We generated a set of star-forming galaxies from observational scaling relations for an array of formation redshifts, sampling their masses $M_{*}$, SFRs,  and average metallicities. Here below, we summarize such scaling relations. 

We adopted the star-forming galaxy stellar mass function derived by \cite{Chruslinska2019}. For each formation redshift, we kept the comoving volume fixed $V \sim \left( 100 \,{}\mathrm{cMpc}\right)^{3}$ and  sampled a number of star-forming galaxies $\mathcal{N}_{\rm gal} \left( z_{\rm form} \right)$ proportional to the galaxy number density. 
We sampled galaxy masses in the range $\left[ M_{\rm min}, 10^{12} \mathrm{M}_{\odot}\right]$. The minimum galaxy mass is a free parameter in our models, we tested different values: $M_{\rm min}= 10^6$, $10^7$, and  $10^8 \mathrm{M}_{\odot}$. 
We assigned a SFR to every sampled galaxy following a double lognormal distribution, with a first peak centered on the galaxy main sequence, and a second one, on the starbust sequence \citep{Boco2021}. The parameters adopted are described in Appendix \ref{sec:osr_appendix}. 

The slope of the galaxy main sequence is highly debated, especially moving toward low-mass galaxies ($\lesssim 10^{8.5} \mathrm{M_{\odot}}$) and high redshifts. For this reason, we tested several galaxy main sequence definitions. Specifically, we considered the definitions given by \citet[][hereafter S14]{Speagle2014}, \citet[][hereafter B18]{Boogaard2018}, and \citet[][hereafter P23]{Popesso2023}. 
In the following equations, the SFR is expressed in $\mathrm{M}_{\odot} \mathrm{yr}^{-1}$, and all masses are given in solar units. 
\citetalias{Speagle2014} define the galaxy main sequence as:
\begin{equation}
    \log \mathrm{SFR} \left( M_\ast, t\right) = (0.84 - 0.026 t)\log M_\ast - 6.51 - 0.11 t,
\end{equation}
where $t$ is the age of the Universe in Gyr. 
The definition by \citetalias{Boogaard2018} is as follows:
\begin{equation}
    \log \mathrm{SFR} \left( M_\ast, z\right) = 0.83 \log \left( \frac{M_\ast}{ M_0} \right) - 0.83 + 1.74 \log \left( \frac{1+z}{ 1+ z_0} \right),
\end{equation}
where $M_0 = 10^{8.5} \mathrm{M}_{\odot}$ and $z_0 = 0.55$ are the median values of the data. 
The fit by \citetalias{Popesso2023} is:
\begin{equation} \label{eq:p23}
    \log \mathrm{SFR} \left( M_\ast, t\right) = a_0 + a_1 t - \log \left[ 1 + \left( M_\ast / 10^{a_2 + a_3 t}\right)^{-a_4} \right],
\end{equation}
with $a_0 = 2.693 \pm 0.012$, $a_1=-0.186 \pm 0.009$, $a_2=10.85 \pm 0.05$, $a_3=-0.0729 \pm 0.0024$ and $a_4=0.99 \pm 0.01$.

Moreover, we built two additional parametric models that explore different slopes of the main sequence relation at low-masses ($M_*<10^9$~M$_\odot$). In this range, in fact, the data is scarce \citep[see e.g.,][]{Popesso2023} and very uncertain. With this approach, we were able to bracket all possible behaviors of the galaxy main sequence at low-masses. 
For these models, we cut the fit in \citetalias{Popesso2023} at $10^9 \mathrm{M}_{\odot}$, and adopted a constant slope $a$ across cosmic times for lower galaxy masses:
\begin{equation} \label{eq:slope}
\log \mathrm{SFR}\left( M_\ast, t\right) =
\begin{cases}
a \log M_\ast + b & \text{for }  M_{*} < 10^9 \;\mathrm{M}_{\odot} \\
\text{Eq. \ref{eq:p23}} & \text{for }  M_{*} \geq 10^9 \; \mathrm{M}_{\odot}.
\end{cases}
\end{equation}
We assumed $a=0.5$, for the SHALLOW model, and $a=1$ for the STEEP model. We decided such values for the SHALLOW (STEEP) slope to always keep it as the  shallowest (steepest) main sequence slope at all redshifts in the low-mass galaxies' tail. The continuity assumption between the two equations sets the value of the second parameter $b$. See Appendix \ref{sec:mainsequence} for more details on the main sequence definitions adopted.

By integrating the galaxies SFR for each redshift bin, we get the SFR density as a function of redshift, shown in Fig.~\ref{fig:SFRDs}. We compare the predicted SFR density for all the adopted galaxy models. The model by \citetalias{Boogaard2018}  predicts a lower SFR density with respect to the observational data. This is likely a result of the data sample utilized in B18 reaching up to $z\sim 0.9$, wheareas here we extrapolate their  relation to much higher redshifts. All the other galaxy main sequence models predict SFR densities that are consistent with UV and far-IR data \citep[see][and references therein]{Madau2014}.
The fit by \citetalias{Speagle2014} adopts galaxy main sequence data, with galaxy stellar masses within $10^{9.5} < M_\ast < 10^{11.5} \mathrm{M}_{\odot}$. \citetalias{Popesso2023} expand the work by \citetalias{Speagle2014}, by incorporating the most updated and recent galaxy main sequence data, encompassing the widest range of galaxy stellar mass ($10^{8.5} < M_\ast < 10^{11.5} \mathrm{M}_{\odot}$) and redshift ($0<z<6$) available to date. Thus, we adopted equation \ref{eq:p23} \citepalias{Popesso2023} as our fiducial SFR--$M_\ast$ relation. 
Moreover, varying the lower mass limit of the galaxy main sequence (represented by different line styles) does not significantly impact the SFR density. This is not surprising, as the SFR from low-mass galaxies is suppressed by the galaxy main sequence relation. Therefore, even if these small galaxies are more numerous, the contribution to the total SFR density is negligible compared to the contribution from higher mass galaxies ($M_{\ast} \geq 10^8 M_{\odot}$).

For each galaxy, we sampled an average metallicity based on the fundamental metallicity relation, that links the average metallicity of a galaxy with its stellar mass and SFR \citep{Mannucci2010,Maiolino2019}. In this work, we compared three different definitions of the fundamental metallicity relation, corresponding to the fits by \citet{Mannucci2011}, \citet{Curti2020}, and \citet{Andrews2013}. 
The recent work by \citet{Nakajima2023} compares known models of mass--metallicity relations with a comprehensive, up-to-date sample of James Webb Space Telescope (JWST) observations of galaxies at  redshift $4$ to $10$. Their data (see their figure 12) show the fundamental metallicity relation by \cite{Andrews2013} to best fit these new data up to $z\sim 8$. For this reason, we adopted the fundamental metallicity relation by \cite{Andrews2013} as our fiducial model.
We assumed that the metallicity is distributed following a lognormal function within each galaxy, with a scatter $\sigma_{\rm Z}=0.14$. 
Such an observed scatter (amounting to $\lesssim 0.1$ dex in the local Universe and $\lesssim 0.2$ dex at high redshift) is small since it refers to galaxies at a given redshift with the same stellar mass and SFR. However, the metallicity at a given $z$ is substantially more dispersed over the whole galaxy populations; for instance, averaging the fundamental metallicity relation over stellar mass (via the stellar mass function) yields a mass-metallicity relation characterized by a large and asymmetric dispersion of $\gtrsim 0.35$ dex, in agreement with observational findings \citep{Chruslinska2021, Boco2021}. 
We describe the adopted observational scaling relations in more detail in Appendix  \ref{sec:osr_appendix} and review all the models' parameters in Section \ref{sec:parameters}.

We also tested a different approach based on the average evolution of the cosmic SFR and metallicity with \textsc{cosmo$\mathcal{R}$ate} \citep{Santoliquido2020}.
In this case, the cosmic SFR density $\psi(z)$ and the average metallicity evolution $\mu(z) = \langle Z/Z_{\odot} \rangle$ are modeled using the fitting formulae provided by \citet{Madau2017}:
\begin{equation}
    \psi(z) = 0.01 \frac{ \left(1+z \right)^{2.6}}{1+ [(1+z)/3.2]^{6.2}} \mathrm{M}_{\odot} \mathrm{Mpc}^{-3} \mathrm{yr}^{-1},
\end{equation}
\begin{equation}
    \log \mu(z) = 0.153 - 0.074 z^{1.34}.
\end{equation}
We spread the metallicities around this average value following a normal distribution with standard deviation $\Tilde{\sigma}_{\rm Z}$:
\begin{equation}
    p(Z \vert z) = \frac{1}{\sqrt{2 \pi \Tilde{\sigma}_{\rm Z}^2}} \exp \left[ -\frac{[\log (Z/Z_{\odot}) - \langle 
 \log (Z/Z_{\odot}) \rangle]^{2}}{2 \Tilde{\sigma}_{\rm Z}^{2}}\right], 
\end{equation}
where $\langle \log{Z/Z_{\odot}}\rangle = \mu(z) - \ln{10} \Tilde{\sigma}_{\rm Z}^{2}/2$.
With this simplified approach, we can change the spread of metallicity at a given redshift just by tuning the parameter $\Tilde{\sigma}_Z$.

\subsection{Merger rate density} \label{sec:mrd_eq}

We evaluate the merger rate density with \textsc{galaxy$\mathcal{R}$ate} as 
\citet{Santoliquido2022}:
\begin{equation} \label{eq:mrd}
    \mathcal{R}(z) = \frac{1}{V^3} \int_{z_{\rm max}}^{z} 
    \left[ \sum_{ \rm{i} = 1}^{\mathcal{N}_{\rm gal}(z')}  \int_{Z_{\rm min}}^{Z_{\rm max}} \mathcal{S}_{\rm i} \left( z', Z \right) \mathcal{F}\left( z', z, Z \right) dZ \right] \frac{dt(z')}{dz'} dz',
\end{equation}
where we sum over all of the formation galaxies, $\mathcal{N}_{\rm gal}(z')$, at redshift $z'$ (i.e., the galaxies where the binary systems form at $z'\geq z$), and then integrate over the redshift between the BBH merger redshift $z$ (i.e., the redshift at which the BBH merges) and the maximum considered formation redshift $z_{\rm max} = 8$. 
For the $i-$th galaxy $\mathcal{S}_{\rm i} \left( z', Z \right) = \psi_{\rm i} \left( z'\right) p_{\rm i}\left( Z \vert z' \right)$ , where $\psi_{\rm i} \left( z'\right)$ is its SFR and $p_{\rm i}\left( Z \vert z' \right)$ is its metallicity distribution at a given redshift $z'$. 
The second term in the integral accounts for the distribution of binaries in our catalogs:
\begin{equation} \label{eq:catalogs}
    \mathcal{F}\left( z', z, Z \right) = \frac{1}{M_{\rm sim}} \frac{\mathcal{N}_{\rm BBH} \left( z', z, Z \right)}{dt} f_{\rm bin} f_{\rm corr},
\end{equation}
where $\mathcal{N}_{\rm BBH} \left( z', z, Z \right) / dt$ is the rate of BBHs progenitors that form at redshift $z'$ and merge at redshift $z$, for a given metallicity $Z$. The factor $f_{\rm bin}=0.5$  corrects for the fraction of binaries, and $f_{\rm corr}=0.251$ takes into account the cut at lower masses from our initial sampling conditions (see Section \ref{sec:initialconditions}).

\begin{figure*}
    \centering
    \includegraphics[width=\linewidth]{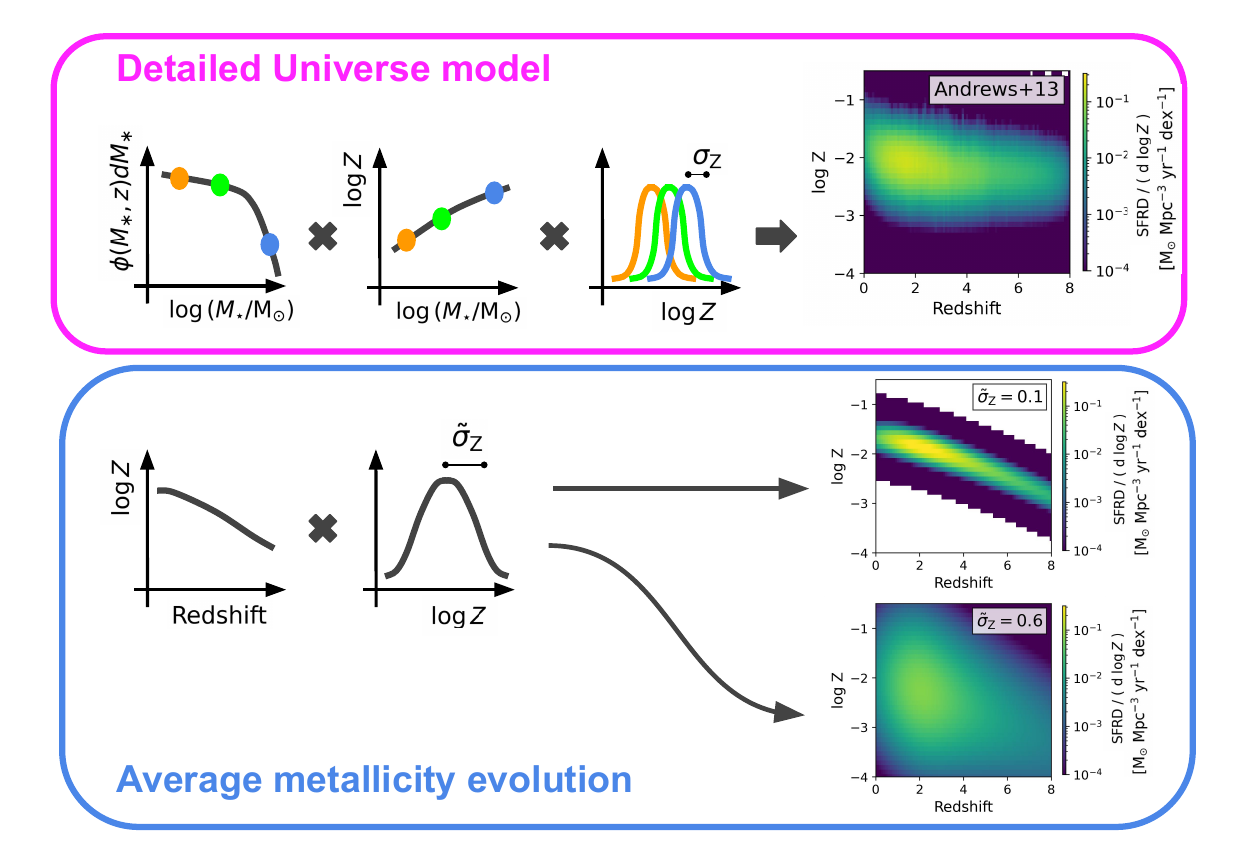}
    \caption{ Schematic overview of the different metal-dependent SFR models adopted here. Upper panel: overview of the detailed Universe model, based on observational scaling relations. Lower panel: overview of the model based on the average SFR and metallicity evolution across cosmic time. The plots on the right illustrate the metallicity distributions as a function of redshift for our different prescriptions. For the average metallicity evolution model, we show the distributions assuming both $\Tilde{\sigma}_{\rm Z} = 0.1$ and $0.6$.}
    \label{fig:model_simple}
\end{figure*}

\subsection{Summary of model parameters} \label{sec:parameters}

\begin{table}
\caption{Summary of the grid of parameters varied in our study.}
\centering
\begin{tabular}{c|c|c}
 & Parameter & Values \\
\hline
\hline
\multirow{3}{*}{Binary  evolution}   
                                    & $\alpha$ & 0.5, \textbf{1}, 3, 5 \\
                                    & \multirow{2}{*}{SN kick}  & \textbf{GM20}\tablefootmark{a} \\
                                                                && H05\tablefootmark{b} \\
\hline
\multirow{9}{*}{Galaxy modeling}   & \multirow{5}{*}{SFR--$M_\ast$} & B18\tablefootmark{c} \\
                                                                    && S14\tablefootmark{d} \\
                                                                    && \textbf{P23}\tablefootmark{e} \\
                                                                    && STEEP (eq. \ref{eq:slope}, a=1) \\
                                                                    && SHALLOW (eq. \ref{eq:slope}, a=0.5) \\
                                    & $M_{\rm min}$ [M$_\odot$] & $10^6$, $10^7$, $10^8$ \\
                                    & \multirow{3}{*}{FMR}          & Mannucci+11\tablefootmark{f} \\
                                                                    && \textbf{Andrews+13}\tablefootmark{g} \\
                                                                    && Curti+20\tablefootmark{h} \\
\hline
\hline
\end{tabular}
\label{tab:models}
\tablefoot{The parameters highlighted in \textbf{bold} face are adopted in our fiducial model.\\
\tablefoottext{a}{\citet{Giacobbo2020}}
\tablefoottext{b}{\citet{Hobbs2005}}
\tablefoottext{c}{\citet{Boogaard2018}}
\tablefoottext{d}{\citet{Speagle2014}}
\tablefoottext{e}{\citet{Popesso2023}}
\tablefoottext{f}{\citet{Mannucci2011}}
\tablefoottext{g}{\citet{Andrews2013}}
\tablefoottext{h}{\citet{Curti2020}}
}
\end{table}

We ran our \textsc{sevn} and \textsc{galaxy$\mathcal{R}$ate} simulations over an extensive grid of model parameters, exploring both binary evolution and galaxy modeling uncertainties. We summarize the grid of models in Table~\ref{tab:models}. Specifically, we ran sets of simulations with \textsc{sevn} varying the CE efficiency parameter $\alpha = 0.5$, $1$, $3$, and $5$. 
We consider two natal-kick models: \citetalias{Giacobbo2020}, and \citetalias{Hobbs2005}, as defined in Section~\ref{sec:sevn}. 

As for the synthetic galaxy models, we vary the minimum galaxy mass assuming  $M_{\rm min} = 10^6$, $10^7$ and $10^8 \,{}\mathrm{M}_{\odot}$.
We explore the galaxy main-sequence definitions from \citetalias{Speagle2014}, \citetalias{Boogaard2018}, \citetalias{Popesso2023}, as well as our SHALLOW and STEEP models (equation \ref{eq:slope} with $a=0.5$ and $a=1$, respectively). For the fundamental metallicity relations, we compare the fits by \cite{Mannucci2011}, \cite{Andrews2013}, and \cite{Curti2020}. 
Our fiducial model incorporates the most updated main sequence definition using \citetalias{Popesso2023}, the fundamental metallicity relation by \citet{Andrews2013}, and a minimum galaxy stellar mass $M_{\rm min} = 10^7 \mathrm{M}_{\odot}$.

\section{A comparative analysis of Universe models} \label{sec:universecomparison}
Figure \ref{fig:model_simple} provides a schematic overview and comparison of the two approaches to  modeling a synthetic Universe. The detailed method offers several improvements over the conventional averaged approach.
We note that $\Tilde{\sigma}_{\rm Z}$, used in the averaged model, and $\sigma_{\rm Z}$, used in the detailed model, have two different physical meanings. $\sigma_{\rm Z}$ represents the metallicity dispersion around the FMR (i.e., at fixed galaxy stellar mass and SFR), that is then convoluted with the distribution of galaxy masses and SFR to obtain the spread in metallicity at a given redshift. Therefore, even if we assume the distribution around the FMR to be log-normal and with a small scatter $\sim 0.14$, the overall distribution of metallicities is asymmetric and can have a large scatter, driven by the observed galaxy distribution at that specific redshift.

In contrast, models relying on averaged metallicity evolution require the total scatter ($\Tilde{\sigma}_{\rm Z}$) to artificially encompass the full metallicity distribution at a given redshift. For this reason typical $\Tilde{\sigma}_{\rm Z}$ values of $\sim 0.5$ are required. In this work, we adopt $\Tilde{\sigma}_{\rm Z}=0.6$ as fiducial value and $\Tilde{\sigma}_{\rm Z}=0.1$, even if not realistic, to demonstrate the effect of reducing the low metallicity tail. Moreover, the distribution adopted is symmetric around the average metallicity. Therefore, even if the spread in the averaged model is large enough to mimic the amount of SFR at low-metallicities of the detailed approach, it inevitably produces an unrealistic high-metallicity tail, that is not supported by observations \citep[see e.g.,][]{Chruslinska2024}.

We note that the distribution of metallicities of the averaged models could be improved by adopting a skewed distribution around the mean value \citep[see e.g.,][]{VanSon2023, Fishbach2025}. Such models allow for a more realistic representation of metallicities, when properly fitted to the data. We do not consider them here as they would ultimately lead to results consistent with our detailed approach.

\section{Results}
\subsection{Merger efficiency and metallicity}

\begin{figure}
    \centering
    \includegraphics[width=.5\textwidth]{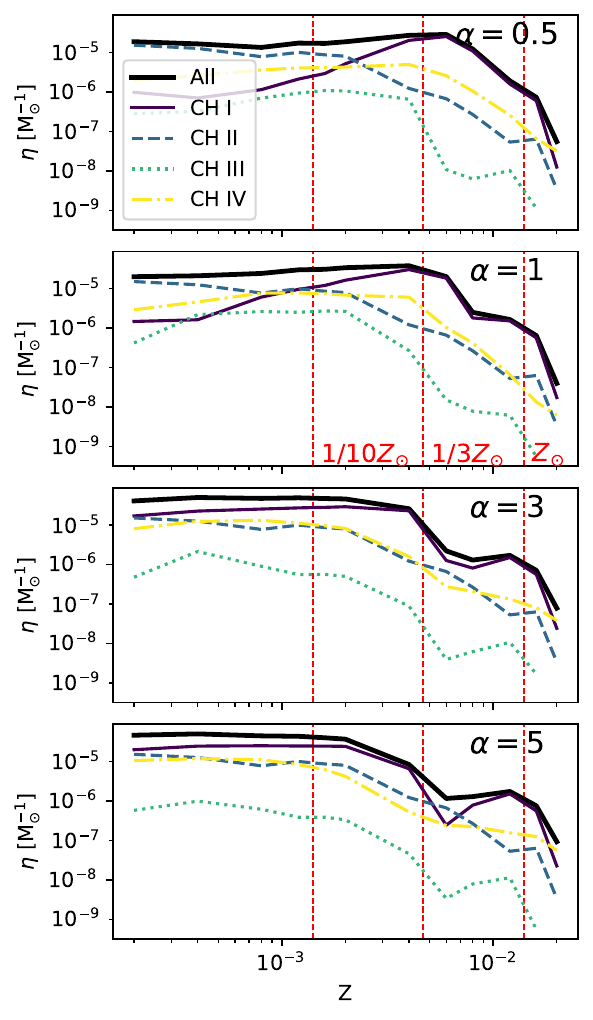}
    \caption{Merger efficiency $\eta$ for BBHs as a function of metallicity $Z$. From top to bottom $\alpha=0.5$, 1, 3, 5. The different colors refer to   formation channels I-IV (Section \ref{sec:formationchannels}). 
    The vertical red lines highlight metallicity values of $Z=1/100$, $1/10$, and $1/3$~Z$_{\odot}$.}
    \label{fig:Efficiency}
\end{figure}

\begin{figure*}
    \centering
    \includegraphics[width=\textwidth]{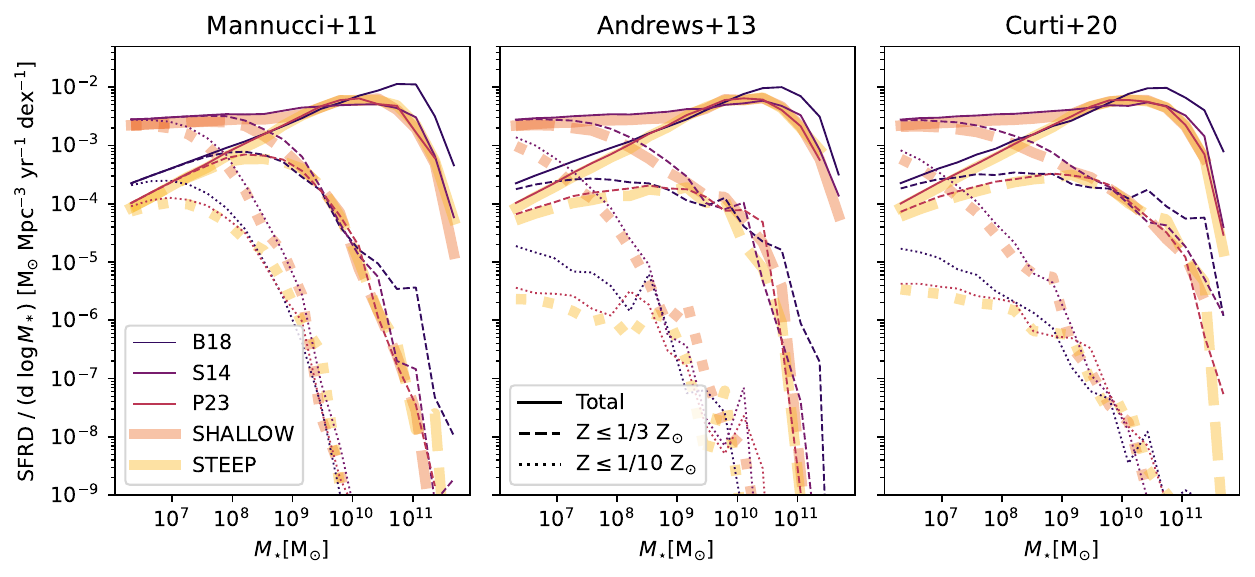}
    \caption{Contribution of galaxies of mass $M_\ast$ to the SFR density evaluated for different metallicity thresholds at $z=0$. The x axis represents the galaxy stellar mass $M_\ast$, while the lines show the total contribution to the SFR density from all galaxies in a given stellar mass bin, with a metallicity $Z \leq Z_{\rm th}$. The different line styles refer to different metallicity thresholds $Z_{\rm th}$; solid lines:$Z_{\rm th} =  1/3 {\rm Z}_{\odot}$, dashed: $Z_{\rm th} =   1/10 {\rm Z}_{\odot}$, and dotted:$Z_{\rm th} =   1/100 {\rm Z}_{\odot}$. The different colors represent models with different SFR--$M_\ast$ relations, blue: B18, violet: S14, pink: P23, orange: SHALLOW, yellow: STEEP. Each panel shows the results for a different FMR, from left to right: \citet{Mannucci2011, Andrews2013, Curti2020}. }
    \label{fig:SFRbymass}
\end{figure*}

The BBH merger efficiency (i.e., the number of BHs that merge per total initial stellar mass)  
depends on metallicity, being higher for metal-poor stars \citep{Dominik2012, Giacobbo2018, Klencki2018}. With \textsc{sevn}, we find an abrupt drop of three orders of magnitude in the BBH merger efficiency, above $\sim{1/3\,{}Z_{\odot}}$ (Fig.  \ref{fig:Efficiency}). The drop mildly depends on the $\alpha$ parameter: it is close to $1/3$~Z$_{\odot}$ for $\alpha \leq 1$ and shifts to lower metallicities ($\sim 1/10$~Z$_{\odot}$) for $\alpha = 3$ or $5$. 
For this reason, the fraction of SFR at low metallicity is crucial for determining the BBH merger rate. 


\subsection{The impact of low-mass galaxies} \label{sec:smallgalaxies}

Figure \ref{fig:SFRbymass} shows the relative contribution of different galaxy masses to the total SFR density and to the SFR density below a certain $Z$ threshold. We consider two reference metallicity thresholds: $Z \leq 1/3 \,{}{\rm Z}_\odot$ and $\leq{}1/10 \,{}{\rm Z}_{\odot}$. We compare the results for different main-sequence definitions. In Figure \ref{fig:SFRbymass}, we emphasize the SHALLOW and STEEP models (wider lines),
because they represent the extreme cases:
the former has the shallowest trend, whereas the latter has the steepest decrease in the SFR density at low galaxy masses.  This Figure shows the results for galaxies at $z=0$; however, the following considerations  hold  for every considered redshift.

We see from Fig.~\ref{fig:SFRbymass} that only with the shallowest SFR--$M_\ast$ relations the contribution from low-mass galaxies is important, for metallicities $Z \leq 1/3$~Z$_{\odot}$. 
For the steepest galaxy main-sequence models (e.g., the STEEP model and \citetalias{Popesso2023}), most of the SFR for metallicities below $1/3$~Z$_{\odot}$ originates from galaxies with masses in the range of $(10^9 - 10^{10}) \mathrm{M}_\odot$ for the fundamental metallicity relations by \cite{Andrews2013} and \cite{Curti2020}, while the contribution from $10^6 \mathrm{M}_\odot$ galaxies is about half. The peak in SFR shifts toward galaxy masses of $(10^8 - 10^9) \mathrm{M}_\odot$ adopting the models by \cite{Mannucci2011}. On the other end, shallower main sequence slopes (e.g., the SHALLOW model, \citetalias{Speagle2014}) show a similar SFR contribution from low mass galaxies up to $\sim 10^8 \mathrm{M}_{\odot}$. Model \citetalias{Boogaard2018} falls in between these two extreme cases.

If we restrict our attention to even lower metallicities ($Z<1/10$~Z$_{\odot}$), the contribution from low-mass galaxies becomes important for all the SFR--$M_\ast$ relations. As a result, for higher $\alpha$ parameters (for which the merging efficiency peaks at smaller metallicities, see Fig.~\ref{fig:Efficiency}) the contribution from low-mass galaxies becomes important, regardless of the chosen SFR--$M_\ast$ relation. 
While the trend described above is true for all the fundamental metallicity relations, we see from the different panels that the prescriptions by \cite{Curti2020} and \cite{Andrews2013} predict on average higher metallicities (Fig.~\ref{fig:SFRD_by_M}).

Several studies have shown how the contribution of starburst galaxies might increase at high redshifts compared to our current treatment. Both \citet{Bisigello2018} and \citet{Rinaldi2022} find that starburst galaxies might contribute to $\gtrsim 50 \%$ of the cosmic SFR at redshifts $z \sim 4-5$. \citet{Bisigello2018} finds that starburst galaxies might constitute the $\sim 16 \%$ of total galaxies at redshifts $2 < z < 3$. 
A higher percentage of starburst galaxies at high redshift would have two main effects. First, we would see an increase in the global SFR, as starburst galaxies have an higher SFR at fixed mass. Second, starburst galaxies specifically increase the SFR at low metallicity. In fact, because starburst galaxies have a higher SFR at fixed mass, they typically exhibit lower metallicity due to the FMR. 
Both factors contribute to an increase in the BBH merger rates, although the impact at $z=0$ is smaller, as this effect primarily affects redshifts $z \gtrsim 2-3$.

\subsection{Cosmic merger rate density}\label{sec:mrd}

\begin{figure*}
    \centering
    \includegraphics[width=\textwidth]{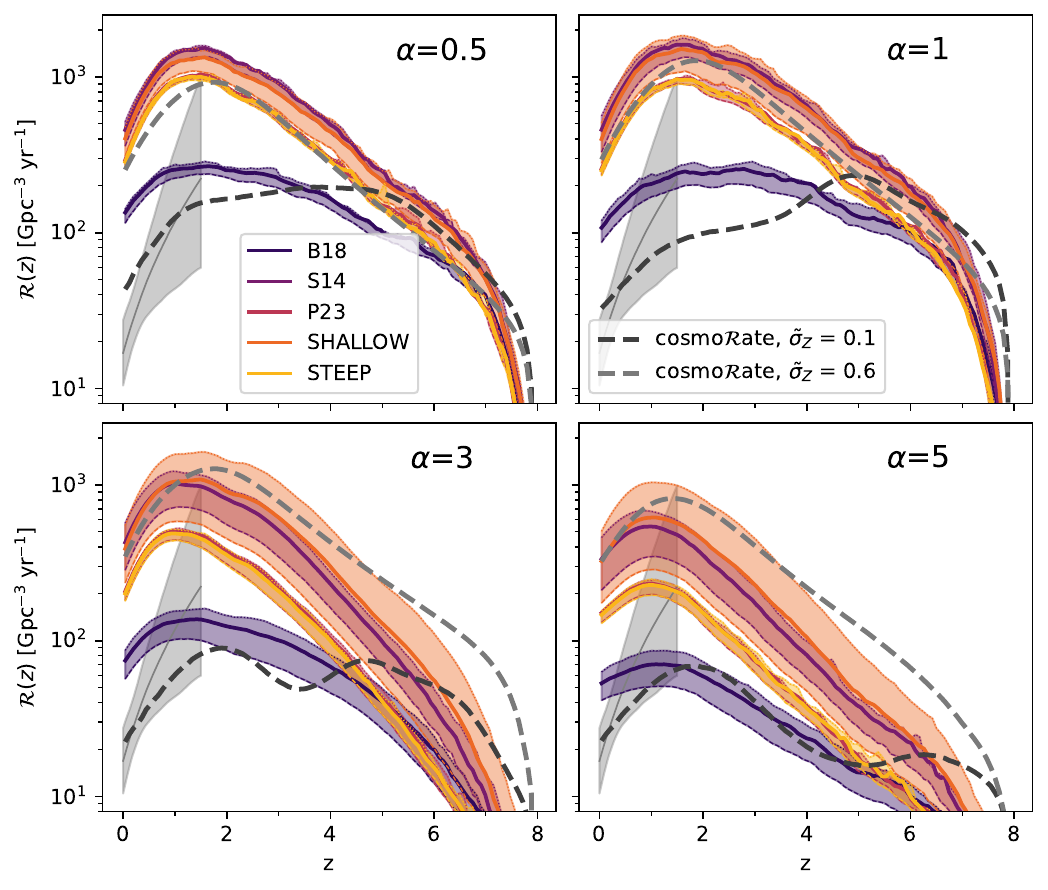}
    \caption{BBH merger rate density as a function of redshift for different observational scaling relations  (different colors) and CE $\alpha$ parameters (different panels). Each color is associated with a different SFR--$M_\ast$ relation, as summarized in the legend in the top left-hand panel. The solid lines show the model with $M_\mathrm{min}=10^7$~M$_\odot$. The shaded area encompasses the difference between $M_\mathrm{min}=10^{6} \mathrm{M}_{\odot}$ (upper boundary), and  $M_\mathrm{min}=10^{8} \mathrm{M}_{\odot}$ (lower boundary).
    All the models assume the fiducial fundamental metallicity relation by \cite{Andrews2013}.
    The thick dashed lines show the merger rate density obtained with \textsc{cosmo$\mathcal{R}$ate} (simplified model), 
    assuming the cosmic SFR density by \cite{Madau2014} and a lognormal distribution for the  metallicity  with a spread $\Tilde{\sigma}_{\rm Z}=0.1$ (dark gray) and $0.6$ (light gray). The gray-shaded regions show the MRD inferred by the LVK collaboration: $\mathcal{R}(z) \propto (1+z)^k$\citep{Abbott2023population}. 
    }
    \label{fig:MRD_all}
\end{figure*}

Figure \ref{fig:MRD_all} shows the BBH cosmic merger rate density we obtain assuming different SFR--$M_\ast$ relations, adopting the  fundamental metallicity plane by \cite{Andrews2013}. The results for the fundamental metallicity relations by \citet{Mannucci2011} and \citet{Curti2020} are shown in Appendix \ref{sec:mrd_allfmr}.
We compare our merger rate densities with the inferred fit $\mathcal{R}(z) \propto (1+z)^k$ by the LVK collaboration up to the third observing run \citep{Abbott2023population}.

All predicted BBH merger rate densities exceed the LVK estimates across all our galaxy-derived models. The B18 model yields the lowest merger rate density. This is expected as B18 results in the lowest SFR density (see Figure \ref{fig:SFRDs}). As discussed in Section~\ref{sec:osr}, this is likely due to our extrapolation of the relation in \citetalias{Boogaard2018} to high redshifts, whereas the data sample considered there is limited to $z\lesssim 0.9$.

The merger rate densities evaluated by setting different minimum galaxy masses differ at most by a factor of $\sim 2 $. Even removing all the galaxies with mass below $10^8 \mathrm{M}_{\odot}$ (lower bound of the shaded areas in Fig. \ref{fig:MRD_all}), the merger rate density is still well above the 90\% credible interval from LVK. The magnitude of the difference between assuming $M_{\rm min} = 10^6$, $10^7$ and $10^8 \mathrm{M}_{\odot}$  depends on the main sequence definition adopted. As expected, steeper SFR--$M_\ast$ relations result in smaller differences with the minimum galaxy stellar mass adopted. The differences between $M_{\rm min} = 10^6$, $10^7$, and $10^8 \,{}\mathrm{M}_{\odot}$ are bigger for $\alpha=3$ and $\alpha=5$, compared to $\alpha=1$ or $\alpha=0.5$, as expected from the results described in Section \ref{sec:smallgalaxies}). 

Figure~\ref{fig:MRD_all} also compares the merger rate density we  obtain through our host-galaxy models (i.e., with \textsc{galaxy$\mathcal{R}$ate}) with the merger-rate density calculated adopting an average SFR density and metallicity evolution of the Universe (i.e., with \textsc{cosmo$\mathcal{R}$ate}). 
The merger rate density we obtain modeling the host galaxies with observational scaling relations and the one we estimate by assuming an average SFR density agree only when we assume a large metallicity spread ($\Tilde{\sigma}_{\rm Z}=0.6$, see Sec.~\ref{sec:osr}) for the latter. 
Moreover, even assuming a large value of $\Tilde{\sigma}_{\rm Z}$, the predicted shape and steepness of the merger rate density differ in the two models, especially at high redshift and for models with $\alpha>1$.

\subsection{Formation channels} \label{sec:channels}

\begin{figure*}
    \centering
    \includegraphics[width=\textwidth]{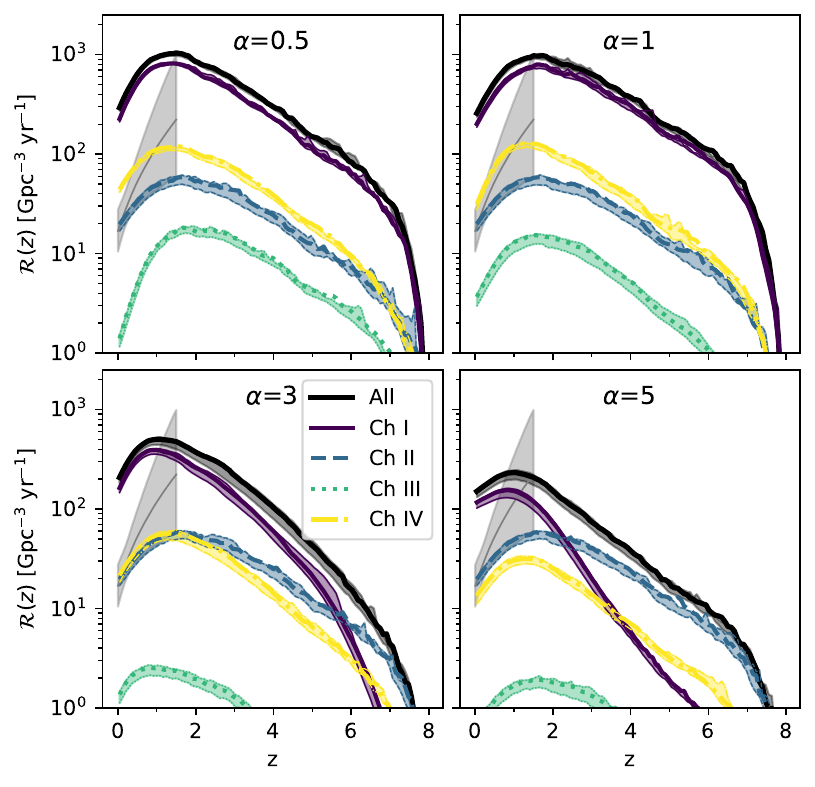}
    \caption{BBH merger rate density as a function of redshift for different BBH formation channels. The solid black lines show the merger rate density of the whole BBH population.
    The shaded area encompasses the difference between $M_\mathrm{min}=10^{6} \mathrm{M}_{\odot}$ (upper boundary), and  $M_\mathrm{min}=10^{8} \mathrm{M}_{\odot}$ (lower boundary).
    The different panels show the results obtained assuming different CE $\alpha$ parameters. All the models assume the fiducial SFR--$M_*$ relation \citetalias{Popesso2023} and the fundamental metallicity relation by \cite{Andrews2013}.
    The gray-shaded regions show the MRD inferred by the LVK collaboration \citep{Abbott2023population}.
    }
    \label{fig:MRD}
\end{figure*}

Figure~\ref{fig:MRD} shows the merger rate density for each BBH formation channel, as defined in \ref{sec:formationchannels}, through equation \ref{eq:mrd}. 
Channel I (traditional CE scenario, with at least one CE phase after the formation of the first BH) gives the main contribution to the full BBH merger rate density at all redshifts, if $\alpha \leq 1$.  Channel I is  subdominant only at high redshift and when $\alpha > 1$. This is the result of longer delay time distributions for channel I, compared to channels II and IV. As a consequence, even if the merger efficiency of channel I is higher at all metallicities (Figure \ref{fig:Efficiency}), channels II and IV dominate the merger rate density at high redshift.
Moreover, the longer delay times result in a much steeper shape of the merger rate density for channel I, especially if we adopt a large value of~$\alpha$.

At low redshift, the merger-rate density associated with channel I is always largely above the LVK 90\% credible interval.
In contrast, the merger rate densities associated with channels II (stable mass transfer), III and IV lie within the observed range reported by LVK for all the considered $\alpha$ parameters. Channel III is by far the least common formation pathway.

\section{Discussion} \label{sec:discussion}
\begin{figure}
    \centering
    \includegraphics[width=0.5\textwidth]{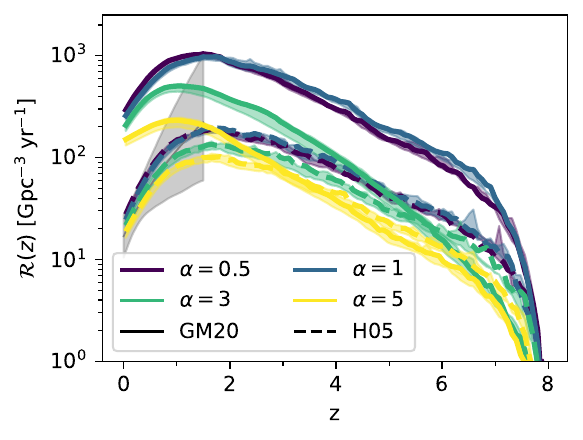}
    \caption{Merger rate density as a function of redshift, for the natal-kick models \protect\citetalias{Giacobbo2020}  (solid lines) and \protect\citetalias{Hobbs2005} (dashed lines). The line colors show the results obtained assuming different CE efficiency parameters ($\alpha$). The upper boundaries of the shaded areas show the models obtained with minimum galaxy mass $M_{\ast}=10^{6} \mathrm{M}_{\odot}$, whereas for the lower boundary the minimum mass is  $10^{8} \mathrm{M}_{\odot}$. All the models assume the fiducial observational scaling relations.
    The gray-shaded region shows the MRD inferred by the LVK collaboration \citep{Abbott2023population}.
    }
    \label{fig:nofallback}
\end{figure}

Predictions of BBH merger rate are highly uncertain, as they strongly depend on binary evolution physics and the metal-dependent SFR. \citet{Mandel2022} show that predicted rates can vary by more than three orders of magnitude depending on the adopted assumptions. Binary population synthesis codes generally tend to overpredict BBH merger rates, suggesting that careful calibration of binary parameters is necessary to bridge the gap with observations. This trend is evident across multiple studies using different population synthesis codes, including \textsc{startrack} \citep{Klencki2018, Chruslinska2019MRD, Olejak2021, Olejak2022, Romagnolo2023}, \textsc{compas} \citep{Neijssel2019, Broekgaarden2021, Riley2021, Broekgaarden2022, Stevenson2022, Rauf2023, Boesky2024MRD}, \textsc{cosmic} \citep{Zevin2020, Srinivasan2023}, \textsc{bpass} \citep{Eldridge2018, Tang2020}, and \textsc{posydon} \citep{Bavera2021, Romangarza2021}.
For similar variations in binary parameters, \citet{Boesky2024MRD} find rates that are consistent with our results, adopting the \textsc{compas} binary population synthesis code \citep{Riley2022}.
Several studies have explored the contribution of different formation channels to the BBH population. 
\citet{Bavera2021} and \citet{Romangarza2021} identify the evolution through at least one CE as the dominant channel in most models. 
The model reported by \citet{Olejak2021} predicts a higher contribution to the BBH merger rates from channel~I; however, their study highlights that different treatments of mass transfer stability can significantly alter the relative contributions of each channel. Similarly, \citet{Dorozsmai2024} find the stability of mass transfer to be a key factor in enhancing the contribution from the stable mass transfer channel. Notably, in the simulations run with \textsc{compas} \citep[][]{Neijssel2019, Broekgaarden2021} channel~II is dominant compared to channel~I because of the different stability criteria. 
Finally, the choice of the metal-dependent SFR can have a large impact on the BBH merger rates, as shown by   \citet{Neijssel2019, Broekgaarden2021, Broekgaarden2022, Srinivasan2023}. Specifically, \citet{Tang2020} show that a narrower metallicity distribution leads to lower merger rates, in agreement with our findings.

In this work, we have modeled the host galaxies of BBHs, their SFR and metallicity with the most up-to-date observational scaling relations. Nevertheless, as we improve 
the description of the metal-dependent SFR, the BBH merger rate density we obtain at low redshift  ($10^2 - 10^3\,{} \mathrm{Gpc}^{-3} \mathrm{yr}^{-1}$)  is always well above the 90\% credible interval inferred by LVK after the third observing run (Fig.~\ref{fig:MRD_all}). This is true for all of our models, considering different SFR--$M_\ast$ relations, different fundamental metallicity relations, and different values of the efficiency parameter for CE ejection. 

As discussed above, we can exclude that the overestimate of the BBH merger rate density is a feature of our population-synthesis code \textsc{sevn} \citep{Iorio2023}. Rather, it is a problem common to current population synthesis codes. 

In the following, we discuss several potential solutions to reconcile theoretical predictions and GW data. 
One possible explanation is that the CE phase occurs less frequently than previously assumed, or that BBH progenitors do not survive the CE phase as often as currently believed. Conversely, the stable mass transfer (channel II) may play a considerably more significant role in BBH formation than previously expected, as already proposed by several authors \citep{Pavlovskii2017, Neijssel2019, Misra2020, Klencki2020, Klencki2021, Olejak2021, Shao2021, Bavera2021, Marchant2021, VanSon2022}. In binary population synthesis codes, the CE channel tends to dominate over the others; in contrast, stable mass transfer is prevalent in detailed stellar-structure calculations  \citep[e.g.,][]{Gallegos2021}. One of the main factors of uncertainty lies in the criteria for the onset of a CE phase. As proposed by \cite{Ge2020}, mass transfer might be stable for a wider range of mass ratios than previously believed, at least under the assumption of adiabatic, conservative mass transfer. Another slightly different interpretation is that the traditional description of semi-major axis evolution  during a CE phase (based on the $\alpha$ parameter, \citealt{Hurley2002}) is too simplified, and does not capture the actual evolution of the system \citep{Hirai2022, Trani2022, Distefano2023, Everson2024}. If the orbital separation during a CE phase shrinks less than predicted by the traditional $\alpha$ formalism, then the resulting merger rate density is likely going to be lower \citep[e.g.,][]{Hirai2022, Lau2022}. 

Natal kicks  also significantly influence the merger rate density  \citep{Mapelli2018, Iorio2023, Boesky2024MRD}. In Fig.~\ref{fig:nofallback}, we compare our fiducial model with a model in which  natal kicks are drawn from a Maxwellian distribution with root-mean square velocity $\sigma=265$ km s$^{-1}$, without accounting for the fallback \citep{Hobbs2005}. This model yields substantially higher  kicks, leading to lower BBH merger rate densities across all considered $\alpha$ parameters.
Thus, another possible way to reconcile observed and predicted merger-rate density is to assume that natal kicks are higher than usually expected for BBHs. However, this also has an impact on the BH mass function, favoring the merger of more massive binary systems \citep{Iorio2023}. The recent work by \citet{Nagarajan2024} reveal the existence of both strong ($\gtrsim 100$ km s$^{-1}$) and weak ($\lesssim 50$ km s$^{-1}$) black hole kicks, potentially indicating a bimodal distribution \citep[see also][]{Verbunt2017}.
Although the uncertainties remain too large to conclusively confirm this scenario, these preliminary results suggest that the black hole kick distribution may lie between the two extreme models proposed in our work.

Finally, recent work by \cite{Schneider2021} shows that envelope stripping can affect the final fate and compact remnant mass of massive stars, increasing the minimum zero-age main sequence mass of BH progenitors from $\sim30$ M$_\odot$ to $\sim 70 \mathrm{M}_{\odot}$. While population calculations still need to be done for this scenario, this might result in a decrease in the BBH merger rate density as well.

A few caveats should be added. Recent work by \cite{Rinaldi2022} suggests that starburst galaxies contribute to 60$-$90\% of the total SFR. \cite{Rinaldi2022} find a clear bimodality between the galaxy main sequence and the starburst sequence for higher massive galaxies, $M_\ast \geq 10^{9} \mathrm{M}_{\odot}$. Our current approach likely underestimates the impact of starburst galaxies \citep[see also][]{Caputi2017, Bisigello2018}. Improved modeling of starburst galaxies is beyond the scope of this work. Nevertheless, an higher contribution of starburst galaxies would further increase the predicted BBH merger rates as they tend to be more metal-poor than main sequence galaxies.
We would like to remark that models based on the evolution of the average SFR and metallicity over cosmic time \citep[e.g.,]{Santoliquido2021} broadly capture our fiducial BBH merger rate densities (Fig. \ref{fig:MRD_all}), when accounting for a wide spread in metallicity for a given redshift bin ($\Tilde{\sigma}_{\rm Z} = 0.6$). This assumption is needed to mimic the complex distribution of metallicities that in our detailed approach naturally results from the spread in galaxy masses and SFRs. 
We do not vary the $q_{\rm crit}$ prescriptions in this study. However, Figure 22 in \citet{Iorio2023} demonstrates that different $q_{\rm crit}$ prescriptions affect the BBH merger rate density by a factor of $\sim 2$. Therefore, given the prescriptions available to date, we do not expect this parameter to significantly influence our results. Nevertheless, different conditions for mass-transfer stability  may have an impact on the relative contribution from different formation channels \citep[see e.g.,][]{Olejak2021, Olejak2022, Dorozsmai2024}.
The parameter $f_{\rm bin}$, takes into account the fraction of binary systems in the Universe. This value strongly varies with stellar mass, being close to 1 for massive stars \citep[$M_1 \geq 16$M$_\odot$, see e.g.,][]{Sana2012, Moe2017} and decreasing for lower primary masses, reaching $\sim 0.4$ for $M_1 \sim 2$M$_\odot$ \citep{Moe2017}. We adopt a constant $f_{\rm bin} = 0.5$ as a reference value, to account for the whole population of binary systems. It should be noted that $f_{\rm bin}$ only acts as a renormalization factor, rigidly shifting the merger rates upward or downward. Adopting a different $f_{\rm bin}$ value in the range $[0.4,1]$ would lead to at most a factor of 2 difference in the merger rates.

As a final remark,  differences between the BBH mass function obtained from population-synthesis calculations and the one assumed to infer the LVK rates, might bias the comparison between the two merger rate densities. However, the discrepancy between predictions and LVK rates (at least a factor of $\sim{10}$) is too large to be fully accounted for by such BBH mass function differences.

Finally, here we have only considered  isolated binary evolution. Including the dynamical formation of BBHs in star clusters and galactic nuclei would only lead to a further increase of the merger rate density because dynamics is known to boost the merger rate density even in a metal-rich environment \citep{Rodriguez2018, Dicarlo2020, Mapelli2022, Barber2024}. 

\section{Summary} \label{sec:summary}

We have studied the merger rate density of BBHs across cosmic time over 360 combinations of parameters, exploring the effects of different galaxy observational scaling relations and binary evolution models. Furthermore, we have studied the relative impact of different isolated BBH formation channels on the merger rate density. We can summarize our main results as follows.

\begin{itemize}
    \item 
    Current models of binary population synthesis predict values of the BBH merger rate density at redshift $z\leq{1}$ that are well above the 90\% credible interval inferred from LVK data \citep{Abbott2023population}.
    
    \item The impact of low-mass galaxies ($M_\ast<10^8$~M$_\odot$) on the BBH merger rate density is highly dependent on the steepness of the SFR--$M_\ast$ relation. For steep SFR--$M_\ast$ relations \citepalias{Popesso2023}, galaxies with mass $M_\ast\lesssim 10^7 \mathrm{M}_{\odot}$ do not contribute significantly to the merger rate density. 

    \item Even removing all the galaxies with mass below $10^8$~M$_\odot$ (hence, with lower metallicity), the merger rate density is still well above the 90\% credible interval from LVK. 
    Thus, uncertainties in the number of low-mass galaxies do not dramatically affect the merger rate density and do not solve the tension between predicted and observed BBH merger rate density.

    \item Models based on the average SFR and metallicity evolution need to account for the whole metallicity spread at a given redshift $\Tilde{\sigma}_{\rm Z} \sim 0.6$ in order to mirror detailed Universe prescriptions. In this case, the merger rate densities predicted by the two approaches broadly agree, with differences that become more noticeable toward high redshifts.

    \item The merger rate density of BBHs is dominated by Channel I, where the systems evolve through at least one CE phase only after the first compact object formation. 

\end{itemize}

Overall, our results clearly indicate that there is a tension between current models of BBH merger rates and values inferred from GW data.
We cannot explain this tension through uncertainties on the cosmic SFR density, as we have adopted state-of-the-art models, robustly grounded in observations.
Future studies are requested to understand the origin of this tension, addressing -e.g.,- the nature of core collapse, the stability of mass transfer, and the magnitude of natal kicks.

\begin{acknowledgements}
    The authors are grateful to Dylan Nelson and Annalisa Pillepich for their helpful insights. The authors thank the anonymous referee for the constructive report.
    MM acknowledges financial support
from the European Research Council for the ERC Consolidator grant DEMOBLACK, under contract no. 770017. MM also acknowledges financial support from the German Excellence Strategy via the Heidelberg Cluster of Excellence (EXC 2181 - 390900948) STRUCTURES.  
GI also received the support of a fellowship from the ”la Caixa” Foundation (ID 100010434)”. The fellowship code is LCF/BQ/PI24/12040020.
GI also acknowledges financial support under the National Recovery and Resilience Plan (NRRP), Mission 4, Component 2, Investment 1.4, - Call for tender No. 3138 of 18/12/2021 of Italian Ministry of University and Research funded by the European Union – NextGenerationEU.
MCA acknowledges financial support from  FONDECYT Iniciaci\'on 11240540. and ANID BASAL project FB210003.
The authors acknowledge support by the state of Baden-W\"urttemberg through bwHPC and the German Research Foundation (DFG) through grants INST 35/1597-1 FUGG and INST 35/1503-1 FUGG. \newline

\textit{Software.}
We use the \textsc{sevn} version V 2.10.1 (commit \href{https://gitlab.com/sevncodes/sevn/-/tree/22c923637bd6a7fe7546bf4585455f9a1d97b71c}{22c9236}) to generate our BBHs catalogs. \textsc{sevn} is publicly available at the gitlab repository \url{https://gitlab.com/sevncodes/sevn}. We use \textsc{trackcruncher} (\url{https://gitlab.com/sevncodes/trackcruncher}) \citep{Iorio2023} to produce the tables needed for the interpolation in \textsc{sevn}. We use the code \textsc{galaxy$\mathcal{R}$ate} to evolve our BBHs in a Universe model.  \textsc{galaxy$\mathcal{R}$ate} can be found at \url{https://gitlab.com/Filippo.santoliquido/galaxy_rate_open}. We use the code \textsc{cosmo$\mathcal{R}$ate} to model the averaged Universe models. \textsc{cosmo$\mathcal{R}$ate} can be found at \url{https://gitlab.com/Filippo.santoliquido/cosmo_rate_public}. 
    This research made use of \textsc{NumPy} \citep{Harris20}, \textsc{SciPy} \citep{SciPy2020}, \textsc{Pandas} \citep{Pandas2024} and \textsc{Astropy} \citep{astropy:2013, astropy:2018, astropy:2022}. For the plots we used \textsc{Matplotlib} \citep{Hunter2007}. 
\end{acknowledgements}

\bibliographystyle{aa}
\bibliography{ref}

\begin{thebibliography}{138}
\expandafter\ifx\csname natexlab\endcsname\relax\def\natexlab#1{#1}\fi

\bibitem[{{Abac} {et~al.}(2024){Abac}, {Abbott}, {Abouelfettouh}, {Acernese}, {Ackley}, {Adhicary}, {Adhikari}, {Adhikari}, {Adkins}, {Agarwal}, {Agathos}, {Abchouyeh}, {Aguiar}, {Aguilar}, {Aiello}, {Ain}, {Ajith}, {Ak{\c{c}}ay}, {Akutsu}, {Albanesi}, {Alfaidi}, {Al-Jodah}, {All{\'e}n{\'e}}, {Allocca}, {Al-Shammari}, {Altin}, {Alvarez-Lopez}, {Amato}, {Amez-Droz}, {Amorosi}, {Amra}, {Ananyeva}, {Anderson}, {Anderson}, {Andia}, {Ando}, {Andrade}, {Andres}, {Andr{\'e}s-Carcasona}, {Andri{\'c}}, {Anglin}, {Ansoldi}, {Antelis}, {Antier}, {Aoumi}, {Appavuravther}, {Appert}, {Apple}, {Arai}, {Araya}, {Araya}, {Areeda}, {Argianas}, {Aritomi}, {Armato}, {Arnaud}, {Arogeti}, {Aronson}, {Arun}, {Ashton}, {Aso}, {Assiduo}, {de Souza Melo}, {Aston}, {Astone}, {Attadio}, {Aubin}, {Aultoneal}, {Avallone}, {Azrad}, {Babak}, {Badaracco}, {Badger}, {Bae}, {Bagnasco}, {Bagui}, {Baier}, {Baiotti}, {Bajpai}, {Baka}, {Ball}, {Ballardin}, {Ballmer}, {Banagiri}, {Banerjee}, {Bankar}, {Baral}, {Barayoga}, {Barish}, {Barker},
  {Barneo}, {Barone}, {Barr}, {Barsotti}, {Barsuglia}, {Barta}, {Bartoletti}, {Barton}, {Bartos}, {Basak}, {Basalaev}, {Bassiri}, {Basti}, {Bates}, {Bawaj}, {Baxi}, {Bayley}, {Baylor}, {Baynard}, {Bazzan}, {Bedakihale}, {Beirnaert}, {Bejger}, {Belardinelli}, {Bell}, {Benedetto}, {Benoit}, {Bentara}, {Bentley}, {Ben Yaala}, {Bera}, {Berbel}, {Bergamin}, {Berger}, {Bernuzzi}, {Beroiz}, {Berry}, {Bersanetti}, {Bertolini}, {Betzwieser}, {Beveridge}, {Bevins}, {Bhandare}, {Bhardwaj}, {Bhatt}, {Bhattacharjee}, {Bhaumik}, {Bhowmick}, {Bianchi}, {Bilenko}, {Billingsley}, {Binetti}, {Bini}, {Birnholtz}, {Biscoveanu}, {Bisht}, {Bitossi}, {Bizouard}, {Blackburn}, {Blagg}, {Blair}, {Blair}, {Bobba}, {Bode}, {Boileau}, {Boldrini}, {Bolingbroke}, {Bolliand}, {Bonavena}, {Bondarescu}, {Bondu}, {Bonilla}, {Bonilla}, {Bonino}, {Bonnand}, {Booker}, {Borchers}, {Boschi}, {Bose}, {Bossilkov}, {Boudart}, {Boudon}, {Bozzi}, {Bradaschia}, {Brady}, {Braglia}, {Branch}, {Branchesi}, {Brandt}, {Braun}, {Breschi}, {Briant}, {Brillet},
  {Brinkmann}, {Brockill}, {Brockmueller}, {Brooks}, {Brown}, {Brown}, {Brozzetti}, {Brunett}, {Bruno}, {Bruntz}, {Bryant}, {Bucci}, {Buchanan}, {Bulashenko}, {Bulik}, {Bulten}, \& {Buonanno}}]{LVK2024_gap}
{Abac}, A.~G., {Abbott}, R., {Abouelfettouh}, I., {et~al.} 2024, \apjl, 970, L34

\bibitem[{Abbott {et~al.}(2016{\natexlab{a}})Abbott, Abbott, Abbott, Abernathy, Acernese, Ackley, Adams, Adams, Addesso, Adhikari, Adya, Affeldt, Agathos, Agatsuma, Aggarwal, Aguiar, Aiello, Ain, Ajith, Allen, Allocca, Altin, Anderson, Anderson, Arai, Arain, Araya, Arceneaux, Areeda, Arnaud, Arun, Ascenzi, Ashton, Ast, Aston, Astone, Aufmuth, Aulbert, Babak, Bacon, Bader, Baker, Baldaccini, Ballardin, Ballmer, Barayoga, Barclay, Barish, Barker, Barone, Barr, Barsotti, Barsuglia, Barta, Bartlett, Barton, Bartos, Bassiri, Basti, Batch, Baune, Bavigadda, Bazzan, Behnke, Bejger, Belczynski, Bell, Bell, Berger, Bergman, Bergmann, Berry, Bersanetti, Bertolini, Betzwieser, Bhagwat, Bhandare, Bilenko, Billingsley, Birch, Birney, Birnholtz, Biscans, Bisht, Bitossi, Biwer, Bizouard, Blackburn, Blair, Blair, Blair, Bloemen, Bock, Bodiya, Boer, Bogaert, Bogan, Bohe, Bojtos, Bond, Bondu, Bonnand, Boom, Bork, Boschi, Bose, Bouffanais, Bozzi, Bradaschia, Brady, Braginsky, Branchesi, Brau, Briant, Brillet, Brinkmann,
  Brisson, Brockill, Brooks, Brown, Brown, Brown, Buchanan, Buikema, Bulik, Bulten, Buonanno, Buskulic, Buy, Byer, Cabero, Cadonati, Cagnoli, Cahillane, Bustillo, Callister, Calloni, Camp, Cannon, Cao, Capano, Capocasa, Carbognani, Caride, Diaz, Casentini, Caudill, Cavagli\`a, Cavalier, Cavalieri, Cella, Cepeda, Baiardi, Cerretani, Cesarini, Chakraborty, Chalermsongsak, Chamberlin, Chan, Chao, Charlton, Chassande-Mottin, Chen, Chen, Cheng, Chincarini, Chiummo, Cho, Cho, Chow, Christensen, Chu, Chua, Chung, Ciani, Clara, Clark, Cleva, Coccia, Cohadon, Colla, Collette, Cominsky, Constancio, Conte, Conti, Cook, Corbitt, Cornish, Corsi, Cortese, Costa, Coughlin, Coughlin, Coulon, Countryman, Couvares, Cowan, Coward, Cowart, Coyne, Coyne, Craig, Creighton, Creighton, Cripe, Crowder, Cruise, Cumming, Cunningham, Cuoco, Canton, Danilishin, D'Antonio, Danzmann, Darman, Da~Silva~Costa, Dattilo, Dave, Daveloza, Davier, Davies, Daw, Day, De, DeBra, Debreczeni, Degallaix, De~Laurentis, Del\'eglise, Del~Pozzo, Denker,
  Dent, Dereli, Dergachev, DeRosa, De~Rosa, DeSalvo, Dhurandhar, D\'{\i}az, Di~Fiore, Di~Giovanni, Di~Lieto, Di~Pace, Di~Palma, Di~Virgilio, Dojcinoski, Dolique, Donovan, Dooley, Doravari, Douglas, Downes, Drago, Drever, Driggers, Du, Ducrot, Dwyer, Edo, Edwards, Effler, Eggenstein, Ehrens, Eichholz, Eikenberry, Engels, Essick, Etzel, Evans, Evans, Everett, Factourovich, Fafone, Fair, Fairhurst, Fan, Fang, Farinon, Farr, Farr, Favata, Fays, Fehrmann, Fejer, Feldbaum, Ferrante, Ferreira, Ferrini, Fidecaro, Finn, Fiori, Fiorucci, Fisher, Flaminio, Fletcher, Fong, Fournier, Franco, Frasca, Frasconi, Frede, Frei, Freise, Frey, Frey, Fricke, Fritschel, Frolov, Fulda, Fyffe, Gabbard, Gair, Gammaitoni, Gaonkar, Garufi, Gatto, Gaur, Gehrels, Gemme, Gendre, Genin, Gennai, George, Gergely, Germain, Ghosh, Ghosh, Ghosh, Giaime, Giardina, Giazotto, Gill, Glaefke, Gleason, Goetz, Goetz, Gondan, Gonz\'alez, Castro, Gopakumar, Gordon, Gorodetsky, Gossan, Gosselin, Gouaty, Graef, Graff, Granata, Grant, Gras, Gray, Greco,
  Green, Greenhalgh, Groot, Grote, Grunewald, Guidi, Guo, Gupta, Gupta, Gushwa, Gustafson, Gustafson, Hacker, Hall, Hall, Hammond, Haney, Hanke, Hanks, Hanna, Hannam, Hanson, Hardwick, Harms, Harry, Harry, Hart, Hartman, Haster, Haughian, Healy, Heefner, Heidmann, Heintze, Heinzel, Heitmann, Hello, Hemming, Hendry, Heng, Hennig, Heptonstall, Heurs, Hild, Hoak, Hodge, Hofman, Hollitt, Holt, Holz, Hopkins, Hosken, Hough, Houston, Howell, Hu, Huang, Huerta, Huet, Hughey, Husa, Huttner, Huynh-Dinh, Idrisy, Indik, Ingram, Inta, Isa, Isac, Isi, Islas, Isogai, Iyer, Izumi, Jacobson, Jacqmin, Jang, Jani, Jaranowski, Jawahar, Jim\'enez-Forteza, Johnson, Johnson-McDaniel, Jones, Jones, Jonker, Ju, Haris, Kalaghatgi, Kalogera, Kandhasamy, Kang, Kanner, Karki, Kasprzack, Katsavounidis, Katzman, Kaufer, Kaur, Kawabe, Kawazoe, K\'ef\'elian, Kehl, Keitel, Kelley, Kells, Kennedy, Keppel, Key, Khalaidovski, Khalili, Khan, Khan, Khan, Khazanov, Kijbunchoo, Kim, Kim, Kim, Kim, Kim, Kim, King, King, Kinzel, Kissel, Kleybolte,
  Klimenko, Koehlenbeck, Kokeyama, Koley, Kondrashov, Kontos, Koranda, Korobko, Korth, Kowalska, Kozak, Kringel, Krishnan, Kr\'olak, Krueger, Kuehn, Kumar, Kumar, Kuo, Kutynia, Kwee, Lackey, Landry, Lange, Lantz, Lasky, Lazzarini, Lazzaro, Leaci, Leavey, Lebigot, Lee, Lee, Lee, Lee, Lenon, Leonardi, Leong, Leroy, Letendre, Levin, Levine, Li, Libson, Littenberg, Lockerbie, Logue, Lombardi, London, Lord, Lorenzini, Loriette, Lormand, Losurdo, Lough, Lousto, Lovelace, L\"uck, Lundgren, Luo, Lynch, Ma, MacDonald, Machenschalk, MacInnis, Macleod, Maga\~na Sandoval, Magee, Mageswaran, Majorana, Maksimovic, Malvezzi, Man, Mandel, Mandic, Mangano, Mansell, Manske, Mantovani, Marchesoni, Marion, M\'arka, M\'arka, Markosyan, Maros, Martelli, Martellini, Martin, Martin, Martynov, Marx, Mason, Masserot, Massinger, Masso-Reid, Matichard, Matone, Mavalvala, Mazumder, Mazzolo, McCarthy, McClelland, McCormick, McGuire, McIntyre, McIver, McManus, McWilliams, Meacher, Meadors, Meidam, Melatos, Mendell, Mendoza-Gandara, Mercer,
  Merilh, Merzougui, Meshkov, Messenger, Messick, Meyers, Mezzani, Miao, Michel, Middleton, Mikhailov, Milano, Miller, Millhouse, Minenkov, Ming, Mirshekari, Mishra, Mitra, Mitrofanov, Mitselmakher, Mittleman, Moggi, Mohan, Mohapatra, Montani, Moore, Moore, Moraru, Moreno, Morriss, Mossavi, Mours, Mow-Lowry, Mueller, Mueller, Muir, Mukherjee, Mukherjee, Mukherjee, Mukund, Mullavey, Munch, Murphy, Murray, Mytidis, Nardecchia, Naticchioni, Nayak, Necula, Nedkova, Nelemans, Neri, Neunzert, Newton, Nguyen, Nielsen, Nissanke, Nitz, Nocera, Nolting, Normandin, Nuttall, Oberling, Ochsner, O'Dell, Oelker, Ogin, Oh, Oh, Ohme, Oliver, Oppermann, Oram, O'Reilly, O'Shaughnessy, Ott, Ottaway, Ottens, Overmier, Owen, Pai, Pai, Palamos, Palashov, Palomba, Pal-Singh, Pan, Pan, Pankow, Pannarale, Pant, Paoletti, Paoli, Papa, Paris, Parker, Pascucci, Pasqualetti, Passaquieti, Passuello, Patricelli, Patrick, Pearlstone, Pedraza, Pedurand, Pekowsky, Pele, Penn, Perreca, Pfeiffer, Phelps, Piccinni, Pichot, Pickenpack,
  Piergiovanni, Pierro, Pillant, Pinard, Pinto, Pitkin, Poeld, Poggiani, Popolizio, Post, Powell, Prasad, Predoi, Premachandra, Prestegard, Price, Prijatelj, Principe, Privitera, Prix, Prodi, Prokhorov, Puncken, Punturo, Puppo, P\"urrer, Qi, Qin, Quetschke, Quintero, Quitzow-James, Raab, Rabeling, Radkins, Raffai, Raja, Rakhmanov, Ramet, Rapagnani, Raymond, Razzano, Re, Read, Reed, Regimbau, Rei, Reid, Reitze, Rew, Reyes, Ricci, Riles, Robertson, Robie, Robinet, Rocchi, Rolland, Rollins, Roma, Romano, Romano, Romanov, Romie, Rosi\ifmmode~\acute{n}\else \'{n}\fi{}ska, Rowan, R\"udiger, Ruggi, Ryan, Sachdev, Sadecki, Sadeghian, Salconi, Saleem, Salemi, Samajdar, Sammut, Sampson, Sanchez, Sandberg, Sandeen, Sanders, Sanders, Sassolas, Sathyaprakash, Saulson, Sauter, Savage, Sawadsky, Schale, Schilling, Schmidt, Schmidt, Schnabel, Schofield, Sch\"onbeck, Schreiber, Schuette, Schutz, Scott, Scott, Sellers, Sengupta, Sentenac, Sequino, Sergeev, Serna, Setyawati, Sevigny, Shaddock, Shaffer, Shah, Shahriar, Shaltev,
  Shao, Shapiro, Shawhan, Sheperd, Shoemaker, Shoemaker, Siellez, Siemens, Sigg, Silva, Simakov, Singer, Singer, Singh, Singh, Singhal, Sintes, Slagmolen, Smith, Smith, Smith, Smith, Son, Sorazu, Sorrentino, Souradeep, Srivastava, Staley, Steinke, Steinlechner, Steinlechner, Steinmeyer, Stephens, Stevenson, Stone, Strain, Straniero, Stratta, Strauss, Strigin, Sturani, Stuver, Summerscales, Sun, Sutton, Swinkels, Szczepa\ifmmode~\acute{n}\else \'{n}\fi{}czyk, Tacca, Talukder, Tanner, T\'apai, Tarabrin, Taracchini, Taylor, Theeg, Thirugnanasambandam, Thomas, Thomas, Thomas, Thorne, Thorne, Thrane, Tiwari, Tiwari, Tokmakov, Tomlinson, Tonelli, Torres, Torrie, T\"oyr\"a, Travasso, Traylor, Trifir\`o, Tringali, Trozzo, Tse, Turconi, Tuyenbayev, Ugolini, Unnikrishnan, Urban, Usman, Vahlbruch, Vajente, Valdes, Vallisneri, van Bakel, van Beuzekom, van~den Brand, Van Den~Broeck, Vander-Hyde, van~der Schaaf, van Heijningen, van Veggel, Vardaro, Vass, Vas\'uth, Vaulin, Vecchio, Vedovato, Veitch, Veitch, Venkateswara,
  Verkindt, Vetrano, Vicer\'e, Vinciguerra, Vine, Vinet, Vitale, Vo, Vocca, Vorvick, Voss, Vousden, Vyatchanin, Wade, Wade, Wade, Waldman, Walker, Wallace, Walsh, Wang, Wang, Wang, Wang, Wang, Ward, Ward, Warner, Was, Weaver, Wei, Weinert, Weinstein, Weiss, Welborn, Wen, We\ss{}els, Westphal, Wette, Whelan, Whitcomb, White, Whiting, Wiesner, Wilkinson, Willems, Williams, Williams, Williamson, Willis, Willke, Wimmer, Winkelmann, Winkler, Wipf, Wiseman, Wittel, Woan, Worden, Wright, Wu, Yablon, Yakushin, Yam, Yamamoto, Yancey, Yap, Yu, Yvert, Zadro\ifmmode~\dot{z}\else \.{z}\fi{}ny, Zangrando, Zanolin, Zendri, Zevin, Zhang, Zhang, Zhang, Zhang, Zhao, Zhou, Zhou, Zhu, Zucker, Zuraw, \& Zweizig}]{Abbott2016a}
Abbott, B.~P., Abbott, R., Abbott, T.~D., {et~al.} 2016{\natexlab{a}}, Phys. Rev. Lett., 116, 061102

\bibitem[{Abbott {et~al.}(2016{\natexlab{b}})Abbott, Abbott, Abbott, Abernathy, Acernese, Ackley, Adams, Adams, Addesso, Adhikari, Adya, Affeldt, Agathos, Agatsuma, Aggarwal, Aguiar, Aiello, Ain, Ajith, Allen, Allocca, Altin, Anderson, Anderson, Arai, Araya, Arceneaux, Areeda, Arnaud, Arun, Ascenzi, Ashton, Ast, Aston, Astone, Aufmuth, Aulbert, Babak, Bacon, Bader, Baker, Baldaccini, Ballardin, Ballmer, Barayoga, Barclay, Barish, Barker, Barone, Barr, Barsotti, Barsuglia, Barta, Bartlett, Bartos, Bassiri, Basti, Batch, Baune, Bavigadda, Bazzan, Bejger, Bell, Berger, Bergmann, Berry, Bersanetti, Bertolini, Betzwieser, Bhagwat, Bhandare, Bilenko, Billingsley, Birch, Birney, Birnholtz, Biscans, Bisht, Bitossi, Biwer, Bizouard, Blackburn, Blair, Blair, Blair, Bloemen, Bock, Boer, Bogaert, Bogan, Bohe, Bond, Bondu, Bonnand, Boom, Bork, Boschi, Bose, Bouffanais, Bozzi, Bradaschia, Brady, Braginsky, Branchesi, Brau, Briant, Brillet, Brinkmann, Brisson, Brockill, Broida, Brooks, Brown, Brown, Brown, Brunett,
  Buchanan, Buikema, Bulik, Bulten, Buonanno, Buskulic, Buy, Byer, Cabero, Cadonati, Cagnoli, Cahillane, Calder\'on~Bustillo, Callister, Calloni, Camp, Cannon, Cao, Capano, Capocasa, Carbognani, Caride, Casanueva~Diaz, Casentini, Caudill, Cavagli\`a, Cavalier, Cavalieri, Cella, Cepeda, Cerboni~Baiardi, Cerretani, Cesarini, Chamberlin, Chan, Chao, Charlton, Chassande-Mottin, Cheeseboro, Chen, Chen, Cheng, Chincarini, Chiummo, Cho, Cho, Chow, Christensen, Chu, Chua, Chung, Ciani, Clara, Clark, Cleva, Coccia, Cohadon, Colla, Collette, Cominsky, Constancio, Conte, Conti, Cook, Corbitt, Cornish, Corsi, Cortese, Costa, Coughlin, Coughlin, Coulon, Countryman, Couvares, Cowan, Coward, Cowart, Coyne, Coyne, Craig, Creighton, Cripe, Crowder, Cumming, Cunningham, Cuoco, Dal~Canton, Danilishin, D'Antonio, Danzmann, Darman, Dasgupta, Da~Silva~Costa, Dattilo, Dave, Davier, Davies, Daw, Day, De, DeBra, Debreczeni, Degallaix, De~Laurentis, Del\'eglise, Del~Pozzo, Denker, Dent, Dergachev, De~Rosa, DeRosa, DeSalvo, Devine,
  Dhurandhar, D\'{\i}az, Di~Fiore, Di~Giovanni, Di~Girolamo, Di~Lieto, Di~Pace, Di~Palma, Di~Virgilio, Dolique, Donovan, Dooley, Doravari, Douglas, Downes, Drago, Drever, Driggers, Ducrot, Dwyer, Edo, Edwards, Effler, Eggenstein, Ehrens, Eichholz, Eikenberry, Engels, Essick, Etzel, Evans, Evans, Everett, Factourovich, Fafone, Fair, Fairhurst, Fan, Fang, Farinon, Farr, Farr, Favata, Fays, Fehrmann, Fejer, Fenyvesi, Ferrante, Ferreira, Ferrini, Fidecaro, Fiori, Fiorucci, Fisher, Flaminio, Fletcher, Fong, Fournier, Frasca, Frasconi, Frei, Freise, Frey, Frey, Fritschel, Frolov, Fulda, Fyffe, Gabbard, Gair, Gammaitoni, Gaonkar, Garufi, Gaur, Gehrels, Gemme, Geng, Genin, Gennai, George, Gergely, Germain, Ghosh, Ghosh, Ghosh, Giaime, Giardina, Giazotto, Gill, Glaefke, Goetz, Goetz, Gondan, Gonz\'alez, Gonzalez~Castro, Gopakumar, Gordon, Gorodetsky, Gossan, Gosselin, Gouaty, Grado, Graef, Graff, Granata, Grant, Gras, Gray, Greco, Green, Groot, Grote, Grunewald, Guidi, Guo, Gupta, Gupta, Gushwa, Gustafson, Gustafson,
  Hacker, Hall, Hall, Hamilton, Hammond, Haney, Hanke, Hanks, Hanna, Hannam, Hanson, Hardwick, Harms, Harry, Harry, Hart, Hartman, Haster, Haughian, Healy, Heidmann, Heintze, Heitmann, Hello, Hemming, Hendry, Heng, Hennig, Henry, Heptonstall, Heurs, Hild, Hoak, Hofman, Holt, Holz, Hopkins, Hough, Houston, Howell, Hu, Huang, Huerta, Huet, Hughey, Husa, Huttner, Huynh-Dinh, Indik, Ingram, Inta, Isa, Isac, Isi, Isogai, Iyer, Izumi, Jacqmin, Jang, Jani, Jaranowski, Jawahar, Jian, Jim\'enez-Forteza, Johnson, Johnson-McDaniel, Jones, Jones, Jonker, Ju, K, Kalaghatgi, Kalogera, Kandhasamy, Kang, Kanner, Kapadia, Karki, Karvinen, Kasprzack, Katsavounidis, Katzman, Kaufer, Kaur, Kawabe, K\'ef\'elian, Kehl, Keitel, Kelley, Kells, Kennedy, Key, Khalili, Khan, Khan, Khan, Khazanov, Kijbunchoo, Kim, Kim, Kim, Kim, Kim, Kim, Kim, Kimbrell, King, King, Kissel, Klein, Kleybolte, Klimenko, Koehlenbeck, Koley, Kondrashov, Kontos, Korobko, Korth, Kowalska, Kozak, Kringel, Krishnan, Kr\'olak, Krueger, Kuehn, Kumar, Kumar, Kuo,
  Kutynia, Lackey, Landry, Lange, Lantz, Lasky, Laxen, Lazzarini, Lazzaro, Leaci, Leavey, Lebigot, Lee, Lee, Lee, Lee, Lenon, Leonardi, Leong, Leroy, Letendre, Levin, Lewis, Li, Libson, Littenberg, Lockerbie, Lombardi, London, Lord, Lorenzini, Loriette, Lormand, Losurdo, Lough, Lousto, L\"uck, Lundgren, Lynch, Ma, Machenschalk, MacInnis, Macleod, Maga\~na Sandoval, Maga\~na Zertuche, Magee, Majorana, Maksimovic, Malvezzi, Man, Mandel, Mandic, Mangano, Mansell, Manske, Mantovani, Marchesoni, Marion, M\'arka, M\'arka, Markosyan, Maros, Martelli, Martellini, Martin, Martynov, Marx, Mason, Masserot, Massinger, Masso-Reid, Mastrogiovanni, Matichard, Matone, Mavalvala, Mazumder, McCarthy, McClelland, McCormick, McGuire, McIntyre, McIver, McManus, McRae, McWilliams, Meacher, Meadors, Meidam, Melatos, Mendell, Mercer, Merilh, Merzougui, Meshkov, Messenger, Messick, Metzdorff, Meyers, Mezzani, Miao, Michel, Middleton, Mikhailov, Milano, Miller, Miller, Miller, Miller, Millhouse, Minenkov, Ming, Mirshekari, Mishra,
  Mitra, Mitrofanov, Mitselmakher, Mittleman, Moggi, Mohan, Mohapatra, Montani, Moore, Moore, Moraru, Moreno, Morriss, Mossavi, Mours, Mow-Lowry, Mueller, Muir, Mukherjee, Mukherjee, Mukherjee, Mukund, Mullavey, Munch, Murphy, Murray, Mytidis, Nardecchia, Naticchioni, Nayak, Nedkova, Nelemans, Nelson, Neri, Neunzert, Newton, Nguyen, Nielsen, Nissanke, Nitz, Nocera, Nolting, Normandin, Nuttall, Oberling, Ochsner, O'Dell, Oelker, Ogin, Oh, Oh, Ohme, Oliver, Oppermann, Oram, O'Reilly, O'Shaughnessy, Ottaway, Overmier, Owen, Pai, Pai, Palamos, Palashov, Palomba, Pal-Singh, Pan, Pankow, Pannarale, Pant, Paoletti, Paoli, Papa, Paris, Parker, Pascucci, Pasqualetti, Passaquieti, Passuello, Patricelli, Patrick, Pearlstone, Pedraza, Pedurand, Pekowsky, Pele, Penn, Perreca, Perri, Pfeiffer, Phelps, Piccinni, Pichot, Piergiovanni, Pierro, Pillant, Pinard, Pinto, Pitkin, Poe, Poggiani, Popolizio, Post, Powell, Prasad, Predoi, Prestegard, Price, Prijatelj, Principe, Privitera, Prix, Prodi, Prokhorov, Puncken, Punturo,
  Puppo, P\"urrer, Qi, Qin, Qiu, Quetschke, Quintero, Quitzow-James, Raab, Rabeling, Radkins, Raffai, Raja, Rajan, Rakhmanov, Rapagnani, Raymond, Razzano, Re, Read, Reed, Regimbau, Rei, Reid, Reitze, Rew, Reyes, Ricci, Riles, Rizzo, Robertson, Robie, Robinet, Rocchi, Rolland, Rollins, Roma, Romano, Romano, Romanov, Romie, Rosi\ifmmode~\acute{n}\else \'{n}\fi{}ska, Rowan, R\"udiger, Ruggi, Ryan, Sachdev, Sadecki, Sadeghian, Sakellariadou, Salconi, Saleem, Salemi, Samajdar, Sammut, Sanchez, Sandberg, Sandeen, Sanders, Sassolas, Sathyaprakash, Saulson, Sauter, Savage, Sawadsky, Schale, Schilling, Schmidt, Schmidt, Schnabel, Schofield, Sch\"onbeck, Schreiber, Schuette, Schutz, Scott, Scott, Sellers, Sengupta, Sentenac, Sequino, Sergeev, Setyawati, Shaddock, Shaffer, Shahriar, Shaltev, Shapiro, Shawhan, Sheperd, Shoemaker, Shoemaker, Siellez, Siemens, Sieniawska, Sigg, Silva, Singer, Singer, Singh, Singh, Singhal, Sintes, Slagmolen, Smith, Smith, Smith, Son, Sorazu, Sorrentino, Souradeep, Srivastava, Staley,
  Steinke, Steinlechner, Steinlechner, Steinmeyer, Stephens, Stevenson, Stone, Strain, Straniero, Stratta, Strauss, Strigin, Sturani, Stuver, Summerscales, Sun, Sunil, Sutton, Swinkels, Szczepa\ifmmode~\acute{n}\else \'{n}\fi{}czyk, Tacca, Talukder, Tanner, T\'apai, Tarabrin, Taracchini, Taylor, Theeg, Thirugnanasambandam, Thomas, Thomas, Thomas, Thorne, Thrane, Tiwari, Tiwari, Tokmakov, Toland, Tomlinson, Tonelli, Tornasi, Torres, Torrie, T\"oyr\"a, Travasso, Traylor, Trifir\`o, Tringali, Trozzo, Tse, Turconi, Tuyenbayev, Ugolini, Unnikrishnan, Urban, Usman, Vahlbruch, Vajente, Valdes, Vallisneri, van Bakel, van Beuzekom, van~den Brand, Van Den~Broeck, Vander-Hyde, van~der Schaaf, van Heijningen, van Veggel, Vardaro, Vass, Vas\'uth, Vaulin, Vecchio, Vedovato, Veitch, Veitch, Venkateswara, Verkindt, Vetrano, Vicer\'e, Vinciguerra, Vine, Vinet, Vitale, Vo, Vocca, Vorvick, Voss, Vousden, Vyatchanin, Wade, Wade, Wade, Walker, Wallace, Walsh, Wang, Wang, Wang, Wang, Wang, Ward, Warner, Was, Weaver, Wei, Weinert,
  Weinstein, Weiss, Wen, We\ss{}els, Westphal, Wette, Whelan, Whiting, Williams, Williamson, Willis, Willke, Wimmer, Winkler, Wipf, Wittel, Woan, Woehler, Worden, Wright, Wu, Wu, Yablon, Yam, Yamamoto, Yancey, Yu, Yvert, Zadro\ifmmode~\dot{z}\else \.{z}\fi{}ny, Zangrando, Zanolin, Zendri, Zevin, Zhang, Zhang, Zhang, Zhao, Zhou, Zhou, Zhu, Zucker, Zuraw, Zweizig, Boyle, Hemberger, Kidder, Lovelace, Ossokine, Scheel, Szilagyi, \& Teukolsky}]{Abbott2016b}
Abbott, B.~P., Abbott, R., Abbott, T.~D., {et~al.} 2016{\natexlab{b}}, Phys. Rev. Lett., 116, 241103

\bibitem[{{Abbott} {et~al.}(2020{\natexlab{a}}){Abbott}, {Abbott}, {Abbott}, {Abraham}, {Acernese}, {Ackley}, {Adams}, {Adhikari}, {Adya}, {Affeldt}, {Agathos}, {Agatsuma}, {Aggarwal}, {Aguiar}, {Aiello}, {Ain}, {Ajith}, {Allen}, {Allocca}, {Aloy}, {Altin}, {Amato}, {Anand}, {Ananyeva}, {Anderson}, {Anderson}, {Angelova}, {Antier}, {Appert}, {Arai}, {Araya}, {Areeda}, {Ar{\`e}ne}, {Arnaud}, {Aronson}, {Arun}, {Ascenzi}, {Ashton}, {Aston}, {Astone}, {Aubin}, {Aufmuth}, {AultONeal}, {Austin}, {Avendano}, {Avila-Alvarez}, {Babak}, {Bacon}, {Badaracco}, {Bader}, {Bae}, {Baird}, {Baker}, {Baldaccini}, {Ballardin}, {Ballmer}, {Bals}, {Banagiri}, {Barayoga}, {Barbieri}, {Barclay}, {Barish}, {Barker}, {Barkett}, {Barnum}, {Barone}, {Barr}, {Barsotti}, {Barsuglia}, {Barta}, {Bartlett}, {Bartos}, {Bassiri}, {Basti}, {Bawaj}, {Bayley}, {Baylor}, {Bazzan}, {B{\'e}csy}, {Bejger}, {Belahcene}, {Bell}, {Beniwal}, {Benjamin}, {Berger}, {Bergmann}, {Bernuzzi}, {Berry}, {Bersanetti}, {Bertolini}, {Betzwieser}, {Bhandare},
  {Bidler}, {Biggs}, {Bilenko}, {Bilgili}, {Billingsley}, {Birney}, {Birnholtz}, {Biscans}, {Bischi}, {Biscoveanu}, {Bisht}, {Bitossi}, {Bizouard}, {Blackburn}, {Blackman}, {Blair}, {Blair}, {Blair}, {Bloemen}, {Bobba}, {Bode}, {Boer}, {Boetzel}, {Bogaert}, {Bondu}, {Bonnand}, {Booker}, {Boom}, {Bork}, {Boschi}, {Bose}, {Bossilkov}, {Bosveld}, {Bouffanais}, {Bozzi}, {Bradaschia}, {Brady}, {Bramley}, {Branchesi}, {Brau}, {Breschi}, {Briant}, {Briggs}, {Brighenti}, {Brillet}, {Brinkmann}, {Brockill}, {Brooks}, {Brooks}, {Brown}, {Brunett}, {Buikema}, {Bulik}, {Bulten}, {Buonanno}, {Buskulic}, {Buy}, {Byer}, {Cabero}, {Cadonati}, {Cagnoli}, {Cahillane}, {Calder{\'o}n Bustillo}, {Callister}, {Calloni}, {Camp}, {Campbell}, {Canepa}, {Cannon}, {Cao}, {Cao}, {Carapella}, {Carbognani}, {Caride}, {Carney}, {Carullo}, {Casanueva Diaz}, {Casentini}, {Caudill}, {Cavagli{\`a}}, {Cavalier}, {Cavalieri}, {Cella}, {Cerd{\'a}-Dur{\'a}n}, {Cesarini}, {Chaibi}, {Chakravarti}, {Chamberlin}, {Chan}, {Chao}, {Charlton}, {Chase},
  {Chassande-Mottin}, {Chatterjee}, {Chaturvedi}, {Chatziioannou}, {Cheeseboro}, {Chen}, {Chen}, {Chen}, {Cheng}, {Cheong}, {Chia}, {Chiadini}, {Chincarini}, {Chiummo}, {Cho}, \& {Cho}}]{LVK2020_gap_2}
{Abbott}, B.~P., {Abbott}, R., {Abbott}, T.~D., {et~al.} 2020{\natexlab{a}}, \apjl, 892, L3

\bibitem[{Abbott {et~al.}(2017{\natexlab{a}})Abbott, Abbott, Abbott, Acernese, Ackley, Adams, Adams, Addesso, Adhikari, Adya, Affeldt, Afrough, Agarwal, Agathos, Agatsuma, Aggarwal, Aguiar, Aiello, Ain, Ajith, Allen, Allen, Allocca, Altin, Amato, Ananyeva, Anderson, Anderson, Angelova, Antier, Appert, Arai, Araya, Areeda, Arnaud, Arun, Ascenzi, Ashton, Ast, Aston, Astone, Atallah, Aufmuth, Aulbert, AultONeal, Austin, Avila-Alvarez, Babak, Bacon, Bader, Bae, Bailes, Baker, Baldaccini, Ballardin, Ballmer, Banagiri, Barayoga, Barclay, Barish, Barker, Barkett, Barone, Barr, Barsotti, Barsuglia, Barta, Barthelmy, Bartlett, Bartos, Bassiri, Basti, Batch, Bawaj, Bayley, Bazzan, B\'ecsy, Beer, Bejger, Belahcene, Bell, Berger, Bergmann, Bernuzzi, Bero, Berry, Bersanetti, Bertolini, Betzwieser, Bhagwat, Bhandare, Bilenko, Billingsley, Billman, Birch, Birney, Birnholtz, Biscans, Biscoveanu, Bisht, Bitossi, Biwer, Bizouard, Blackburn, Blackman, Blair, Blair, Blair, Bloemen, Bock, Bode, Boer, Bogaert, Bohe, Bondu,
  Bonilla, Bonnand, Boom, Bork, Boschi, Bose, Bossie, Bouffanais, Bozzi, Bradaschia, Brady, Branchesi, Brau, Briant, Brillet, Brinkmann, Brisson, Brockill, Broida, Brooks, Brown, Brown, Brunett, Buchanan, Buikema, Bulik, Bulten, Buonanno, Buskulic, Buy, Byer, Cabero, Cadonati, Cagnoli, Cahillane, Calder\'on~Bustillo, Callister, Calloni, Camp, Canepa, Canizares, Cannon, Cao, Cao, Capano, Capocasa, Carbognani, Caride, Carney, Carullo, Casanueva~Diaz, Casentini, Caudill, Cavagli\`a, Cavalier, Cavalieri, Cella, Cepeda, Cerd\'a-Dur\'an, Cerretani, Cesarini, Chamberlin, Chan, Chao, Charlton, Chase, Chassande-Mottin, Chatterjee, Chatziioannou, Cheeseboro, Chen, Chen, Chen, Cheng, Chia, Chincarini, Chiummo, Chmiel, Cho, Cho, Chow, Christensen, Chu, Chua, Chua, Chung, Chung, Ciani, Ciolfi, Cirelli, Cirone, Clara, Clark, Clearwater, Cleva, Cocchieri, Coccia, Cohadon, Cohen, Colla, Collette, Cominsky, Constancio, Conti, Cooper, Corban, Corbitt, Cordero-Carri\'on, Corley, Cornish, Corsi, Cortese, Costa, Coughlin,
  Coughlin, Coulon, Countryman, Couvares, Covas, Cowan, Coward, Cowart, Coyne, Coyne, Creighton, Creighton, Cripe, Crowder, Cullen, Cumming, Cunningham, Cuoco, Dal~Canton, D\'alya, Danilishin, D'Antonio, Danzmann, Dasgupta, Da~Silva~Costa, Dattilo, Dave, Davier, Davis, Daw, Day, De, DeBra, Degallaix, De~Laurentis, Del\'eglise, Del~Pozzo, Demos, Denker, Dent, De~Pietri, Dergachev, De~Rosa, DeRosa, De~Rossi, DeSalvo, de~Varona, Devenson, Dhurandhar, D\'{\i}az, Dietrich, Di~Fiore, Di~Giovanni, Di~Girolamo, Di~Lieto, Di~Pace, Di~Palma, Di~Renzo, Doctor, Dolique, Donovan, Dooley, Doravari, Dorrington, Douglas, Dovale~\'Alvarez, Downes, Drago, Dreissigacker, Driggers, Du, Ducrot, Dudi, Dupej, Dwyer, Edo, Edwards, Effler, Eggenstein, Ehrens, Eichholz, Eikenberry, Eisenstein, Essick, Estevez, Etienne, Etzel, Evans, Evans, Factourovich, Fafone, Fair, Fairhurst, Fan, Farinon, Farr, Farr, Fauchon-Jones, Favata, Fays, Fee, Fehrmann, Feicht, Fejer, Fernandez-Galiana, Ferrante, Ferreira, Ferrini, Fidecaro, Finstad, Fiori,
  Fiorucci, Fishbach, Fisher, Fitz-Axen, Flaminio, Fletcher, Fong, Font, Forsyth, Forsyth, Fournier, Frasca, Frasconi, Frei, Freise, Frey, Frey, Fries, Fritschel, Frolov, Fulda, Fyffe, Gabbard, Gadre, Gaebel, Gair, Gammaitoni, Ganija, Gaonkar, Garcia-Quiros, Garufi, Gateley, Gaudio, Gaur, Gayathri, Gehrels, Gemme, Genin, Gennai, George, George, Gergely, Germain, Ghonge, Ghosh, Ghosh, Ghosh, Giaime, Giardina, Giazotto, Gill, Glover, Goetz, Goetz, Gomes, Goncharov, Gonz\'alez, Gonzalez~Castro, Gopakumar, Gorodetsky, Gossan, Gosselin, Gouaty, Grado, Graef, Granata, Grant, Gras, Gray, Greco, Green, Gretarsson, Groot, Grote, Grunewald, Gruning, Guidi, Guo, Gupta, Gupta, Gushwa, Gustafson, Gustafson, Halim, Hall, Hall, Hamilton, Hammond, Haney, Hanke, Hanks, Hanna, Hannam, Hannuksela, Hanson, Hardwick, Harms, Harry, Harry, Hart, Haster, Haughian, Healy, Heidmann, Heintze, Heitmann, Hello, Hemming, Hendry, Heng, Hennig, Heptonstall, Heurs, Hild, Hinderer, Ho, Hoak, Hofman, Holt, Holz, Hopkins, Horst, Hough, Houston,
  Howell, Hreibi, Hu, Huerta, Huet, Hughey, Husa, Huttner, Huynh-Dinh, Indik, Inta, Intini, Isa, Isac, Isi, Iyer, Izumi, Jacqmin, Jani, Jaranowski, Jawahar, Jim\'enez-Forteza, Johnson, Johnson-McDaniel, Jones, Jones, Jonker, Ju, Junker, Kalaghatgi, Kalogera, Kamai, Kandhasamy, Kang, Kanner, Kapadia, Karki, Karvinen, Kasprzack, Kastaun, Katolik, Katsavounidis, Katzman, Kaufer, Kawabe, K\'ef\'elian, Keitel, Kemball, Kennedy, Kent, Key, Khalili, Khan, Khan, Khan, Khazanov, Kijbunchoo, Kim, Kim, Kim, Kim, Kim, Kim, Kimbrell, King, King, Kinley-Hanlon, Kirchhoff, Kissel, Kleybolte, Klimenko, Knowles, Koch, Koehlenbeck, Koley, Kondrashov, Kontos, Korobko, Korth, Kowalska, Kozak, Kr\"amer, Kringel, Krishnan, Kr\'olak, Kuehn, Kumar, Kumar, Kumar, Kuo, Kutynia, Kwang, Lackey, Lai, Landry, Lang, Lange, Lantz, Lanza, Larson, Lartaux-Vollard, Lasky, Laxen, Lazzarini, Lazzaro, Leaci, Leavey, Lee, Lee, Lee, Lee, Lee, Lehmann, Lenon, Leon, Leonardi, Leroy, Letendre, Levin, Li, Linker, Littenberg, Liu, Liu, Lo, Lockerbie,
  London, Lord, Lorenzini, Loriette, Lormand, Losurdo, Lough, Lousto, Lovelace, L\"uck, Lumaca, Lundgren, Lynch, Ma, Macas, Macfoy, Machenschalk, MacInnis, Macleod, Maga\~na Hernandez, Maga\~na Sandoval, Maga\~na Zertuche, Magee, Majorana, Maksimovic, Man, Mandic, Mangano, Mansell, Manske, Mantovani, Marchesoni, Marion, M\'arka, M\'arka, Markakis, Markosyan, Markowitz, Maros, Marquina, Marsh, Martelli, Martellini, Martin, Martin, Martynov, Marx, Mason, Massera, Masserot, Massinger, Masso-Reid, Mastrogiovanni, Matas, Matichard, Matone, Mavalvala, Mazumder, McCarthy, McClelland, McCormick, McCuller, McGuire, McIntyre, McIver, McManus, McNeill, McRae, McWilliams, Meacher, Meadors, Mehmet, Meidam, Mejuto-Villa, Melatos, Mendell, Mercer, Merilh, Merzougui, Meshkov, Messenger, Messick, Metzdorff, Meyers, Miao, Michel, Middleton, Mikhailov, Milano, Miller, Miller, Miller, Millhouse, Milovich-Goff, Minazzoli, Minenkov, Ming, Mishra, Mitra, Mitrofanov, Mitselmakher, Mittleman, Moffa, Moggi, Mogushi, Mohan, Mohapatra,
  Molina, Montani, Moore, Moraru, Moreno, Morisaki, Morriss, Mours, Mow-Lowry, Mueller, Muir, Mukherjee, Mukherjee, Mukherjee, Mukund, Mullavey, Munch, Mu\~niz, Muratore, Murray, Nagar, Napier, Nardecchia, Naticchioni, Nayak, Neilson, Nelemans, Nelson, Nery, Neunzert, Nevin, Newport, Newton, Ng, Nguyen, Nguyen, Nichols, Nielsen, Nissanke, Nitz, Noack, Nocera, Nolting, North, Nuttall, Oberling, O'Dea, Ogin, Oh, Oh, Ohme, Okada, Oliver, Oppermann, Oram, O'Reilly, Ormiston, Ortega, O'Shaughnessy, Ossokine, Ottaway, Overmier, Owen, Pace, Page, Page, Pai, Pai, Palamos, Palashov, Palomba, Pal-Singh, Pan, Pan, Pang, Pang, Pankow, Pannarale, Pant, Paoletti, Paoli, Papa, Parida, Parker, Pascucci, Pasqualetti, Passaquieti, Passuello, Patil, Patricelli, Pearlstone, Pedraza, Pedurand, Pekowsky, Pele, Penn, Perez, Perreca, Perri, Pfeiffer, Phelps, Piccinni, Pichot, Piergiovanni, Pierro, Pillant, Pinard, Pinto, Pirello, Pitkin, Poe, Poggiani, Popolizio, Porter, Post, Powell, Prasad, Pratt, Pratten, Predoi, Prestegard,
  Prijatelj, Principe, Privitera, Prix, Prodi, Prokhorov, Puncken, Punturo, Puppo, P\"urrer, Qi, Quetschke, Quintero, Quitzow-James, Raab, Rabeling, Radkins, Raffai, Raja, Rajan, Rajbhandari, Rakhmanov, Ramirez, Ramos-Buades, Rapagnani, Raymond, Razzano, Read, Regimbau, Rei, Reid, Reitze, Ren, Reyes, Ricci, Ricker, Rieger, Riles, Rizzo, Robertson, Robie, Robinet, Rocchi, Rolland, Rollins, Roma, Romano, Romano, Romel, Romie, Rosi\ifmmode~\acute{n}\else \'{n}\fi{}ska, Ross, Rowan, R\"udiger, Ruggi, Rutins, Ryan, Sachdev, Sadecki, Sadeghian, Sakellariadou, Salconi, Saleem, Salemi, Samajdar, Sammut, Sampson, Sanchez, Sanchez, Sanchis-Gual, Sandberg, Sanders, Sassolas, Sathyaprakash, Saulson, Sauter, Savage, Sawadsky, Schale, Scheel, Scheuer, Schmidt, Schmidt, Schnabel, Schofield, Sch\"onbeck, Schreiber, Schuette, Schulte, Schutz, Schwalbe, Scott, Scott, Seidel, Sellers, Sengupta, Sentenac, Sequino, Sergeev, Shaddock, Shaffer, Shah, Shahriar, Shaner, Shao, Shapiro, Shawhan, Sheperd, Shoemaker, Shoemaker, Siellez,
  Siemens, Sieniawska, Sigg, Silva, Singer, Singh, Singhal, Sintes, Slagmolen, Smith, Smith, Smith, Somala, Son, Sonnenberg, Sorazu, Sorrentino, Souradeep, Spencer, Srivastava, Staats, Staley, Steinke, Steinlechner, Steinlechner, Steinmeyer, Stevenson, Stone, Stops, Strain, Stratta, Strigin, Strunk, Sturani, Stuver, Summerscales, Sun, Sunil, Suresh, Sutton, Swinkels, Szczepa\ifmmode~\acute{n}\else \'{n}\fi{}czyk, Tacca, Tait, Talbot, Talukder, Tanner, T\'apai, Taracchini, Tasson, Taylor, Taylor, Tewari, Theeg, Thies, Thomas, Thomas, Thomas, Thorne, Thorne, Thrane, Tiwari, Tiwari, Tokmakov, Toland, Tonelli, Tornasi, Torres-Forn\'e, Torrie, T\"oyr\"a, Travasso, Traylor, Trinastic, Tringali, Trozzo, Tsang, Tse, Tso, Tsukada, Tsuna, Tuyenbayev, Ueno, Ugolini, Unnikrishnan, Urban, Usman, Vahlbruch, Vajente, Valdes, Vallisneri, van Bakel, van Beuzekom, van~den Brand, Van Den~Broeck, Vander-Hyde, van~der Schaaf, van Heijningen, van Veggel, Vardaro, Varma, Vass, Vas\'uth, Vecchio, Vedovato, Veitch, Veitch,
  Venkateswara, Venugopalan, Verkindt, Vetrano, Vicer\'e, Viets, Vinciguerra, Vine, Vinet, Vitale, Vo, Vocca, Vorvick, Vyatchanin, Wade, Wade, Wade, Walet, Walker, Wallace, Walsh, Wang, Wang, Wang, Wang, Wang, Ward, Warner, Was, Watchi, Weaver, Wei, Weinert, Weinstein, Weiss, Wen, Wessel, We\ss{}els, Westerweck, Westphal, Wette, Whelan, Whitcomb, Whiting, Whittle, Wilken, Williams, Williams, Williamson, Willis, Willke, Wimmer, Winkler, Wipf, Wittel, Woan, Woehler, Wofford, Wong, Worden, Wright, Wu, Wysocki, Xiao, Yamamoto, Yancey, Yang, Yap, Yazback, Yu, Yu, Yvert, Zadro\ifmmode~\dot{z}\else \.{z}\fi{}ny, Zanolin, Zelenova, Zendri, Zevin, Zhang, Zhang, Zhang, Zhang, Zhao, Zhou, Zhou, Zhu, Zhu, Zimmerman, Zucker, \& Zweizig}]{Abbott2017a}
Abbott, B.~P., Abbott, R., Abbott, T.~D., {et~al.} 2017{\natexlab{a}}, Phys. Rev. Lett., 119, 161101

\bibitem[{Abbott {et~al.}(2017{\natexlab{b}})Abbott, Abbott, Abbott, Acernese, Ackley, Adams, Adams, Addesso, Adhikari, Adya, Affeldt, Afrough, Agarwal, Agathos, Agatsuma, Aggarwal, Aguiar, Aiello, Ain, Ajith, Allen, Allen, Allocca, Altin, Amato, Ananyeva, Anderson, Anderson, Antier, Appert, Arai, Araya, Areeda, Arnaud, Arun, Ascenzi, Ashton, Ast, Aston, Astone, Aufmuth, Aulbert, AultONeal, Avila-Alvarez, Babak, Bacon, Bader, Bae, Baker, Baldaccini, Ballardin, Ballmer, Banagiri, Barayoga, Barclay, Barish, Barker, Barone, Barr, Barsotti, Barsuglia, Barta, Bartlett, Bartos, Bassiri, Basti, Batch, Baune, Bawaj, Bazzan, B\'ecsy, Beer, Bejger, Belahcene, Bell, Berger, Bergmann, Berry, Bersanetti, Bertolini, Betzwieser, Bhagwat, Bhandare, Bilenko, Billingsley, Billman, Birch, Birney, Birnholtz, Biscans, Bisht, Bitossi, Biwer, Bizouard, Blackburn, Blackman, Blair, Blair, Blair, Bloemen, Bock, Bode, Boer, Bogaert, Bohe, Bondu, Bonnand, Boom, Bork, Boschi, Bose, Bouffanais, Bozzi, Bradaschia, Brady, Braginsky,
  Branchesi, Brau, Briant, Brillet, Brinkmann, Brisson, Brockill, Broida, Brooks, Brown, Brown, Brown, Brunett, Buchanan, Buikema, Bulik, Bulten, Buonanno, Buskulic, Buy, Byer, Cabero, Cadonati, Cagnoli, Cahillane, Calder\'on~Bustillo, Callister, Calloni, Camp, Canepa, Canizares, Cannon, Cao, Cao, Capano, Capocasa, Carbognani, Caride, Carney, Casanueva~Diaz, Casentini, Caudill, Cavagli\`a, Cavalier, Cavalieri, Cella, Cepeda, Cerboni~Baiardi, Cerretani, Cesarini, Chamberlin, Chan, Chao, Charlton, Chassande-Mottin, Chatterjee, Chatziioannou, Cheeseboro, Chen, Chen, Cheng, Chincarini, Chiummo, Chmiel, Cho, Cho, Chow, Christensen, Chu, Chua, Chua, Chung, Chung, Ciani, Ciolfi, Cirelli, Cirone, Clara, Clark, Cleva, Cocchieri, Coccia, Cohadon, Colla, Collette, Cominsky, Constancio, Conti, Cooper, Corban, Corbitt, Corley, Cornish, Corsi, Cortese, Costa, Coughlin, Coughlin, Coulon, Countryman, Couvares, Covas, Cowan, Coward, Cowart, Coyne, Coyne, Creighton, Creighton, Cripe, Crowder, Cullen, Cumming, Cunningham,
  Cuoco, Dal~Canton, Danilishin, D'Antonio, Danzmann, Dasgupta, Da~Silva~Costa, Dattilo, Dave, Davier, Davis, Daw, Day, De, DeBra, Deelman, Degallaix, De~Laurentis, Del\'eglise, Del~Pozzo, Denker, Dent, Dergachev, De~Rosa, DeRosa, DeSalvo, Devenson, Devine, Dhurandhar, D\'{\i}az, Di~Fiore, Di~Giovanni, Di~Girolamo, Di~Lieto, Di~Pace, Di~Palma, Di~Renzo, Doctor, Dolique, Donovan, Dooley, Doravari, Dorrington, Douglas, Dovale~\'Alvarez, Downes, Drago, Drever, Driggers, Du, Ducrot, Duncan, Dwyer, Edo, Edwards, Effler, Eggenstein, Ehrens, Eichholz, Eikenberry, Eisenstein, Essick, Etienne, Etzel, Evans, Evans, Factourovich, Fafone, Fair, Fairhurst, Fan, Farinon, Farr, Farr, Fauchon-Jones, Favata, Fays, Fehrmann, Feicht, Fejer, Fernandez-Galiana, Ferrante, Ferreira, Ferrini, Fidecaro, Fiori, Fiorucci, Fisher, Flaminio, Fletcher, Fong, Forsyth, Forsyth, Fournier, Frasca, Frasconi, Frei, Freise, Frey, Frey, Fries, Fritschel, Frolov, Fulda, Fyffe, Gabbard, Gabel, Gadre, Gaebel, Gair, Gammaitoni, Ganija, Gaonkar,
  Garufi, Gaudio, Gaur, Gayathri, Gehrels, Gemme, Genin, Gennai, George, George, Gergely, Germain, Ghonge, Ghosh, Ghosh, Ghosh, Giaime, Giardina, Giazotto, Gill, Glover, Goetz, Goetz, Gomes, Gonz\'alez, Gonzalez~Castro, Gopakumar, Gorodetsky, Gossan, Gosselin, Gouaty, Grado, Graef, Granata, Grant, Gras, Gray, Greco, Green, Groot, Grote, Grunewald, Gruning, Guidi, Guo, Gupta, Gupta, Gushwa, Gustafson, Gustafson, Hall, Hall, Hammond, Haney, Hanke, Hanks, Hanna, Hannam, Hannuksela, Hanson, Hardwick, Harms, Harry, Harry, Hart, Haster, Haughian, Healy, Heidmann, Heintze, Heitmann, Hello, Hemming, Hendry, Heng, Hennig, Henry, Heptonstall, Heurs, Hild, Hoak, Hofman, Holt, Holz, Hopkins, Horst, Hough, Houston, Howell, Hu, Huerta, Huet, Hughey, Husa, Huttner, Huynh-Dinh, Indik, Ingram, Inta, Intini, Isa, Isac, Isi, Iyer, Izumi, Jacqmin, Jani, Jaranowski, Jawahar, Jim\'enez-Forteza, Johnson, Johnson-McDaniel, Jones, Jones, Jonker, Ju, Junker, Kalaghatgi, Kalogera, Kandhasamy, Kang, Kanner, Karki, Karvinen, Kasprzack,
  Katolik, Katsavounidis, Katzman, Kaufer, Kawabe, K\'ef\'elian, Keitel, Kemball, Kennedy, Kent, Key, Khalili, Khan, Khan, Khan, Khazanov, Kijbunchoo, Kim, Kim, Kim, Kim, Kim, Kimbrell, King, King, Kirchhoff, Kissel, Kleybolte, Klimenko, Koch, Koehlenbeck, Koley, Kondrashov, Kontos, Korobko, Korth, Kowalska, Kozak, Kr\"amer, Kringel, Krishnan, Kr\'olak, Kuehn, Kumar, Kumar, Kumar, Kuo, Kutynia, Kwang, Lackey, Lai, Landry, Lang, Lange, Lantz, Lanza, Lartaux-Vollard, Lasky, Laxen, Lazzarini, Lazzaro, Leaci, Leavey, Lee, Lee, Lee, Lee, Lee, Lehmann, Lenon, Leonardi, Leroy, Letendre, Levin, Li, Libson, Littenberg, Liu, Lo, Lockerbie, London, Lord, Lorenzini, Loriette, Lormand, Losurdo, Lough, Lovelace, L\"uck, Lumaca, Lundgren, Lynch, Ma, Macfoy, Machenschalk, MacInnis, Macleod, Maga\~na Hernandez, Maga\~na Sandoval, Maga\~na Zertuche, Magee, Majorana, Maksimovic, Man, Mandic, Mangano, Mansell, Manske, Mantovani, Marchesoni, Marion, M\'arka, M\'arka, Markakis, Markosyan, Maros, Martelli, Martellini, Martin,
  Martynov, Marx, Mason, Masserot, Massinger, Masso-Reid, Mastrogiovanni, Matas, Matichard, Matone, Mavalvala, Mayani, Mazumder, McCarthy, McClelland, McCormick, McCuller, McGuire, McIntyre, McIver, McManus, McRae, McWilliams, Meacher, Meadors, Meidam, Mejuto-Villa, Melatos, Mendell, Mercer, Merilh, Merzougui, Meshkov, Messenger, Messick, Metzdorff, Meyers, Mezzani, Miao, Michel, Middleton, Mikhailov, Milano, Miller, Miller, Miller, Miller, Millhouse, Minazzoli, Minenkov, Ming, Mishra, Mitra, Mitrofanov, Mitselmakher, Mittleman, Moggi, Mohan, Mohapatra, Montani, Moore, Moore, Moraru, Moreno, Morriss, Mours, Mow-Lowry, Mueller, Muir, Mukherjee, Mukherjee, Mukherjee, Mukund, Mullavey, Munch, Muniz, Murray, Napier, Nardecchia, Naticchioni, Nayak, Nelemans, Nelson, Neri, Nery, Neunzert, Newport, Newton, Ng, Nguyen, Nichols, Nielsen, Nissanke, Nitz, Noack, Nocera, Nolting, Normandin, Nuttall, Oberling, Ochsner, Oelker, Ogin, Oh, Oh, Ohme, Oliver, Oppermann, Oram, O'Reilly, Ormiston, Ortega, O'Shaughnessy, Ottaway,
  Overmier, Owen, Pace, Page, Page, Pai, Pai, Palamos, Palashov, Palomba, Pal-Singh, Pan, Pang, Pang, Pankow, Pannarale, Pant, Paoletti, Paoli, Papa, Paris, Parker, Pascucci, Pasqualetti, Passaquieti, Passuello, Patricelli, Pearlstone, Pedraza, Pedurand, Pekowsky, Pele, Penn, Perez, Perreca, Perri, Pfeiffer, Phelps, Piccinni, Pichot, Piergiovanni, Pierro, Pillant, Pinard, Pinto, Pitkin, Poggiani, Popolizio, Porter, Post, Powell, Prasad, Pratt, Predoi, Prestegard, Prijatelj, Principe, Privitera, Prodi, Prokhorov, Puncken, Punturo, Puppo, P\"urrer, Qi, Qin, Qiu, Quetschke, Quintero, Quitzow-James, Raab, Rabeling, Radkins, Raffai, Raja, Rajan, Rakhmanov, Ramirez, Rapagnani, Raymond, Razzano, Read, Regimbau, Rei, Reid, Reitze, Rew, Reyes, Ricci, Ricker, Rieger, Riles, Rizzo, Robertson, Robie, Robinet, Rocchi, Rolland, Rollins, Roma, Romano, Romano, Romel, Romie, Rosi\ifmmode~\acute{n}\else \'{n}\fi{}ska, Ross, Rowan, R\"udiger, Ruggi, Ryan, Rynge, Sachdev, Sadecki, Sadeghian, Sakellariadou, Salconi, Saleem,
  Salemi, Samajdar, Sammut, Sampson, Sanchez, Sandberg, Sandeen, Sanders, Sassolas, Sathyaprakash, Saulson, Sauter, Savage, Sawadsky, Schale, Scheuer, Schmidt, Schmidt, Schmidt, Schnabel, Schofield, Sch\"onbeck, Schreiber, Schuette, Schulte, Schutz, Schwalbe, Scott, Scott, Seidel, Sellers, Sengupta, Sentenac, Sequino, Sergeev, Shaddock, Shaffer, Shah, Shahriar, Shao, Shapiro, Shawhan, Sheperd, Shoemaker, Shoemaker, Siellez, Siemens, Sieniawska, Sigg, Silva, Singer, Singer, Singh, Singh, Singhal, Sintes, Slagmolen, Smith, Smith, Smith, Son, Sonnenberg, Sorazu, Sorrentino, Souradeep, Spencer, Srivastava, Staley, Steinke, Steinlechner, Steinlechner, Steinmeyer, Stephens, Stevenson, Stone, Strain, Stratta, Strigin, Sturani, Stuver, Summerscales, Sun, Sunil, Sutton, Swinkels, Szczepa\ifmmode~\acute{n}\else \'{n}\fi{}czyk, Tacca, Talukder, Tanner, T\'apai, Taracchini, Taylor, Taylor, Theeg, Thomas, Thomas, Thomas, Thorne, Thorne, Thrane, Tiwari, Tiwari, Tokmakov, Toland, Tonelli, Tornasi, Torrie, T\"oyr\"a,
  Travasso, Traylor, Trifir\`o, Trinastic, Tringali, Trozzo, Tsang, Tse, Tso, Tuyenbayev, Ueno, Ugolini, Unnikrishnan, Urban, Usman, Vahi, Vahlbruch, Vajente, Valdes, Vallisneri, van Bakel, van Beuzekom, van~den Brand, Van Den~Broeck, Vander-Hyde, van~der Schaaf, van Heijningen, van Veggel, Vardaro, Varma, Vass, Vas\'uth, Vecchio, Vedovato, Veitch, Veitch, Venkateswara, Venugopalan, Verkindt, Vetrano, Vicer\'e, Viets, Vinciguerra, Vine, Vinet, Vitale, Vo, Vocca, Vorvick, Voss, Vousden, Vyatchanin, Wade, Wade, Wade, Wald, Walet, Walker, Wallace, Walsh, Wang, Wang, Wang, Wang, Wang, Wang, Ward, Warner, Was, Watchi, Weaver, Wei, Weinert, Weinstein, Weiss, Wen, Wessel, We\ss{}els, Westphal, Wette, Whelan, Whiting, Whittle, Williams, Williams, Williamson, Willis, Willke, Wimmer, Winkler, Wipf, Wittel, Woan, Woehler, Wofford, Wong, Worden, Wright, Wu, Wu, Yam, Yamamoto, Yancey, Yap, Yu, Yu, Yvert, Zadro\ifmmode~\dot{z}\else \.{z}\fi{}ny, Zanolin, Zelenova, Zendri, Zevin, Zhang, Zhang, Zhang, Zhang, Zhao, Zhou,
  Zhou, Zhu, Zimmerman, Zucker, \& Zweizig}]{Abbott2017b}
Abbott, B.~P., Abbott, R., Abbott, T.~D., {et~al.} 2017{\natexlab{b}}, Phys. Rev. Lett., 118, 221101

\bibitem[{Abbott {et~al.}(2021)Abbott, Abbott, Abraham, Acernese, Ackley, Adams, Adams, Adhikari, Adya, Affeldt, Agathos, Agatsuma, Aggarwal, Aguiar, Aiello, Ain, Ajith, Akcay, Allen, Allocca, Altin, Amato, Anand, Ananyeva, Anderson, Anderson, Angelova, Ansoldi, Antelis, Antier, Appert, Arai, Araya, Areeda, Ar\`ene, Arnaud, Aronson, Arun, Asali, Ascenzi, Ashton, Aston, Astone, Aubin, Aufmuth, AultONeal, Austin, Avendano, Babak, Badaracco, Bader, Bae, Baer, Bagnasco, Baird, Ball, Ballardin, Ballmer, Bals, Balsamo, Baltus, Banagiri, Bankar, Bankar, Barayoga, Barbieri, Barish, Barker, Barneo, Barnum, Barone, Barr, Barsotti, Barsuglia, Barta, Bartlett, Bartos, Bassiri, Basti, Bawaj, Bayley, Bazzan, Becher, B\'ecsy, Bedakihale, Bejger, Belahcene, Beniwal, Benjamin, Bennett, Bentley, Bergamin, Berger, Bergmann, Bernuzzi, Berry, Bersanetti, Bertolini, Betzwieser, Bhandare, Bhandari, Bhattacharjee, Bidler, Bilenko, Billingsley, Birney, Birnholtz, Biscans, Bischi, Biscoveanu, Bisht, Bitossi, Bizouard, Blackburn,
  Blackman, Blair, Blair, Blair, Blanch, Bobba, Bode, Boer, Boetzel, Bogaert, Boldrini, Bondu, Bonilla, Bonnand, Booker, Boom, Bork, Boschi, Bose, Bossilkov, Boudart, Bouffanais, Bozzi, Bradaschia, Brady, Bramley, Branchesi, Brau, Breschi, Briant, Briggs, Brighenti, Brillet, Brinkmann, Brockill, Brooks, Brooks, Brown, Brunett, Bruno, Bruntz, Buikema, Bulik, Bulten, Buonanno, Buscicchio, Buskulic, Byer, Cabero, Cadonati, Caesar, Cagnoli, Cahillane, Calder\'on~Bustillo, Callaghan, Callister, Calloni, Camp, Canepa, Cannon, Cao, Cao, Carapella, Carbognani, Carney, Carpinelli, Carullo, Carver, Casanueva~Diaz, Casentini, Caudill, Cavagli\`a, Cavalier, Cavalieri, Cella, Cerd\'a-Dur\'an, Cesarini, Chaibi, Chakravarti, Chan, Chan, Chandra, Chanial, Chao, Charlton, Chase, Chassande-Mottin, Chatterjee, Chattopadhyay, Chaturvedi, Chatziioannou, Chen, Chen, Chen, Chen, Cheng, Cheong, Chia, Chiadini, Chierici, Chincarini, Chiummo, Cho, Cho, Cho, Choate, Christensen, Chu, Chua, Chung, Chung, Ciani, Ciecielag,
  Cie\ifmmode~\acute{s}\else \'{s}\fi{}lar, Cifaldi, Ciobanu, Ciolfi, Cipriano, Cirone, Clara, Clark, Clark, Clarke, Clearwater, Clesse, Cleva, Coccia, Cohadon, Cohen, Colleoni, Collette, Collins, Colpi, Constancio, Conti, Cooper, Corban, Corbitt, Cordero-Carri\'on, Corezzi, Corley, Cornish, Corre, Corsi, Cortese, Costa, Cotesta, Coughlin, Coughlin, Coulon, Countryman, Cousins, Couvares, Covas, Coward, Cowart, Coyne, Coyne, Creighton, Creighton, Croquette, Crowder, Cudell, Cullen, Cumming, Cummings, Cunningham, Cuoco, Cury\l{}o, Canton, D\'alya, Dana, DaneshgaranBajastani, D'Angelo, Danila, Danilishin, D'Antonio, Danzmann, Darsow-Fromm, Dasgupta, Datrier, Dattilo, Dave, Davier, Davies, Davis, Daw, Dean, DeBra, Deenadayalan, Degallaix, De~Laurentis, Del\'eglise, Del~Favero, De~Lillo, De~Lillo, Del~Pozzo, DeMarchi, De~Matteis, D'Emilio, Demos, Denker, Dent, Depasse, De~Pietri, De~Rosa, De~Rossi, DeSalvo, de~Varona, Dhurandhar, D\'{\i}az, Diaz-Ortiz, Didio, Dietrich, Di~Fiore, DiFronzo, Di~Giorgio, Di~Giovanni,
  Di~Giovanni, Di~Girolamo, Di~Lieto, Ding, Di~Pace, Di~Palma, Di~Renzo, Divakarla, Dmitriev, Doctor, D'Onofrio, Donovan, Dooley, Doravari, Dorrington, Downes, Drago, Driggers, Du, Ducoin, Dupej, Durante, D'Urso, Duverne, Dwyer, Easter, Eddolls, Edelman, Edo, Edy, Effler, Eichholz, Eikenberry, Eisenmann, Eisenstein, Ejlli, Errico, Essick, Estell\'es, Estevez, Etienne, Etzel, Evans, Evans, Ewing, Fafone, Fair, Fairhurst, Fan, Farah, Farinon, Farr, Farr, Fauchon-Jones, Favata, Fays, Fazio, Feicht, Fejer, Feng, Fenyvesi, Ferguson, Fernandez-Galiana, Ferrante, Ferreira, Fidecaro, Figura, Fiori, Fiorucci, Fishbach, Fisher, Fishner, Fittipaldi, Fitz-Axen, Fiumara, Flaminio, Floden, Flynn, Fong, Font, Forsyth, Fournier, Frasca, Frasconi, Frei, Freise, Frey, Frey, Fritschel, Frolov, Fronz\'e, Fulda, Fyffe, Gabbard, Gadre, Gaebel, Gair, Gais, Galaudage, Gamba, Ganapathy, Ganguly, Gaonkar, Garaventa, Garc\'{\i}a-Quir\'os, Garufi, Gateley, Gaudio, Gayathri, Gemme, Gennai, George, George, George, Gergely, Ghonge, Ghosh,
  Ghosh, Ghosh, Giacomazzo, Giacoppo, Giaime, Giardina, Gibson, Gier, Gill, Giri, Glanzer, Gleckl, Godwin, Goetz, Goetz, Gohlke, Goncharov, Gonz\'alez, Gopakumar, Gossan, Gosselin, Gouaty, Grace, Grado, Granata, Granata, Grant, Gras, Grassia, Gray, Gray, Greco, Green, Green, Gretarsson, Griggs, Grignani, Grimaldi, Grimes, Grimm, Grote, Grunewald, Gruning, Guerrero, Guidi, Guimaraes, Guix\'e, Gulati, Guo, Gupta, Gupta, Gupta, Gustafson, Gustafson, Guzman, Haegel, Halim, Hall, Hamilton, Hammond, Haney, Hanke, Hanks, Hanna, Hannam, Hannuksela, Hannuksela, Hansen, Hansen, Hanson, Harder, Hardwick, Haris, Harms, Harry, Harry, Hartwig, Hasskew, Haster, Haughian, Hayes, Healy, Heidmann, Heintze, Heinze, Heinzel, Heitmann, Hellman, Hello, Helmling-Cornell, Hemming, Hendry, Heng, Hennes, Hennig, Hennig, Hernandez~Vivanco, Heurs, Hild, Hill, Hines, Hochheim, Hofgard, Hofman, Hohmann, Holgado, Holland, Hollows, Holmes, Holt, Holz, Hopkins, Horst, Hough, Howell, Hoy, Hoyland, Huang, H\"ubner, Huddart, Huerta, Hughey,
  Hui, Husa, Huttner, Hutzler, Huxford, Huynh-Dinh, Idzkowski, Iess, Imperato, Inchauspe, Ingram, Intini, Isi, Iyer, JaberianHamedan, Jacqmin, Jadhav, Jadhav, James, Jani, Janssens, Janthalur, Jaranowski, Jariwala, Jaume, Jenkins, Jeunon, Jiang, Johns, Johnson-McDaniel, Jones, Jones, Jones, Jones, Jones, Jonker, Ju, Junker, Kalaghatgi, Kalogera, Kamai, Kandhasamy, Kang, Kanner, Kapadia, Kapasi, Karathanasis, Karki, Kashyap, Kasprzack, Kastaun, Katsanevas, Katsavounidis, Katzman, Kawabe, K\'ef\'elian, Keitel, Key, Khadka, Khalili, Khan, Khan, Khazanov, Khetan, Khursheed, Kijbunchoo, Kim, Kim, Kim, Kim, Kim, Kim, Kimball, King, Kinley-Hanlon, Kirchhoff, Kissel, Kleybolte, Klimenko, Knowles, Knyazev, Koch, Koehlenbeck, Koekoek, Koley, Kolstein, Komori, Kondrashov, Kontos, Koper, Korobko, Korth, Kovalam, Kozak, Kr\"amer, Kringel, Krishnendu, Kr\'olak, Kuehn, Kumar, Kumar, Kumar, Kumar, Kuns, Kwang, Lackey, Laghi, Lalande, Lam, Lamberts, Landry, Lane, Lang, Lange, Lantz, Lanza, La~Rosa, Lartaux-Vollard, Lasky,
  Laxen, Lazzarini, Lazzaro, Leaci, Leavey, Lecoeuche, Lee, Lee, Lee, Lee, Lehmann, Leon, Leroy, Letendre, Levin, Li, Li, Li, Li, Li, Linde, Linker, Linley, Littenberg, Liu, Liu, Llorens-Monteagudo, Lo, Lockwood, London, Longo, Lorenzini, Loriette, Lormand, Losurdo, Lough, Lousto, Lovelace, L\"uck, Lumaca, Lundgren, Ma, Macas, MacInnis, Macleod, MacMillan, Macquet, Maga\~na Hernandez, Maga\~na Sandoval, Magazz\`u, Magee, Majorana, Maksimovic, Maliakal, Malik, Man, Mandic, Mangano, Mansell, Manske, Mantovani, Mapelli, Marchesoni, Marion, M\'arka, M\'arka, Markakis, Markosyan, Markowitz, Maros, Marquina, Marsat, Martelli, Martin, Martin, Martinez, Martinez, Martynov, Masalehdan, Mason, Massera, Masserot, Massinger, Masso-Reid, Mastrogiovanni, Matas, Mateu-Lucena, Matichard, Matiushechkina, Mavalvala, Maynard, McCann, McCarthy, McClelland, McCormick, McCuller, McGuire, McIsaac, McIver, McManus, McRae, McWilliams, Meacher, Meadors, Mehmet, Mehta, Melatos, Melchor, Mendell, Menendez-Vazquez, Mercer, Mereni,
  Merfeld, Merilh, Merritt, Merzougui, Meshkov, Messenger, Messick, Metzdorff, Meyers, Meylahn, Mhaske, Miani, Miao, Michaloliakos, Michel, Middleton, Milano, Miller, Millhouse, Mills, Milotti, Milovich-Goff, Minazzoli, Minenkov, Mir, Mishkin, Mishra, Mistry, Mitra, Mitrofanov, Mitselmakher, Mittleman, Mo, Mogushi, Mohapatra, Mohite, Molina, Molina-Ruiz, Mondin, Montani, Moore, Moraru, Morawski, Moreno, Morisaki, Mours, Mow-Lowry, Mozzon, Muciaccia, Mukherjee, Mukherjee, Mukherjee, Mukherjee, Mukund, Mullavey, Munch, Mu\~niz, Murray, Nadji, Nagar, Nardecchia, Naticchioni, Nayak, Neil, Neilson, Nelemans, Nelson, Nery, Neunzert, Nitz, Ng, Ng, Nguyen, Nguyen, Nguyen, Nichols, Nissanke, Nocera, Noh, North, Nothard, Nuttall, Oberling, O'Brien, O'Dell, Oganesyan, Ogin, Oh, Oh, Ohme, Ohta, Okada, Olivetto, Oppermann, Oram, O'Reilly, Ormiston, Ortega, O'Shaughnessy, Ossokine, Osthelder, Ottaway, Overmier, Owen, Pace, Pagano, Page, Pagliaroli, Pai, Pai, Palamos, Palashov, Palomba, Pan, Panda, Pang, Pankow, Pannarale,
  Pant, Paoletti, Paoli, Paolone, Parker, Pascucci, Pasqualetti, Passaquieti, Passuello, Patel, Patricelli, Payne, Pechsiri, Pedraza, Pegoraro, Pele, Penn, Perego, Perez, P\'erigois, Perreca, Perri\`es, Petermann, Petterson, Pfeiffer, Pham, Phukon, Piccinni, Pichot, Piendibene, Piergiovanni, Pierini, Pierro, Pillant, Pilo, Pinard, Pinto, Piotrzkowski, Pirello, Pitkin, Placidi, Plastino, Pluchar, Poggiani, Polini, Pong, Ponrathnam, Popolizio, Porter, Poverman, Powell, Pracchia, Prajapati, Prasai, Prasanna, Pratten, Prestegard, Principe, Prodi, Prokhorov, Prosposito, Prudenzi, Puecher, Punturo, Puosi, Puppo, P\"urrer, Qi, Quetschke, Quinonez, Quitzow-James, Raab, Raaijmakers, Radkins, Radulesco, Raffai, Rafferty, Rail, Raja, Rajan, Rajbhandari, Rakhmanov, Ramirez, Ramirez, Ramos-Buades, Rana, Rao, Rapagnani, Rapol, Ratto, Raymond, Razzano, Read, Regimbau, Rei, Reid, Reitze, Rettegno, Ricci, Richardson, Richardson, Richardson, Ricker, Riemenschneider, Riles, Rizzo, Robertson, Robinet, Rocchi, Rocha, Rodriguez,
  Rodriguez-Soto, Rolland, Rollins, Roma, Romanelli, Romano, Romel, Romero, Romero-Shaw, Romie, Ronchini, Rose, Rose, Rose, Rosell, Rosi\ifmmode~\acute{n}\else \'{n}\fi{}ska, Rosofsky, Ross, Rowan, Rowlinson, Roy, Roy, Ruggi, Ryan, Sachdev, Sadecki, Sadiq, Sakellariadou, Salafia, Salconi, Saleem, Samajdar, Sanchez, Sanchez, Sanchez, Sanchis-Gual, Sanders, Sandles, Santiago, Santos, Saravanan, Sarin, Sassolas, Sathyaprakash, Sauter, Savage, Savant, Sawant, Sayah, Schaetzl, Schale, Scheel, Scheuer, Schindler-Tyka, Schmidt, Schnabel, Schofield, Sch\"onbeck, Schreiber, Schulte, Schutz, Schwarm, Schwartz, Scott, Scott, Seglar-Arroyo, Seidel, Sellers, Sengupta, Sennett, Sentenac, Sequino, Sergeev, Setyawati, Shaffer, Shahriar, Sharifi, Sharma, Sharma, Shawhan, Shen, Shikauchi, Shink, Shoemaker, Shoemaker, Shukla, ShyamSundar, Sieniawska, Sigg, Singer, Singh, Singh, Singha, Singhal, Sintes, Sipala, Skliris, Slagmolen, Slaven-Blair, Smetana, Smith, Smith, Somala, Son, Soni, Soni, Sorazu, Sordini, Sorrentino,
  Sorrentino, Soulard, Souradeep, Sowell, Spencer, Spera, Srivastava, Srivastava, Staats, Stachie, Steer, Steinhoff, Steinke, Steinlechner, Steinlechner, Steinmeyer, Stevenson, Stolle-McAllister, Stops, Stover, Strain, Stratta, Strunk, Sturani, Stuver, S\"udbeck, Sudhagar, Sudhir, Suh, Summerscales, Sun, Sun, Sunil, Sur, Suresh, Sutton, Swinkels, Szczepa\ifmmode~\acute{n}\else \'{n}\fi{}czyk, Tacca, Tait, Talbot, Tanasijczuk, Tanner, Tao, Tapia, Tapia San~Martin, Tasson, Taylor, Tenorio, Terkowski, Thirugnanasambandam, Thomas, Thomas, Thomas, Thompson, Thondapu, Thorne, Thrane, Tiwari, Tiwari, Tiwari, Toland, Tolley, Tonelli, Tornasi, Torres-Forn\'e, Torrie, e~Melo, T\"oyr\"a, Tran, Trapananti, Travasso, Traylor, Tringali, Tripathee, Trovato, Trudeau, Tsai, Tsang, Tse, Tso, Tsukada, Tsuna, Tsutsui, Turconi, Ubhi, Udall, Ueno, Ugolini, Unnikrishnan, Urban, Usman, Utina, Vahlbruch, Vajente, Vajpeyi, Valdes, Valentini, Valsan, van Bakel, van Beuzekom, van~den Brand, Van Den~Broeck, Vander-Hyde, van~der Schaaf,
  van Heijningen, Vardaro, Vargas, Varma, Vass, Vas\'uth, Vecchio, Vedovato, Veitch, Veitch, Venkateswara, Venneberg, Venugopalan, Verkindt, Verma, Veske, Vetrano, Vicer\'e, Viets, Vijaykumar, Villa-Ortega, Vinet, Vitale, Vo, Vocca, Vorvick, Vyatchanin, Wade, Wade, Wade, Walet, Walker, Wallace, Wallace, Walsh, Wang, Wang, Wang, Wang, Ward, Warner, Was, Washington, Watchi, Weaver, Wei, Weinert, Weinstein, Weiss, Wellmann, Wen, We\ss{}els, Westhouse, Wette, Whelan, White, White, Whiting, Whittle, Wilken, Williams, Williams, Williamson, Willis, Willke, Wilson, Wimmer, Winkler, Wipf, Woan, Woehler, Wofford, Wong, Wrangel, Wright, Wu, Wysocki, Xiao, Yamamoto, Yang, Yang, Yang, Yap, Yeeles, Yoon, Yu, Yu, Yuen, Zadro\ifmmode~\dot{z}\else \.{z}\fi{}ny, Zanolin, Zelenova, Zendri, Zevin, Zhang, Zhang, Zhang, Zhang, Zhao, Zhao, Zheng, Zhou, Zhou, Zhu, Zimmerman, Zlochower, Zucker, \& Zweizig}]{Abbott2021gwtc21}
Abbott, R., Abbott, T.~D., Abraham, S., {et~al.} 2021, Phys. Rev. X, 11, 021053

\bibitem[{{Abbott} {et~al.}(2020{\natexlab{b}}){Abbott}, {Abbott}, {Abraham}, {Acernese}, {Ackley}, {Adams}, {Adhikari}, {Adya}, {Affeldt}, {Agathos}, {Agatsuma}, {Aggarwal}, {Aguiar}, {Aich}, {Aiello}, {Ain}, {Ajith}, {Akcay}, {Allen}, {Allocca}, {Altin}, {Amato}, {Anand}, {Ananyeva}, {Anderson}, {Anderson}, {Angelova}, {Ansoldi}, {Antier}, {Appert}, {Arai}, {Araya}, {Areeda}, {Ar{\`e}ne}, {Arnaud}, {Aronson}, {Arun}, {Asali}, {Ascenzi}, {Ashton}, {Aston}, {Astone}, {Aubin}, {Aufmuth}, {AultONeal}, {Austin}, {Avendano}, {Babak}, {Bacon}, {Badaracco}, {Bader}, {Bae}, {Baer}, {Baird}, {Baldaccini}, {Ballardin}, {Ballmer}, {Bals}, {Balsamo}, {Baltus}, {Banagiri}, {Bankar}, {Bankar}, {Barayoga}, {Barbieri}, {Barish}, {Barker}, {Barkett}, {Barneo}, {Barone}, {Barr}, {Barsotti}, {Barsuglia}, {Barta}, {Bartlett}, {Bartos}, {Bassiri}, {Basti}, {Bawaj}, {Bayley}, {Bazzan}, {B{\'e}csy}, {Bejger}, {Belahcene}, {Bell}, {Beniwal}, {Benjamin}, {Bentley}, {Bergamin}, {Berger}, {Bergmann}, {Bernuzzi}, {Berry}, {Bersanetti},
  {Bertolini}, {Betzwieser}, {Bhandare}, {Bhandari}, {Bidler}, {Biggs}, {Bilenko}, {Billingsley}, {Birney}, {Birnholtz}, {Biscans}, {Bischi}, {Biscoveanu}, {Bisht}, {Bissenbayeva}, {Bitossi}, {Bizouard}, {Blackburn}, {Blackman}, {Blair}, {Blair}, {Blair}, {Bobba}, {Bode}, {Boer}, {Boetzel}, {Bogaert}, {Bondu}, {Bonilla}, {Bonnand}, {Booker}, {Boom}, {Bork}, {Boschi}, {Bose}, {Bossilkov}, {Bosveld}, {Bouffanais}, {Bozzi}, {Bradaschia}, {Brady}, {Bramley}, {Branchesi}, {Brau}, {Breschi}, {Briant}, {Briggs}, {Brighenti}, {Brillet}, {Brinkmann}, {Brockill}, {Brooks}, {Brooks}, {Brown}, {Brunett}, {Bruno}, {Bruntz}, {Buikema}, {Bulik}, {Bulten}, {Buonanno}, {Buscicchio}, {Buskulic}, {Byer}, {Cabero}, {Cadonati}, {Cagnoli}, {Cahillane}, {Calder{\'o}n Bustillo}, {Callaghan}, {Callister}, {Calloni}, {Camp}, {Canepa}, {Cannon}, {Cao}, {Cao}, {Carapella}, {Carbognani}, {Caride}, {Carney}, {Carullo}, {Casanueva Diaz}, {Casentini}, {Casta{\~n}eda}, {Caudill}, {Cavagli{\`a}}, {Cavalier}, {Cavalieri}, {Cella},
  {Cerd{\'a}-Dur{\'a}n}, {Cesarini}, {Chaibi}, {Chakravarti}, {Chan}, {Chan}, {Chandra}, {Chao}, {Charlton}, {Chase}, {Chassande-Mottin}, {Chatterjee}, {Chaturvedi}, {Chatziioannou}, {Chen}, \& {Chen}}]{LVK2020_PIgap}
{Abbott}, R., {Abbott}, T.~D., {Abraham}, S., {et~al.} 2020{\natexlab{b}}, \prl, 125, 101102

\bibitem[{Abbott {et~al.}(2020)Abbott, Abbott, Abraham, Acernese, Ackley, Adams, Adhikari, Adya, Affeldt, Agathos, Agatsuma, Aggarwal, Aguiar, Aich, Aiello, Ain, Ajith, Akcay, Allen, Allocca, Altin, Amato, Anand, Ananyeva, Anderson, Anderson, Angelova, Ansoldi, Antier, Appert, Arai, Araya, Areeda, Arène, Arnaud, Aronson, Arun, Asali, Ascenzi, Ashton, Aston, Astone, Aubin, Aufmuth, AultONeal, Austin, Avendano, Babak, Bacon, Badaracco, Bader, Bae, Baer, Baird, Baldaccini, Ballardin, Ballmer, Bals, Balsamo, Baltus, Banagiri, Bankar, Bankar, Barayoga, Barbieri, Barish, Barker, Barkett, Barneo, Barone, Barr, Barsotti, Barsuglia, Barta, Bartlett, Bartos, Bassiri, Basti, Bawaj, Bayley, Bazzan, Bécsy, Bejger, Belahcene, Bell, Beniwal, Benjamin, Benkel, Bentley, Bergamin, Berger, Bergmann, Bernuzzi, Berry, Bersanetti, Bertolini, Betzwieser, Bhandare, Bhandari, Bidler, Biggs, Bilenko, Billingsley, Birney, Birnholtz, Biscans, Bischi, Biscoveanu, Bisht, Bissenbayeva, Bitossi, Bizouard, Blackburn, Blackman, Blair,
  Blair, Blair, Bobba, Bode, Boer, Boetzel, Bogaert, Bondu, Bonilla, Bonnand, Booker, Boom, Bork, Boschi, Bose, Bossilkov, Bosveld, Bouffanais, Bozzi, Bradaschia, Brady, Bramley, Branchesi, Brau, Breschi, Briant, Briggs, Brighenti, Brillet, Brinkmann, Brito, Brockill, Brooks, Brooks, Brown, Brunett, Bruno, Bruntz, Buikema, Bulik, Bulten, Buonanno, Buskulic, Byer, Cabero, Cadonati, Cagnoli, Cahillane, Bustillo, Callaghan, Callister, Calloni, Camp, Canepa, Cannon, Cao, Cao, Carapella, Carbognani, Caride, Carney, Carullo, Diaz, Casentini, Castañeda, Caudill, Cavaglià, Cavalier, Cavalieri, Cella, Cerdá-Durán, Cesarini, Chaibi, Chakravarti, Chan, Chan, Chao, Charlton, Chase, Chassande-Mottin, Chatterjee, Chaturvedi, Chatziioannou, Chen, Chen, Chen, Cheng, Cheong, Chia, Chiadini, Chierici, Chincarini, Chiummo, Cho, Cho, Cho, Christensen, Chu, Chua, Chung, Chung, Ciani, Ciecielag, Cieślar, Ciobanu, Ciolfi, Cipriano, Cirone, Clara, Clark, Clearwater, Clesse, Cleva, Coccia, Cohadon, Cohen, Colleoni, Collette,
  Collins, Colpi, Constancio, Conti, Cooper, Corban, Corbitt, Cordero-Carrión, Corezzi, Corley, Cornish, Corre, Corsi, Cortese, Costa, Cotesta, Coughlin, Coughlin, Coulon, Countryman, Couvares, Covas, Coward, Cowart, Coyne, Coyne, Creighton, Creighton, Cripe, Croquette, Crowder, Cudell, Cullen, Cumming, Cummings, Cunningham, Cuoco, Curylo, Canton, Dálya, Dana, Daneshgaran-Bajastani, D’Angelo, Danilishin, D’Antonio, Danzmann, Darsow-Fromm, Dasgupta, Datrier, Dattilo, Dave, Davier, Davies, Davis, Daw, DeBra, Deenadayalan, Degallaix, Laurentis, Deléglise, Delfavero, Lillo, Pozzo, DeMarchi, D’Emilio, Demos, Dent, Pietri, Rosa, Rossi, DeSalvo, Varona, Dhurandhar, Díaz, Diaz-Ortiz, Dietrich, Fiore, Fronzo, Giorgio, Giovanni, Giovanni, Girolamo, Lieto, Ding, Pace, Palma, Renzo, Divakarla, Dmitriev, Doctor, Donovan, Dooley, Doravari, Dorrington, Downes, Drago, Driggers, Du, Ducoin, Dupej, Durante, D’Urso, Dwyer, Easter, Eddolls, Edelman, Edo, Edy, Effler, Ehrens, Eichholz, Eikenberry, Eisenmann,
  Eisenstein, Ejlli, Errico, Essick, Estelles, Estevez, Etienne, Etzel, Evans, Evans, Ewing, Fafone, Fairhurst, Fan, Farinon, Farr, Farr, Fauchon-Jones, Favata, Fays, Fazio, Feicht, Fejer, Feng, Fenyvesi, Ferguson, Fernandez-Galiana, Ferrante, Ferreira, Ferreira, Fidecaro, Fiori, Fiorucci, Fishbach, Fisher, Fittipaldi, Fitz-Axen, Fiumara, Flaminio, Floden, Flynn, Fong, Font, Forsyth, Fournier, Frasca, Frasconi, Frei, Freise, Frey, Frey, Fritschel, Frolov, Fronzè, Fulda, Fyffe, Gabbard, Gadre, Gaebel, Gair, Galaudage, Ganapathy, Ganguly, Gaonkar, García-Quirós, Garufi, Gateley, Gaudio, Gayathri, Gemme, Genin, Gennai, George, George, Gergely, Ghonge, Ghosh, Ghosh, Ghosh, Giacomazzo, Giaime, Giardina, Gibson, Gier, Gill, Glanzer, Gniesmer, Godwin, Goetz, Goetz, Gohlke, Goncharov, González, Gopakumar, Gossan, Gosselin, Gouaty, Grace, Grado, Granata, Grant, Gras, Grassia, Gray, Gray, Greco, Green, Green, Gretarsson, Griggs, Grignani, Grimaldi, Grimm, Grote, Grunewald, Gruning, Guidi, Guimaraes, Guixé, Gulati,
  Guo, Gupta, Gupta, Gupta, Gustafson, Gustafson, Haegel, Halim, Hall, Hamilton, Hammond, Haney, Hanke, Hanks, Hanna, Hannam, Hannuksela, Hansen, Hanson, Harder, Hardwick, Haris, Harms, Harry, Harry, Hasskew, Haster, Haughian, Hayes, Healy, Heidmann, Heintze, Heinze, Heitmann, Hellman, Hello, Hemming, Hendry, Heng, Hennes, Hennig, Heurs, Hild, Hinderer, Hoback, Hochheim, Hofgard, Hofman, Holgado, Holland, Holt, Holz, Hopkins, Horst, Hough, Howell, Hoy, Huang, Hübner, Huerta, Huet, Hughey, Hui, Husa, Huttner, Huxford, Huynh-Dinh, Idzkowski, Iess, Inchauspe, Ingram, Intini, Isac, Isi, Iyer, Jacqmin, Jadhav, Jadhav, James, Jani, Janthalur, Jaranowski, Jariwala, Jaume, Jenkins, Jiang, Johns, Johnson-McDaniel, Jones, Jones, Jones, Jones, Jones, Jonker, Ju, Junker, Kalaghatgi, Kalogera, Kamai, Kandhasamy, Kang, Kanner, Kapadia, Karki, Kashyap, Kasprzack, Kastaun, Katsanevas, Katsavounidis, Katzman, Kaufer, Kawabe, Kéfélian, Keitel, Keivani, Kennedy, Key, Khadka, Khalili, Khan, Khan, Khan, Khazanov, Khetan,
  Khursheed, Kijbunchoo, Kim, Kim, Kim, Kim, Kim, Kim, Kim, Kimball, King, Kinley-Hanlon, Kirchhoff, Kissel, Kleybolte, Klimenko, Knowles, Knyazev, Koch, Koehlenbeck, Koekoek, Koley, Kondrashov, Kontos, Koper, Korobko, Korth, Kovalam, Kozak, Kringel, Krishnendu, Królak, Krupinski, Kuehn, Kumar, Kumar, Kumar, Kumar, Kumar, Kuo, Kutynia, Lackey, Laghi, Lalande, Lam, Lamberts, Landry, Landry, Lane, Lang, Lange, Lantz, Lanza, Rosa, Lartaux-Vollard, Lasky, Laxen, Lazzarini, Lazzaro, Leaci, Leavey, Lecoeuche, Lee, Lee, Lee, Lee, Lee, Lehmann, Leroy, Letendre, Levin, Li, Li, li, Li, Li, Linde, Linker, Linley, Littenberg, Liu, Liu, Llorens-Monteagudo, Lo, Lockwood, London, Longo, Lorenzini, Loriette, Lormand, Losurdo, Lough, Lousto, Lovelace, Lück, Lumaca, Lundgren, Ma, Macas, Macfoy, MacInnis, Macleod, MacMillan, Macquet, Hernandez, Magaña-Sandoval, Magee, Majorana, Maksimovic, Malik, Man, Mandic, Mangano, Mansell, Manske, Mantovani, Mapelli, Marchesoni, Marion, Márka, Márka, Markakis, Markosyan, Markowitz,
  Maros, Marquina, Marsat, Martelli, Martin, Martin, Martinez, Martynov, Masalehdan, Mason, Massera, Masserot, Massinger, Masso-Reid, Mastrogiovanni, Matas, Matichard, Mavalvala, Maynard, McCann, McCarthy, McClelland, McCormick, McCuller, McGuire, McIsaac, McIver, McManus, McRae, McWilliams, Meacher, Meadors, Mehmet, Mehta, Villa, Melatos, Mendell, Mercer, Mereni, Merfeld, Merilh, Merritt, Merzougui, Meshkov, Messenger, Messick, Metzdorff, Meyers, Meylahn, Mhaske, Miani, Miao, Michaloliakos, Michel, Middleton, Milano, Miller, Millhouse, Mills, Milotti, Milovich-Goff, Minazzoli, Minenkov, Mishkin, Mishra, Mistry, Mitra, Mitrofanov, Mitselmakher, Mittleman, Mo, Mogushi, Mohapatra, Mohite, Molina-Ruiz, Mondin, Montani, Moore, Moraru, Morawski, Moreno, Morisaki, Mours, Mow-Lowry, Mozzon, Muciaccia, Mukherjee, Mukherjee, Mukherjee, Mukherjee, Mukund, Mullavey, Munch, Muñiz, Murray, Nagar, Nardecchia, Naticchioni, Nayak, Neil, Neilson, Nelemans, Nelson, Nery, Neunzert, Ng, Ng, Nguyen, Nguyen, Nichols, Nichols,
  Nissanke, Nocera, Noh, North, Nothard, Nuttall, Oberling, O’Brien, Oganesyan, Ogin, Oh, Oh, Ohme, Ohta, Okada, Oliver, Olivetto, Oppermann, Oram, O’Reilly, Ormiston, Ortega, O’Shaughnessy, Ossokine, Osthelder, Ottaway, Overmier, Owen, Pace, Pagano, Page, Pagliaroli, Pai, Pai, Palamos, Palashov, Palomba, Pan, Panda, Pang, Pankow, Pannarale, Pant, Paoletti, Paoli, Parida, Parker, Pascucci, Pasqualetti, Passaquieti, Passuello, Patricelli, Payne, Pearlstone, Pechsiri, Pedersen, Pedraza, Pele, Penn, Perego, Perez, Périgois, Perreca, Perriès, Petermann, Pfeiffer, Phelps, Phukon, Piccinni, Pichot, Piendibene, Piergiovanni, Pierro, Pillant, Pinard, Pinto, Piotrzkowski, Pirello, Pitkin, Plastino, Poggiani, Pong, Ponrathnam, Popolizio, Porter, Powell, Prajapati, Prasai, Prasanna, Pratten, Prestegard, Principe, Prodi, Prokhorov, Punturo, Puppo, Pürrer, Qi, Quetschke, Quinonez, Raab, Raaijmakers, Radkins, Radulesco, Raffai, Rafferty, Raja, Rajan, Rajbhandari, Rakhmanov, Ramirez, Ramos-Buades, Rana, Rao,
  Rapagnani, Raymond, Razzano, Read, Regimbau, Rei, Reid, Reitze, Rettegno, Ricci, Richardson, Richardson, Ricker, Riemenschneider, Riles, Rizzo, Robertson, Robinet, Rocchi, Rodriguez-Soto, Rolland, Rollins, Roma, Romanelli, Romano, Romel, Romero-Shaw, Romie, Rose, Rose, Rose, Rosińska, Rosofsky, Ross, Rowan, Rowlinson, Roy, Roy, Roy, Ruggi, Rutins, Ryan, Sachdev, Sadecki, Sakellariadou, Salafia, Salconi, Saleem, Salemi, Samajdar, Sanchez, Sanchez, Sanchis-Gual, Sanders, Santiago, Santos, Sarin, Sassolas, Sathyaprakash, Sauter, Savage, Savant, Sawant, Sayah, Schaetzl, Schale, Scheel, Scheuer, Schmidt, Schnabel, Schofield, Schönbeck, Schreiber, Schulte, Schutz, Schwarm, Schwartz, Scott, Scott, Seidel, Sellers, Sengupta, Sennett, Sentenac, Sequino, Sergeev, Setyawati, Shaddock, Shaffer, Shahriar, Sharma, Sharma, Shawhan, Shen, Shikauchi, Shink, Shoemaker, Shoemaker, Shukla, ShyamSundar, Siellez, Sieniawska, Sigg, Singer, Singh, Singh, Singha, Singhal, Sintes, Sipala, Skliris, Slagmolen, Slaven-Blair, Smetana,
  Smith, Smith, Somala, Son, Soni, Sorazu, Sordini, Sorrentino, Souradeep, Sowell, Spencer, Spera, Srivastava, Srivastava, Staats, Stachie, Standke, Steer, Steinhoff, Steinke, Steinlechner, Steinlechner, Steinmeyer, Stevenson, Stocks, Stops, Stover, Strain, Stratta, Strunk, Sturani, Stuver, Sudhagar, Sudhir, Summerscales, Sun, Sunil, Sur, Suresh, Sutton, Swinkels, Szczepańczyk, Tacca, Tait, Talbot, Tanasijczuk, Tanner, Tao, Tápai, Tapia, San~Martin, Tasson, Taylor, Tenorio, Terkowski, Thirugnanasambandam, Thomas, Thomas, Thompson, Thondapu, Thorne, Thrane, Tinsman, Saravanan, Tiwari, Tiwari, Tiwari, Toland, Tonelli, Tornasi, Torres-Forné, Torrie, e~Melo, Töyrä, Trail, Travasso, Traylor, Tringali, Tripathee, Trovato, Trudeau, Tsang, Tse, Tso, Tsukada, Tsuna, Tsutsui, Turconi, Ubhi, Ueno, Ugolini, Unnikrishnan, Urban, Usman, Utina, Vahlbruch, Vajente, Valdes, Valentini, Bakel, Beuzekom, Brand, Broeck, Vander-Hyde, Schaaf, Heijningen, Veggel, Vardaro, Varma, Vass, Vasúth, Vecchio, Vedovato, Veitch, Veitch,
  Venkateswara, Venugopalan, Verkindt, Veske, Vetrano, Viceré, Viets, Vinciguerra, Vine, Vinet, Vitale, Vivanco, Vo, Vocca, Vorvick, Vyatchanin, Wade, Wade, Wade, Walet, Walker, Wallace, Wallace, Walsh, Wang, Wang, Wang, Ward, Warden, Warner, Was, Watchi, Weaver, Wei, Weinert, Weinstein, Weiss, Wellmann, Wen, Weßels, Westhouse, Wette, Whelan, Whiting, Whittle, Wilken, Williams, Willis, Willke, Winkler, Wipf, Wittel, Woan, Woehler, Wofford, Wong, Wright, Wu, Wysocki, Xiao, Yamamoto, Yang, Yang, Yang, Yap, Yazback, Yeeles, Yu, Yu, Yuen, Zadrożny, Zadrożny, Zanolin, Zelenova, Zendri, Zevin, Zhang, Zhang, Zhang, Zhao, Zhao, Zhou, Zhou, Zhu, Zimmerman, Zucker, Zweizig, Collaboration, \& Collaboration}]{LVK2020_gap_1}
Abbott, R., Abbott, T.~D., Abraham, S., {et~al.} 2020, The Astrophysical Journal Letters, 896, L44

\bibitem[{Abbott {et~al.}(2023{\natexlab{a}})Abbott, Abbott, Acernese, Ackley, Adams, Adhikari, Adhikari, Adya, Affeldt, Agarwal, Agathos, Agatsuma, Aggarwal, Aguiar, Aiello, Ain, Ajith, Akcay, Akutsu, Albanesi, Allocca, Altin, Amato, Anand, Anand, Ananyeva, Anderson, Anderson, Ando, Andrade, Andres, Andri\ifmmode~\acute{c}\else \'{c}\fi{}, Angelova, Ansoldi, Antelis, Antier, Appert, Arai, Arai, Arai, Araki, Araya, Araya, Areeda, Ar\`ene, Aritomi, Arnaud, Arogeti, Aronson, Arun, Asada, Asali, Ashton, Aso, Assiduo, Aston, Astone, Aubin, Austin, Babak, Badaracco, Bader, Badger, Bae, Bae, Baer, Bagnasco, Bai, Baiotti, Baird, Bajpai, Ball, Ballardin, Ballmer, Balsamo, Baltus, Banagiri, Bankar, Barayoga, Barbieri, Barish, Barker, Barneo, Barone, Barr, Barsotti, Barsuglia, Barta, Bartlett, Barton, Bartos, Bassiri, Basti, Bawaj, Bayley, Baylor, Bazzan, B\'ecsy, Bedakihale, Bejger, Belahcene, Benedetto, Beniwal, Bennett, Bentley, BenYaala, Bergamin, Berger, Bernuzzi, Berry, Bersanetti, Bertolini, Betzwieser,
  Beveridge, Bhandare, Bhardwaj, Bhattacharjee, Bhaumik, Bilenko, Billingsley, Bini, Birney, Birnholtz, Biscans, Bischi, Biscoveanu, Bisht, Biswas, Bitossi, Bizouard, Blackburn, Blair, Blair, Blair, Bobba, Bode, Boer, Bogaert, Boldrini, Bonavena, Bondu, Bonilla, Bonnand, Booker, Boom, Bork, Boschi, Bose, Bose, Bossilkov, Boudart, Bouffanais, Bozzi, Bradaschia, Brady, Bramley, Branch, Branchesi, Brandt, Brau, Breschi, Briant, Briggs, Brillet, Brinkmann, Brockill, Brooks, Brooks, Brown, Brunett, Bruno, Bruntz, Bryant, Bulik, Bulten, Buonanno, Buscicchio, Buskulic, Buy, Byer, Davies, Cadonati, Cagnoli, Cahillane, Bustillo, Callaghan, Callister, Calloni, Cameron, Camp, Canepa, Canevarolo, Cannavacciuolo, Cannon, Cao, Cao, Capocasa, Capote, Carapella, Carbognani, Carlin, Carney, Carpinelli, Carrillo, Carullo, Carver, Diaz, Casentini, Castaldi, Caudill, Cavagli\`a, Cavalier, Cavalieri, Ceasar, Cella, Cerd\'a-Dur\'an, Cesarini, Chaibi, Chakravarti, Subrahmanya, Champion, Chan, Chan, Chan, Chan, Chan, Chandra,
  Chanial, Chao, Chapman-Bird, Charlton, Chase, Chassande-Mottin, Chatterjee, Chatterjee, Chatterjee, Chaturvedi, Chaty, Chatziioannou, Chen, Chen, Chen, Chen, Chen, Chen, Chen, Chen, Cheng, Cheong, Cheung, Chia, Chiadini, Chiang, Chiarini, Chierici, Chincarini, Chiofalo, Chiummo, Cho, Cho, Choudhary, Choudhary, Christensen, Chu, Chu, Chu, Chua, Chung, Ciani, Ciecielag, Cie\ifmmode~\acute{s}\else \'{s}\fi{}lar, Cifaldi, Ciobanu, Ciolfi, Cipriano, Cirone, Clara, Clark, Clark, Clarke, Clearwater, Clesse, Cleva, Coccia, Codazzo, Cohadon, Cohen, Cohen, Colleoni, Collette, Colombo, Colpi, Compton, Constancio, Conti, Cooper, Corban, Corbitt, Cordero-Carri\'on, Corezzi, Corley, Cornish, Corre, Corsi, Cortese, Costa, Cotesta, Coughlin, Coulon, Countryman, Cousins, Couvares, Coward, Cowart, Coyne, Coyne, Creighton, Creighton, Criswell, Croquette, Crowder, Cudell, Cullen, Cumming, Cummings, Cunningham, Cuoco, Cury\l{}o, Dabadie, Canton, Dall'Osso, D\'alya, Dana, DaneshgaranBajastani, D'Angelo, Danila, Danilishin,
  D'Antonio, Danzmann, Darsow-Fromm, Dasgupta, Datrier, Dattilo, Dave, Davier, Davis, Davis, Daw, de~Alarc\'on, Dean, DeBra, Deenadayalan, Degallaix, De~Laurentis, Del\'eglise, Del~Favero, De~Lillo, De~Lillo, Del~Pozzo, DeMarchi, De~Matteis, D'Emilio, Demos, Dent, Depasse, De~Pietri, De~Rosa, De~Rossi, DeSalvo, De~Simone, Dhurandhar, D\'{\i}az, Diaz-Ortiz, Didio, Dietrich, Di~Fiore, Di~Fronzo, Di~Giorgio, Di~Giovanni, Di~Giovanni, Di~Girolamo, Di~Lieto, Ding, Di~Pace, Di~Palma, Di~Renzo, Divakarla, Dmitriev, Doctor, D'Onofrio, Donovan, Dooley, Doravari, Dorrington, Drago, Driggers, Drori, Ducoin, Dupej, Durante, D'Urso, Duverne, Dwyer, Eassa, Easter, Ebersold, Eckhardt, Eddolls, Edelman, Edo, Edy, Effler, Eguchi, Eichholz, Eikenberry, Eisenmann, Eisenstein, Ejlli, Engelby, Enomoto, Errico, Essick, Estell\'es, Estevez, Etienne, Etzel, Evans, Evans, Ewing, Fafone, Fair, Fairhurst, Farah, Farinon, Farr, Farr, Farrow, Fauchon-Jones, Favaro, Favata, Fays, Fazio, Feicht, Fejer, Fenyvesi, Ferguson,
  Fernandez-Galiana, Ferrante, Ferreira, Fidecaro, Figura, Fiori, Fishbach, Fisher, Fittipaldi, Fiumara, Flaminio, Floden, Fong, Font, Fornal, Forsyth, Franke, Frasca, Frasconi, Frederick, Freed, Frei, Freise, Frey, Fritschel, Frolov, Fronz\'e, Fujii, Fujikawa, Fukunaga, Fukushima, Fulda, Fyffe, Gabbard, Gabella, Gadre, Gair, Gais, Galaudage, Gamba, Ganapathy, Ganguly, Gao, Gaonkar, Garaventa, Garc\'{\i}a, Garc\'{\i}a-N\'u\~nez, Garc\'{\i}a-Quir\'os, Garufi, Gateley, Gaudio, Gayathri, Ge, Gemme, Gennai, George, George, Gerberding, Gergely, Gewecke, Ghonge, Ghosh, Ghosh, Ghosh, Ghosh, Giacomazzo, Giacoppo, Giaime, Giardina, Gibson, Gier, Giesler, Giri, Gissi, Glanzer, Gleckl, Godwin, Goetz, Goetz, Gohlke, Golomb, Goncharov, Gonz\'alez, Gopakumar, Gosselin, Gouaty, Gould, Grace, Grado, Granata, Granata, Grant, Gras, Grassia, Gray, Gray, Greco, Green, Green, Gretarsson, Gretarsson, Griffith, Griffiths, Griggs, Grignani, Grimaldi, Grimm, Grote, Grunewald, Gruning, Guerra, Guidi, Guimaraes, Guix\'e, Gulati, Guo,
  Guo, Gupta, Gupta, Gupta, Gustafson, Gustafson, Guzman, Ha, Haegel, Hagiwara, Haino, Halim, Hall, Hamilton, Hammond, Han, Haney, Hanks, Hanna, Hannam, Hannuksela, Hansen, Hansen, Hanson, Harder, Hardwick, Haris, Harms, Harry, Harry, Hartwig, Hasegawa, Haskell, Hasskew, Haster, Hattori, Haughian, Hayakawa, Hayama, Hayes, Healy, Heidmann, Heidt, Heintze, Heinze, Heinzel, Heitmann, Hellman, Hello, Helmling-Cornell, Hemming, Hendry, Heng, Hennes, Hennig, Hennig, Hernandez, Hernandez~Vivanco, Heurs, Hild, Hill, Himemoto, Hines, Hiranuma, Hirata, Hirose, Hochheim, Hofman, Hohmann, Holcomb, Holland, Holley-Bockelmann, Hollows, Holmes, Holt, Holz, Hong, Hopkins, Hough, Hourihane, Howell, Hoy, Hoyland, Hreibi, Hsieh, Hsu, Huang, Huang, Huang, Huang, Huang, Huang, H\"ubner, Huddart, Hughey, Hui, Hui, Husa, Huttner, Huxford, Huynh-Dinh, Ide, Idzkowski, Iess, Ikenoue, Imam, Inayoshi, Ingram, Inoue, Ioka, Isi, Isleif, Ito, Itoh, Iyer, Izumi, JaberianHamedan, Jacqmin, Jadhav, Jadhav, James, Jan, Jani, Janquart, Janssens,
  Janthalur, Jaranowski, Jariwala, Jaume, Jenkins, Jenner, Jeon, Jeunon, Jia, Jin, Johns, Johnson-McDaniel, Jones, Jones, Jones, Jones, Jones, Jonker, Ju, Jung, Jung, Junker, Juste, Kaihotsu, Kajita, Kakizaki, Kalaghatgi, Kalogera, Kamai, Kamiizumi, Kanda, Kandhasamy, Kang, Kanner, Kao, Kapadia, Kapasi, Karat, Karathanasis, Karki, Kashyap, Kasprzack, Kastaun, Katsanevas, Katsavounidis, Katzman, Kaur, Kawabe, Kawaguchi, Kawai, Kawasaki, K\'ef\'elian, Keitel, Key, Khadka, Khalili, Khan, Khazanov, Khetan, Khursheed, Kijbunchoo, Kim, Kim, Kim, Kim, Kim, Kim, Kimball, Kimura, Kinley-Hanlon, Kirchhoff, Kissel, Kita, Kitazawa, Kleybolte, Klimenko, Knee, Knowles, Knyazev, Koch, Koekoek, Kojima, Kokeyama, Koley, Kolitsidou, Kolstein, Komori, Kondrashov, Kong, Kontos, Koper, Korobko, Kotake, Kovalam, Kozak, Kozakai, Kozu, Kringel, Krishnendu, Kr\'olak, Kuehn, Kuei, Kuijer, Kulkarni, Kumar, Kumar, Kumar, Kumar, Kume, Kuns, Kuo, Kuo, Kuromiya, Kuroyanagi, Kusayanagi, Kuwahara, Kwak, Lagabbe, Laghi, Lalande, Lam,
  Lamberts, Landry, Lane, Lang, Lange, Lantz, La~Rosa, Lartaux-Vollard, Lasky, Laxen, Lazzarini, Lazzaro, Leaci, Leavey, Lecoeuche, Lee, Lee, Lee, Lee, Lee, Lee, Lehmann, Lema\^{\i}tre, Leonardi, Leroy, Letendre, Levesque, Levin, Leviton, Leyde, Li, Li, Li, Li, Li, Li, Lin, Lin, Lin, Lin, Lin, Linde, Linker, Linley, Littenberg, Liu, Liu, Liu, Liu, Llamas, Llorens-Monteagudo, Lo, Lockwood, Loh, London, Longo, Lopez, Portilla, Lorenzini, Loriette, Lormand, Losurdo, Lott, Lough, Lousto, Lovelace, Lucaccioni, L\"uck, Lumaca, Lundgren, Luo, Lynam, Macas, MacInnis, Macleod, MacMillan, Macquet, Hernandez, Magazz\`u, Magee, Maggiore, Magnozzi, Mahesh, Majorana, Makarem, Maksimovic, Maliakal, Malik, Man, Mandic, Mangano, Mango, Mansell, Manske, Mantovani, Mapelli, Marchesoni, Marchio, Marion, Mark, M\'arka, M\'arka, Markakis, Markosyan, Markowitz, Maros, Marquina, Marsat, Martelli, Martin, Martin, Martinez, Martinez, Martinez, Martinovic, Martynov, Marx, Masalehdan, Mason, Massera, Masserot, Massinger, Masso-Reid,
  Mastrogiovanni, Matas, Mateu-Lucena, Matichard, Matiushechkina, Mavalvala, McCann, McCarthy, McClelland, McClincy, McCormick, McCuller, McGhee, McGuire, McIsaac, McIver, McRae, McWilliams, Meacher, Mehmet, Mehta, Meijer, Melatos, Melchor, Mendell, Menendez-Vazquez, Menoni, Mercer, Mereni, Merfeld, Merilh, Merritt, Merzougui, Meshkov, Messenger, Messick, Meyers, Meylahn, Mhaske, Miani, Miao, Michaloliakos, Michel, Michimura, Middleton, Milano, Miller, Miller, Miller, Millhouse, Mills, Milotti, Minazzoli, Minenkov, Mio, Mir, Miravet-Ten\'es, Mishra, Mishra, Mistry, Mitra, Mitrofanov, Mitselmakher, Mittleman, Miyakawa, Miyamoto, Miyazaki, Miyo, Miyoki, Mo, Modafferi, Moguel, Mogushi, Mohapatra, Mohite, Molina, Molina-Ruiz, Mondin, Montani, Moore, Moraru, Morawski, More, Moreno, Moreno, Mori, Morisaki, Moriwaki, Morr\'as, Mours, Mow-Lowry, Mozzon, Muciaccia, Mukherjee, Mukherjee, Mukherjee, Mukherjee, Mukherjee, Mukund, Mullavey, Munch, Mu\~niz, Murray, Musenich, Muusse, Nadji, Nagano, Nagano, Nagar, Nakamura,
  Nakano, Nakano, Nakashima, Nakayama, Napolano, Nardecchia, Narikawa, Naticchioni, Nayak, Nayak, Negishi, Neil, Neilson, Nelemans, Nelson, Nery, Neubauer, Neunzert, Ng, Ng, Nguyen, Nguyen, Nguyen, Quynh, Ni, Nichols, Nishizawa, Nissanke, Nitoglia, Nocera, Norman, North, Nozaki, Siles, Nuttall, Oberling, O'Brien, Obuchi, O'Dell, Oelker, Ogaki, Oganesyan, Oh, Oh, Oh, Ohashi, Ohishi, Ohkawa, Ohme, Ohta, Okada, Okutani, Okutomi, Olivetto, Oohara, Ooi, Oram, O'Reilly, Ormiston, Ormsby, Ortega, O'Shaughnessy, O'Shea, Oshino, Ossokine, Osthelder, Otabe, Ottaway, Overmier, Pace, Pagano, Page, Pagliaroli, Pai, Pai, Palamos, Palashov, Palomba, Pan, Pan, Panda, Pang, Pang, Pankow, Pannarale, Pant, Panther, Paoletti, Paoli, Paolone, Parisi, Park, Park, Parker, Pascucci, Pasqualetti, Passaquieti, Passuello, Patel, Pathak, Patricelli, Patron, Paul, Payne, Pedraza, Pegoraro, Pele, Arellano, Penn, Perego, Pereira, Pereira, Perez, P\'erigois, Perkins, Perreca, Perri\`es, Petermann, Petterson, Pfeiffer, Pham, Phukon,
  Piccinni, Pichot, Piendibene, Piergiovanni, Pierini, Pierro, Pillant, Pillas, Pilo, Pinard, Pinto, Pinto, Piotrzkowski, Piotrzkowski, Pirello, Pitkin, Placidi, Planas, Plastino, Pluchar, Poggiani, Polini, Pong, Ponrathnam, Popolizio, Porter, Poulton, Powell, Pracchia, Pradier, Prajapati, Prasai, Prasanna, Pratten, Principe, Prodi, Prokhorov, Prosposito, Prudenzi, Puecher, Punturo, Puosi, Puppo, P\"urrer, Qi, Quetschke, Quitzow-James, Qutob, Raab, Raaijmakers, Radkins, Radulesco, Raffai, Rail, Raja, Rajan, Ramirez, Ramirez, Ramos-Buades, Rana, Rapagnani, Rapol, Ray, Raymond, Raza, Razzano, Read, Rees, Regimbau, Rei, Reid, Reid, Reitze, Relton, Renzini, Rettegno, Reza, Rezac, Ricci, Richards, Richardson, Richardson, Riemenschneider, Riles, Rinaldi, Rink, Rizzo, Robertson, Robie, Robinet, Rocchi, Rodriguez, Rolland, Rollins, Romanelli, Romano, Romel, Romero-Rodr\'{\i}guez, Romero-Shaw, Romie, Ronchini, Rosa, Rose, Rosi\ifmmode~\acute{n}\else \'{n}\fi{}ska, Ross, Rowan, Rowlinson, Roy, Roy, Roy, Rozza, Ruggi,
  Ruiz-Rocha, Ryan, Sachdev, Sadecki, Sadiq, Sago, Saito, Saito, Sakai, Sakai, Sakellariadou, Sakuno, Salafia, Salconi, Saleem, Salemi, Samajdar, Sanchez, Sanchez, Sanchez, Sanchis-Gual, Sanders, Sanuy, Saravanan, Sarin, Sassolas, Satari, Sathyaprakash, Sato, Sato, Sauter, Savage, Sawada, Sawant, Sawant, Sayah, Schaetzl, Scheel, Scheuer, Schiworski, Schmidt, Schmidt, Schnabel, Schneewind, Schofield, Sch\"onbeck, Schulte, Schutz, Schwartz, Scott, Scott, Seglar-Arroyo, Sekiguchi, Sekiguchi, Sellers, Sengupta, Sentenac, Seo, Sequino, Sergeev, Setyawati, Shaffer, Shahriar, Shams, Shao, Sharma, Sharma, Shawhan, Shcheblanov, Shibagaki, Shikauchi, Shimizu, Shimoda, Shimode, Shinkai, Shishido, Shoda, Shoemaker, Shoemaker, ShyamSundar, Sieniawska, Sigg, Singer, Singh, Singh, Singha, Sintes, Sipala, Skliris, Slagmolen, Slaven-Blair, Smetana, Smith, Smith, Soldateschi, Somala, Somiya, Son, Soni, Soni, Sordini, Sorrentino, Sorrentino, Sotani, Soulard, Souradeep, Sowell, Spagnuolo, Spencer, Spera, Srinivasan, Srivastava,
  Srivastava, Staats, Stachie, Steer, Steinhoff, Steinlechner, Steinlechner, Stevenson, Stops, Stover, Strain, Strang, Stratta, Strunk, Sturani, Stuver, Sudhagar, Sudhir, Sugimoto, Suh, Sullivan, Sullivan, Summerscales, Sun, Sun, Sunil, Sur, Suresh, Sutton, Suzuki, Suzuki, Swinkels, Szczepa\ifmmode~\acute{n}\else \'{n}\fi{}czyk, Szewczyk, Tacca, Tagoshi, Tait, Takahashi, Takahashi, Takamori, Takano, Takeda, Takeda, Talbot, Talbot, Tanaka, Tanaka, Tanaka, Tanaka, Tanaka, Tanasijczuk, Tanioka, Tanner, Tao, Tao, Mart\'{\i}n, Taranto, Tasson, Telada, Tenorio, Terhune, Terkowski, Thirugnanasambandam, Thomas, Thomas, Thomas, Thompson, Thondapu, Thorne, Thrane, Tiwari, Tiwari, Tiwari, Toivonen, Toland, Tolley, Tomaru, Tomigami, Tomura, Tonelli, Torres-Forn\'e, Torrie, e~Melo, T\"oyr\"a, Trapananti, Travasso, Traylor, Trevor, Tringali, Tripathee, Troiano, Trovato, Trozzo, Trudeau, Tsai, Tsai, Tsang, Tsang, Tsao, Tse, Tso, Tsubono, Tsuchida, Tsukada, Tsuna, Tsutsui, Tsuzuki, Turbang, Turconi, Tuyenbayev, Ubhi,
  Uchikata, Uchiyama, Udall, Ueda, Uehara, Ueno, Ueshima, Unnikrishnan, Uraguchi, Urban, Ushiba, Utina, Vahlbruch, Vajente, Vajpeyi, Valdes, Valentini, Valsan, van Bakel, van Beuzekom, van~den Brand, Van Den~Broeck, Vander-Hyde, van~der Schaaf, van Heijningen, Vanosky, van Putten, van Remortel, Vardaro, Vargas, Varma, Vas\'uth, Vecchio, Vedovato, Veitch, Veitch, Venneberg, Venugopalan, Verkindt, Verma, Verma, Veske, Vetrano, Vicer\'e, Vidyant, Viets, Vijaykumar, Villa-Ortega, Vinet, Virtuoso, Vitale, Vo, Vocca, von Reis, von Wrangel, Vorvick, Vyatchanin, Wade, Wade, Wagner, Walet, Walker, Wallace, Wallace, Walsh, Wang, Wang, Wang, Ward, Warner, Was, Washimi, Washington, Watchi, Weaver, Webster, Weinert, Weinstein, Weiss, Weller, Weller, Wellmann, Wen, We\ss{}els, Wette, Whelan, White, Whiting, Whittle, Wilken, Williams, Williams, Williams, Williamson, Willis, Willke, Wilson, Winkler, Wipf, Wlodarczyk, Woan, Woehler, Wofford, Wong, Wu, Wu, Wu, Wu, Wysocki, Xiao, Xu, Yamada, Yamamoto, Yamamoto, Yamamoto,
  Yamamoto, Yamashita, Yamazaki, Yang, Yang, Yang, Yang, Yang, Yap, Yeeles, Yelikar, Ying, Yokogawa, Yokoyama, Yokozawa, Yoo, Yoshioka, Yu, Yu, Yuzurihara, Zadro\ifmmode~\dot{z}\else \.{z}\fi{}ny, Zanolin, Zeidler, Zelenova, Zendri, Zevin, Zhan, Zhang, Zhang, Zhang, Zhang, Zhang, Zhao, Zhao, Zhao, Zhao, Zheng, Zhou, Zhou, Zhu, Zhu, Zimmerman, Zlochower, Zucker, \& Zweizig}]{Abbott2023gwtc3}
Abbott, R., Abbott, T.~D., Acernese, F., {et~al.} 2023{\natexlab{a}}, Phys. Rev. X, 13, 041039

\bibitem[{Abbott {et~al.}(2023{\natexlab{b}})Abbott, Abbott, Acernese, Ackley, Adams, Adhikari, Adhikari, Adya, Affeldt, Agarwal, Agathos, Agatsuma, Aggarwal, Aguiar, Aiello, Ain, Ajith, Akutsu, de~Alarc\'on, Akcay, Albanesi, Allocca, Altin, Amato, Anand, Anand, Ananyeva, Anderson, Anderson, Ando, Andrade, Andres, Andri\ifmmode~\acute{c}\else \'{c}\fi{}, Angelova, Ansoldi, Antelis, Antier, Antonini, Appert, Arai, Arai, Arai, Araki, Araya, Araya, Areeda, Ar\`ene, Aritomi, Arnaud, Arogeti, Aronson, Arun, Asada, Asali, Ashton, Aso, Assiduo, Aston, Astone, Aubin, Austin, Babak, Badaracco, Bader, Badger, Bae, Bae, Baer, Bagnasco, Bai, Baiotti, Baird, Bajpai, Ball, Ballardin, Ballmer, Balsamo, Baltus, Banagiri, Bankar, Barayoga, Barbieri, Barish, Barker, Barneo, Barone, Barr, Barsotti, Barsuglia, Barta, Bartlett, Barton, Bartos, Bassiri, Basti, Bawaj, Bayley, Baylor, Bazzan, B\'ecsy, Bedakihale, Bejger, Belahcene, Benedetto, Beniwal, Bennett, Bentley, BenYaala, Bergamin, Berger, Bernuzzi, Berry, Bersanetti,
  Bertolini, Betzwieser, Beveridge, Bhandare, Bhardwaj, Bhattacharjee, Bhaumik, Bilenko, Billingsley, Bini, Birney, Birnholtz, Biscans, Bischi, Biscoveanu, Bisht, Biswas, Bitossi, Bizouard, Blackburn, Blair, Blair, Blair, Bobba, Bode, Boer, Bogaert, Boldrini, Bonavena, Bondu, Bonilla, Bonnand, Booker, Boom, Bork, Boschi, Bose, Bose, Bossilkov, Boudart, Bouffanais, Bozzi, Bradaschia, Brady, Bramley, Branch, Branchesi, Brandt, Brau, Breschi, Briant, Briggs, Brillet, Brinkmann, Brockill, Brooks, Brooks, Brown, Brunett, Bruno, Bruntz, Bryant, Bulik, Bulten, Buonanno, Buscicchio, Buskulic, Buy, Byer, Cadonati, Cagnoli, Cahillane, Bustillo, Callaghan, Callister, Calloni, Cameron, Camp, Canepa, Canevarolo, Cannavacciuolo, Cannon, Cao, Cao, Capocasa, Capote, Carapella, Carbognani, Carlin, Carney, Carpinelli, Carrillo, Carullo, Carver, Diaz, Casentini, Castaldi, Caudill, Cavagli\`a, Cavalier, Cavalieri, Ceasar, Cella, Cerd\'a-Dur\'an, Cesarini, Chaibi, Chakravarti, Subrahmanya, Champion, Chan, Chan, Chan, Chan, Chan,
  Chandra, Chanial, Chao, Chapman-Bird, Charlton, Chase, Chassande-Mottin, Chatterjee, Chatterjee, Chatterjee, Chaturvedi, Chaty, Chatziioannou, Chen, Chen, Chen, Chen, Chen, Chen, Chen, Chen, Cheng, Cheong, Cheung, Chia, Chiadini, Chiang, Chiarini, Chierici, Chincarini, Chiofalo, Chiummo, Cho, Cho, Choudhary, Choudhary, Christensen, Chu, Chu, Chu, Chua, Chung, Ciani, Ciecielag, Cie\ifmmode~\acute{s}\else \'{s}\fi{}lar, Cifaldi, Ciobanu, Ciolfi, Cipriano, Cirone, Clara, Clark, Clark, Clarke, Clearwater, Clesse, Cleva, Coccia, Codazzo, Cohadon, Cohen, Cohen, Colleoni, Collette, Colombo, Colpi, Compton, Constancio, Conti, Cooper, Corban, Corbitt, Cordero-Carri\'on, Corezzi, Corley, Cornish, Corre, Corsi, Cortese, Costa, Cotesta, Coughlin, Coulon, Countryman, Cousins, Couvares, Coward, Cowart, Coyne, Coyne, Creighton, Creighton, Criswell, Croquette, Crowder, Cudell, Cullen, Cumming, Cummings, Cunningham, Cuoco, Cury\l{}o, Dabadie, Canton, Dall'Osso, D\'alya, Dana, DaneshgaranBajastani, D'Angelo, Danila,
  Danilishin, D'Antonio, Danzmann, Darsow-Fromm, Dasgupta, Datrier, Datta, Dattilo, Dave, Davier, Davies, Davis, Davis, Daw, Dean, DeBra, Deenadayalan, Degallaix, De~Laurentis, Del\'eglise, Del~Favero, De~Lillo, De~Lillo, Del~Pozzo, DeMarchi, De~Matteis, D'Emilio, Demos, Dent, Depasse, De~Pietri, De~Rosa, De~Rossi, DeSalvo, De~Simone, Dhurandhar, D\'{\i}az, Diaz-Ortiz, Didio, Dietrich, Di~Fiore, Di~Fronzo, Di~Giorgio, Di~Giovanni, Di~Giovanni, Di~Girolamo, Di~Lieto, Ding, Di~Pace, Di~Palma, Di~Renzo, Divakarla, Dmitriev, Doctor, D'Onofrio, Donovan, Dooley, Doravari, Dorrington, Drago, Driggers, Drori, Ducoin, Dupej, Durante, D'Urso, Duverne, Dwyer, Eassa, Easter, Ebersold, Eckhardt, Eddolls, Edelman, Edo, Edy, Effler, Eguchi, Eichholz, Eikenberry, Eisenmann, Eisenstein, Ejlli, Engelby, Enomoto, Errico, Essick, Estell\'es, Estevez, Etienne, Etzel, Evans, Evans, Ewing, Fafone, Fair, Fairhurst, Farah, Farinon, Farr, Farr, Farrow, Fauchon-Jones, Favaro, Favata, Fays, Fazio, Feicht, Fejer, Fenyvesi, Ferguson,
  Fernandez-Galiana, Ferrante, Ferreira, Fidecaro, Figura, Fiori, Fishbach, Fisher, Fittipaldi, Fiumara, Flaminio, Floden, Fong, Font, Fornal, Forsyth, Franke, Frasca, Frasconi, Frederick, Freed, Frei, Freise, Frey, Fritschel, Frolov, Fronz\'e, Fujii, Fujikawa, Fukunaga, Fukushima, Fulda, Fyffe, Gabbard, Gadre, Gair, Gais, Galaudage, Gamba, Ganapathy, Ganguly, Gao, Gaonkar, Garaventa, Garc\'{\i}a, Garc\'{\i}a-N\'u\~nez, Garc\'{\i}a-Quir\'os, Garufi, Gateley, Gaudio, Gayathri, Ge, Gemme, Gennai, George, George, Gerberding, Gergely, Gewecke, Ghonge, Ghosh, Ghosh, Ghosh, Ghosh, Giacomazzo, Giacoppo, Giaime, Giardina, Gibson, Gier, Giesler, Giri, Gissi, Glanzer, Gleckl, Godwin, Golomb, Goetz, Goetz, Gohlke, Goncharov, Gonz\'alez, Gopakumar, Gosselin, Gouaty, Gould, Grace, Grado, Granata, Granata, Grant, Gras, Grassia, Gray, Gray, Greco, Green, Green, Gretarsson, Gretarsson, Griffith, Griffiths, Griggs, Grignani, Grimaldi, Grimm, Grote, Grunewald, Gruning, Guerra, Guidi, Guimaraes, Guix\'e, Gulati, Guo, Guo,
  Gupta, Gupta, Gupta, Gustafson, Gustafson, Guzman, Ha, Haegel, Hagiwara, Haino, Halim, Hall, Hamilton, Hammond, Han, Haney, Hanks, Hanna, Hannam, Hannuksela, Hansen, Hansen, Hanson, Harder, Hardwick, Haris, Harms, Harry, Harry, Hartwig, Hasegawa, Haskell, Hasskew, Haster, Hattori, Haughian, Hayakawa, Hayama, Hayes, Healy, Heidmann, Heidt, Heintze, Heinze, Heinzel, Heitmann, Hellman, Hello, Helmling-Cornell, Hemming, Hendry, Heng, Hennes, Hennig, Hennig, Hernandez, Vivanco, Heurs, Hild, Hill, Himemoto, Hines, Hiranuma, Hirata, Hirose, Hochheim, Hofman, Hohmann, Holcomb, Holland, Hollows, Holmes, Holt, Holz, Hong, Hopkins, Hough, Hourihane, Howell, Hoy, Hoyland, Hreibi, Hsieh, Hsu, Huang, Huang, Huang, Huang, Huang, Huang, H\"ubner, Huddart, Hughey, Hui, Hui, Husa, Huttner, Huxford, Huynh-Dinh, Ide, Idzkowski, Iess, Ikenoue, Imam, Inayoshi, Ingram, Inoue, Ioka, Isi, Isleif, Ito, Itoh, Iyer, Izumi, JaberianHamedan, Jacqmin, Jadhav, Jadhav, James, Jan, Jani, Janquart, Janssens, Janthalur, Jaranowski, Jariwala,
  Jaume, Jenkins, Jenner, Jeon, Jeunon, Jia, Jin, Johns, Jones, Jones, Jones, Jones, Jones, Jonker, Ju, Jung, Jung, Junker, Juste, Kaihotsu, Kajita, Kakizaki, Kalaghatgi, Kalogera, Kamai, Kamiizumi, Kanda, Kandhasamy, Kang, Kanner, Kao, Kapadia, Kapasi, Karat, Karathanasis, Karki, Kashyap, Kasprzack, Kastaun, Katsanevas, Katsavounidis, Katzman, Kaur, Kawabe, Kawaguchi, Kawai, Kawasaki, K\'ef\'elian, Keitel, Key, Khadka, Khalili, Khan, Khazanov, Khetan, Khursheed, Kijbunchoo, Kim, Kim, Kim, Kim, Kim, Kim, Kimball, Kimura, Kinley-Hanlon, Kirchhoff, Kissel, Kita, Kitazawa, Kleybolte, Klimenko, Knee, Knowles, Knyazev, Koch, Koekoek, Kojima, Kokeyama, Koley, Kolitsidou, Kolstein, Komori, Kondrashov, Kong, Kontos, Koper, Korobko, Kotake, Kovalam, Kozak, Kozakai, Kozu, Kringel, Krishnendu, Kr\'olak, Kuehn, Kuei, Kuijer, Kulkarni, Kumar, Kumar, Kumar, Kumar, Kume, Kuns, Kuo, Kuo, Kuromiya, Kuroyanagi, Kusayanagi, Kuwahara, Kwak, Lagabbe, Laghi, Lalande, Lam, Lamberts, Landry, Landry, Lane, Lang, Lange, Lantz,
  La~Rosa, Lartaux-Vollard, Lasky, Laxen, Lazzarini, Lazzaro, Leaci, Leavey, Lecoeuche, Lee, Lee, Lee, Lee, Lee, Lee, Lehmann, Lema\^{\i}tre, Leonardi, Leroy, Letendre, Levesque, Levin, Leviton, Leyde, Li, Li, Li, Li, Li, Li, Lin, Lin, Lin, Lin, Lin, Linde, Linker, Linley, Littenberg, Liu, Liu, Liu, Liu, Llamas, Llorens-Monteagudo, Lo, Lockwood, Loh, London, Longo, Lopez, Portilla, Lorenzini, Loriette, Lormand, Losurdo, Lott, Lough, Lousto, Lovelace, Lucaccioni, L\"uck, Lumaca, Lundgren, Luo, Lynam, Macas, MacInnis, Macleod, MacMillan, Macquet, Hernandez, Magazz\`u, Magee, Maggiore, Magnozzi, Mahesh, Majorana, Makarem, Maksimovic, Maliakal, Malik, Man, Mandic, Mangano, Mango, Mansell, Manske, Mantovani, Mapelli, Marchesoni, Marchio, Marion, Mark, M\'arka, M\'arka, Markakis, Markosyan, Markowitz, Maros, Marquina, Marsat, Martelli, Martin, Martin, Martinez, Martinez, Martinez, Martinovic, Martynov, Marx, Masalehdan, Mason, Massera, Masserot, Massinger, Masso-Reid, Mastrogiovanni, Matas, Mateu-Lucena, Matichard,
  Matiushechkina, Mavalvala, McCann, McCarthy, McClelland, McClincy, McCormick, McCuller, McGhee, McGuire, McIsaac, McIver, McRae, McWilliams, Meacher, Mehmet, Mehta, Meijer, Melatos, Melchor, Mendell, Menendez-Vazquez, Menoni, Mercer, Mereni, Merfeld, Merilh, Merritt, Merzougui, Meshkov, Messenger, Messick, Meyers, Meylahn, Mhaske, Miani, Miao, Michaloliakos, Michel, Michimura, Middleton, Milano, Miller, Miller, Miller, Miller, Millhouse, Mills, Milotti, Minazzoli, Minenkov, Mio, Mir, Miravet-Ten\'es, Mishra, Mishra, Mistry, Mitra, Mitrofanov, Mitselmakher, Mittleman, Miyakawa, Miyamoto, Miyazaki, Miyo, Miyoki, Mo, Modafferi, Moguel, Mogushi, Mohapatra, Mohite, Molina, Molina-Ruiz, Mondin, Montani, Moore, Moraru, Morawski, More, Moreno, Moreno, Mori, Morisaki, Moriwaki, Morr\'as, Mours, Mow-Lowry, Mozzon, Muciaccia, Mukherjee, Mukherjee, Mukherjee, Mukherjee, Mukherjee, Mukund, Mullavey, Munch, Mu\~niz, Murray, Musenich, Muusse, Nadji, Nagano, Nagano, Nagar, Nakamura, Nakano, Nakano, Nakashima, Nakayama,
  Napolano, Nardecchia, Narikawa, Naticchioni, Nayak, Nayak, Negishi, Neil, Neilson, Nelemans, Nelson, Nery, Neubauer, Neunzert, Ng, Ng, Nguyen, Nguyen, Nguyen, Quynh, Ni, Nichols, Nishizawa, Nissanke, Nitoglia, Nocera, Norman, North, Nozaki, Siles, Nuttall, Oberling, O'Brien, Obuchi, O'Dell, Oelker, Ogaki, Oganesyan, Oh, Oh, Oh, Ohashi, Ohishi, Ohkawa, Ohme, Ohta, Okada, Okutani, Okutomi, Olivetto, Oohara, Ooi, Oram, O'Reilly, Ormiston, Ormsby, Ortega, O'Shaughnessy, O'Shea, Oshino, Ossokine, Osthelder, Otabe, Ottaway, Overmier, Pace, Pagano, Page, Pagliaroli, Pai, Pai, Palamos, Palashov, Palomba, Pan, Pan, Panda, Pang, Pang, Pankow, Pannarale, Pant, Panther, Paoletti, Paoli, Paolone, Parisi, Park, Park, Parker, Pascucci, Pasqualetti, Passaquieti, Passuello, Patel, Pathak, Patricelli, Patron, Paul, Payne, Pedraza, Pegoraro, Pele, Arellano, Penn, Perego, Pereira, Pereira, Perez, P\'erigois, Perkins, Perreca, Perri\`es, Petermann, Petterson, Pfeiffer, Pham, Phukon, Piccinni, Pichot, Piendibene, Piergiovanni,
  Pierini, Pierro, Pillant, Pillas, Pilo, Pinard, Pinto, Pinto, Piotrzkowski, Piotrzkowski, Pirello, Pitkin, Placidi, Planas, Plastino, Pluchar, Poggiani, Polini, Pong, Ponrathnam, Popolizio, Porter, Poulton, Powell, Pracchia, Pradier, Prajapati, Prasai, Prasanna, Pratten, Principe, Prodi, Prokhorov, Prosposito, Prudenzi, Puecher, Punturo, Puosi, Puppo, P\"urrer, Qi, Quetschke, Quitzow-James, Raab, Raaijmakers, Radkins, Radulesco, Raffai, Rail, Raja, Rajan, Ramirez, Ramirez, Ramos-Buades, Rana, Rapagnani, Rapol, Ray, Raymond, Raza, Razzano, Read, Rees, Regimbau, Rei, Reid, Reid, Reitze, Relton, Renzini, Rettegno, Reza, Rezac, Ricci, Richards, Richardson, Richardson, Riemenschneider, Riles, Rinaldi, Rink, Rizzo, Robertson, Robie, Robinet, Rocchi, Rodriguez, Rolland, Rollins, Romanelli, Romano, Romel, Romero-Rodr\'{\i}guez, Romero-Shaw, Romie, Ronchini, Rosa, Rose, Rosi\ifmmode~\acute{n}\else \'{n}\fi{}ska, Ross, Rowan, Rowlinson, Roy, Roy, Roy, Rozza, Ruggi, Ryan, Sachdev, Sadecki, Sadiq, Sago, Saito, Saito,
  Sakai, Sakai, Sakellariadou, Sakuno, Salafia, Salconi, Saleem, Salemi, Samajdar, Sanchez, Sanchez, Sanchez, Sanchis-Gual, Sanders, Sanuy, Saravanan, Sarin, Sassolas, Satari, Sathyaprakash, Sato, Sato, Sauter, Savage, Sawada, Sawant, Sawant, Sayah, Schaetzl, Scheel, Scheuer, Schiworski, Schmidt, Schmidt, Schnabel, Schneewind, Schofield, Sch\"onbeck, Schulte, Schutz, Schwartz, Scott, Scott, Seglar-Arroyo, Sekiguchi, Sekiguchi, Sellers, Sengupta, Sentenac, Seo, Sequino, Sergeev, Setyawati, Shaffer, Shahriar, Shams, Shao, Sharma, Sharma, Shawhan, Shcheblanov, Shibagaki, Shikauchi, Shimizu, Shimoda, Shimode, Shinkai, Shishido, Shoda, Shoemaker, Shoemaker, ShyamSundar, Sieniawska, Sigg, Singer, Singh, Singh, Singha, Sintes, Sipala, Skliris, Slagmolen, Slaven-Blair, Smetana, Smith, Smith, Soldateschi, Somala, Somiya, Son, Soni, Soni, Sordini, Sorrentino, Sorrentino, Sotani, Soulard, Souradeep, Sowell, Spagnuolo, Spencer, Spera, Srinivasan, Srivastava, Srivastava, Staats, Stachie, Steer, Steinhoff, Steinlechner,
  Steinlechner, Stevenson, Stops, Stover, Strain, Strang, Stratta, Strunk, Sturani, Stuver, Sudhagar, Sudhir, Sugimoto, Suh, Sullivan, Summerscales, Sun, Sun, Sunil, Sur, Suresh, Sutton, Suzuki, Suzuki, Swinkels, Szczepa\ifmmode~\acute{n}\else \'{n}\fi{}czyk, Szewczyk, Tacca, Tagoshi, Tait, Takahashi, Takahashi, Takamori, Takano, Takeda, Takeda, Talbot, Talbot, Tanaka, Tanaka, Tanaka, Tanaka, Tanaka, Tanasijczuk, Tanioka, Tanner, Tao, Tao, Mart\'{\i}n, Taranto, Tasson, Telada, Tenorio, Terhune, Terkowski, Thirugnanasambandam, Thomas, Thomas, Thomas, Thompson, Thondapu, Thorne, Thrane, Tiwari, Tiwari, Tiwari, Toivonen, Toland, Tolley, Tomaru, Tomigami, Tomura, Tonelli, Torres-Forn\'e, Torrie, e~Melo, T\"oyr\"a, Trapananti, Travasso, Traylor, Trevor, Tringali, Tripathee, Troiano, Trovato, Trozzo, Trudeau, Tsai, Tsai, Tsang, Tsang, Tsao, Tse, Tso, Tsubono, Tsuchida, Tsukada, Tsuna, Tsutsui, Tsuzuki, Turbang, Turconi, Tuyenbayev, Ubhi, Uchikata, Uchiyama, Udall, Ueda, Uehara, Ueno, Ueshima, Unnikrishnan,
  Uraguchi, Urban, Ushiba, Utina, Vahlbruch, Vajente, Vajpeyi, Valdes, Valentini, Valsan, van Bakel, van Beuzekom, van~den Brand, Van Den~Broeck, Vander-Hyde, van~der Schaaf, van Heijningen, Vanosky, van Putten, van Remortel, Vardaro, Vargas, Varma, Vas\'uth, Vecchio, Vedovato, Veitch, Veitch, Venneberg, Venugopalan, Verkindt, Verma, Verma, Veske, Vetrano, Vicer\'e, Vidyant, Viets, Vijaykumar, Villa-Ortega, Vinet, Virtuoso, Vitale, Vo, Vocca, von Reis, von Wrangel, Vorvick, Vyatchanin, Wade, Wade, Wagner, Walet, Walker, Wallace, Wallace, Walsh, Wang, Wang, Wang, Ward, Warner, Was, Washimi, Washington, Watchi, Weaver, Webster, Weinert, Weinstein, Weiss, Weller, Wellmann, Wen, We\ss{}els, Wette, Whelan, White, Whiting, Whittle, Wilken, Williams, Williams, Williamson, Willis, Willke, Wilson, Winkler, Wipf, Wlodarczyk, Woan, Woehler, Wofford, Wong, Wu, Wu, Wu, Wu, Wysocki, Xiao, Xu, Yamada, Yamamoto, Yamamoto, Yamamoto, Yamamoto, Yamashita, Yamazaki, Yang, Yang, Yang, Yang, Yang, Yap, Yeeles, Yelikar, Ying,
  Yokogawa, Yokoyama, Yokozawa, Yoo, Yoshioka, Yu, Yu, Yuzurihara, Zadro\ifmmode~\dot{z}\else \.{z}\fi{}ny, Zanolin, Zeidler, Zelenova, Zendri, Zevin, Zhan, Zhang, Zhang, Zhang, Zhang, Zhang, Zhao, Zhao, Zhao, Zhao, Zheng, Zhou, Zhou, Zhu, Zhu, Zimmerman, Zlochower, Zucker, \& Zweizig}]{Abbott2023population}
Abbott, R., Abbott, T.~D., Acernese, F., {et~al.} 2023{\natexlab{b}}, Phys. Rev. X, 13, 011048

\bibitem[{Abbott {et~al.}(2024)Abbott, Abbott, Acernese, Ackley, Adams, Adhikari, Adhikari, Adya, Affeldt, Agarwal, Agathos, Agatsuma, Aggarwal, Aguiar, Aiello, Ain, Ajith, Albanesi, Allocca, Altin, Amato, Anand, Anand, Ananyeva, Anderson, Anderson, Andrade, Andres, Andri\ifmmode~\acute{c}\else \'{c}\fi{}, Angelova, Ansoldi, Antelis, Antier, Appert, Arai, Araya, Areeda, Ar\`ene, Arnaud, Aronson, Arun, Asali, Ashton, Assiduo, Aston, Astone, Aubin, Austin, Babak, Badaracco, Bader, Badger, Bae, Baer, Bagnasco, Bai, Baird, Ball, Ballardin, Ballmer, Balsamo, Baltus, Banagiri, Bankar, Barayoga, Barbieri, Barish, Barker, Barneo, Barone, Barr, Barsotti, Barsuglia, Barta, Bartlett, Barton, Bartos, Bassiri, Basti, Bawaj, Bayley, Baylor, Bazzan, B\'ecsy, Bedakihale, Bejger, Belahcene, Benedetto, Beniwal, Bennett, Bentley, BenYaala, Bergamin, Berger, Bernuzzi, Berry, Bersanetti, Bertolini, Betzwieser, Beveridge, Bhandare, Bhardwaj, Bhattacharjee, Bhaumik, Bilenko, Billingsley, Bini, Birney, Birnholtz, Biscans, Bischi,
  Biscoveanu, Bisht, Biswas, Bitossi, Bizouard, Blackburn, Blair, Blair, Blair, Bobba, Bode, Boer, Bogaert, Boldrini, Bonavena, Bondu, Bonilla, Bonnand, Booker, Boom, Bork, Boschi, Bose, Bose, Bossilkov, Boudart, Bouffanais, Bozzi, Bradaschia, Brady, Bramley, Branch, Branchesi, Brau, Breschi, Briant, Briggs, Brillet, Brinkmann, Brockill, Brooks, Brooks, Brown, Brunett, Bruno, Bruntz, Bryant, Bulik, Bulten, Buonanno, Buscicchio, Buskulic, Buy, Byer, Cadonati, Cagnoli, Cahillane, Bustillo, Callaghan, Callister, Calloni, Cameron, Camp, Canepa, Canevarolo, Cannavacciuolo, Cannon, Cao, Capote, Carapella, Carbognani, Carlin, Carney, Carpinelli, Carrillo, Carullo, Carver, Diaz, Casentini, Castaldi, Caudill, Cavagli\`a, Cavalier, Cavalieri, Ceasar, Cella, Cerd\'a-Dur\'an, Cesarini, Chaibi, Chakravarti, Subrahmanya, Champion, Chan, Chan, Chan, Chan, Chandra, Chanial, Chao, Charlton, Chase, Chassande-Mottin, Chatterjee, Chatterjee, Chatterjee, Chattopadhyay, Chaturvedi, Chaty, Chatziioannou, Chen, Chen, Chen, Chen,
  Chen, Cheng, Cheong, Cheung, Chia, Chiadini, Chiarini, Chierici, Chincarini, Chiofalo, Chiummo, Cho, Cho, Choudhary, Choudhary, Christensen, Chu, Chua, Chung, Ciani, Ciecielag, Cie\ifmmode~\acute{s}\else \'{s}\fi{}lar, Cifaldi, Ciobanu, Ciolfi, Cipriano, Cirone, Clara, Clark, Clark, Clarke, Clearwater, Clesse, Cleva, Coccia, Codazzo, Cohadon, Cohen, Cohen, Colleoni, Collette, Colombo, Colpi, Compton, Constancio, Conti, Cooper, Corban, Corbitt, Cordero-Carri\'on, Corezzi, Corley, Cornish, Corre, Corsi, Cortese, Costa, Cotesta, Coughlin, Coulon, Countryman, Cousins, Couvares, Coward, Cowart, Coyne, Coyne, Creighton, Creighton, Criswell, Croquette, Crowder, Cudell, Cullen, Cumming, Cummings, Cunningham, Cuoco, Cury\l{}o, Dabadie, Canton, Dall'Osso, D\'alya, Dana, DaneshgaranBajastani, D'Angelo, Danila, Danilishin, D'Antonio, Danzmann, Darsow-Fromm, Dasgupta, Datrier, Datta, Dattilo, Dave, Davier, Davies, Davis, Davis, Daw, Dean, DeBra, Deenadayalan, Degallaix, De~Laurentis, Del\'eglise, Del~Favero, De~Lillo,
  De~Lillo, Del~Pozzo, DeMarchi, De~Matteis, D'Emilio, Demos, Dent, Depasse, De~Pietri, De~Rosa, De~Rossi, DeSalvo, De~Simone, Dhurandhar, D\'{\i}az, Diaz-Ortiz, Didio, Dietrich, Di~Fiore, Di~Fronzo, Di~Giorgio, Di~Giovanni, Di~Giovanni, Di~Girolamo, Di~Lieto, Ding, Di~Pace, Di~Palma, Di~Renzo, Divakarla, Divyajyoti, Dmitriev, Doctor, D'Onofrio, Donovan, Dooley, Doravari, Dorrington, Drago, Driggers, Drori, Ducoin, Dupej, Durante, D'Urso, Duverne, Dwyer, Eassa, Easter, Ebersold, Eckhardt, Eddolls, Edelman, Edo, Edy, Effler, Eichholz, Eikenberry, Eisenmann, Eisenstein, Ejlli, Engelby, Errico, Essick, Estell\'es, Estevez, Etienne, Etzel, Evans, Evans, Ewing, Fafone, Fair, Fairhurst, Fanning, Farah, Farinon, Farr, Farr, Farrow, Fauchon-Jones, Favaro, Favata, Fays, Fazio, Feicht, Fejer, Fenyvesi, Ferguson, Fernandez-Galiana, Ferrante, Ferreira, Fidecaro, Figura, Fiori, Fishbach, Fisher, Fittipaldi, Fiumara, Flaminio, Floden, Fong, Font, Fornal, Forsyth, Franke, Frasca, Frasconi, Frederick, Freed, Frei, Freise,
  Frey, Fritschel, Frolov, Fronz\'e, Fulda, Fyffe, Gabbard, Gabella, Gadre, Gair, Gais, Galaudage, Gamba, Ganapathy, Ganguly, Gaonkar, Garaventa, Garc\'{\i}a, Garc\'{\i}a-N\'u\~nez, Garc\'{\i}a-Quir\'os, Garufi, Gateley, Gaudio, Gayathri, Gemme, Gennai, George, George, Gerberding, Gergely, Gewecke, Ghonge, Ghosh, Ghosh, Ghosh, Ghosh, Giacomazzo, Giacoppo, Giaime, Giardina, Gibson, Gier, Giesler, Giri, Gissi, Glanzer, Gleckl, Godwin, Goetz, Goetz, Gohlke, Goncharov, Gonz\'alez, Gopakumar, Gosselin, Gouaty, Gould, Grace, Grado, Granata, Granata, Grant, Gras, Grassia, Gray, Gray, Greco, Green, Green, Gretarsson, Gretarsson, Griffith, Griffiths, Griggs, Grignani, Grimaldi, Grimm, Grote, Grunewald, Gruning, Guerra, Guidi, Guimaraes, Guix\'e, Gulati, Guo, Guo, Gupta, Gupta, Gupta, Gustafson, Gustafson, Guzman, Haegel, Halim, Hall, Hamilton, Hammond, Haney, Hanks, Hanna, Hannam, Hannuksela, Hansen, Hansen, Hanson, Harder, Hardwick, Haris, Harms, Harry, Harry, Hartwig, Haskell, Hasskew, Haster, Haughian, Hayes,
  Healy, Heidmann, Heidt, Heintze, Heinze, Heinzel, Heitmann, Hellman, Hello, Helmling-Cornell, Hemming, Hendry, Heng, Hennes, Hennig, Hennig, Hernandez, Vivanco, Heurs, Hild, Hill, Hines, Hochheim, Hofman, Hohmann, Holcomb, Holland, Holley-Bockelmann, Hollows, Holmes, Holt, Holz, Hopkins, Hough, Hourihane, Howell, Hoy, Hoyland, Hreibi, Hsu, Huang, H\"ubner, Huddart, Hughey, Hui, Husa, Huttner, Huxford, Huynh-Dinh, Idzkowski, Iess, Ingram, Isi, Isleif, Iyer, JaberianHamedan, Jacqmin, Jadhav, Jadhav, James, Jan, Jani, Janquart, Janssens, Janthalur, Jaranowski, Jariwala, Jaume, Jenkins, Jenner, Jeunon, Jia, Johns, Johnson-McDaniel, Jones, Jones, Jones, Jones, Jones, Jonker, Ju, Junker, Juste, Kalaghatgi, Kalogera, Kamai, Kandhasamy, Kang, Kanner, Kao, Kapadia, Kapasi, Karat, Karathanasis, Karki, Kashyap, Kasprzack, Kastaun, Katsanevas, Katsavounidis, Katzman, Kaur, Kawabe, K\'ef\'elian, Keitel, Key, Khadka, Khalili, Khan, Khazanov, Khetan, Khursheed, Kijbunchoo, Kim, Kim, Kim, Kim, Kim, Kimball, Kinley-Hanlon,
  Kirchhoff, Kissel, Kleybolte, Klimenko, Knee, Knowles, Knyazev, Koch, Koekoek, Koley, Kolitsidou, Kolstein, Komori, Kondrashov, Kontos, Koper, Korobko, Kovalam, Kozak, Kringel, Krishnendu, Kr\'olak, Kuehn, Kuei, Kuijer, Kulkarni, Kumar, Kumar, Kumar, Kumar, Kuns, Kuwahara, Lagabbe, Laghi, Lalande, Lam, Lamberts, Landry, Lane, Lang, Lange, Lantz, La~Rosa, Lartaux-Vollard, Lasky, Laxen, Lazzarini, Lazzaro, Leaci, Leavey, Lecoeuche, Lee, Lee, Lee, Lee, Lehmann, Lema\^{\i}tre, Leroy, Letendre, Levesque, Levin, Leviton, Leyde, Li, Li, Li, Li, Li, Linde, Linker, Linley, Littenberg, Liu, Liu, Liu, Llamas, Llorens-Monteagudo, Lo, Lockwood, London, Longo, Lopez, Portilla, Lorenzini, Loriette, Lormand, Losurdo, Lott, Lough, Lousto, Lovelace, Lucaccioni, L\"uck, Lumaca, Lundgren, Lynam, Macas, MacInnis, Macleod, MacMillan, Macquet, Hernandez, Magazz\`u, Magee, Maggiore, Magnozzi, Mahesh, Majorana, Makarem, Maksimovic, Maliakal, Malik, Man, Mandic, Mangano, Mango, Mansell, Manske, Mantovani, Mapelli, Marchesoni,
  Marion, Mark, M\'arka, M\'arka, Markakis, Markosyan, Markowitz, Maros, Marquina, Marsat, Martelli, Martin, Martin, Martinez, Martinez, Martinez, Martinovic, Martynov, Marx, Masalehdan, Mason, Massera, Masserot, Massinger, Masso-Reid, Mastrogiovanni, Matas, Mateu-Lucena, Matichard, Matiushechkina, Mavalvala, McCann, McCarthy, McClelland, McClincy, McCormick, McCuller, McGhee, McGuire, McIsaac, McIver, McRae, McWilliams, Meacher, Mehmet, Mehta, Meijer, Melatos, Melchor, Mendell, Menendez-Vazquez, Menoni, Mercer, Mereni, Merfeld, Merilh, Merritt, Merzougui, Meshkov, Messenger, Messick, Meyers, Meylahn, Mhaske, Miani, Miao, Michaloliakos, Michel, Middleton, Milano, Miller, Miller, Miller, Millhouse, Mills, Milotti, Minazzoli, Minenkov, Mir, Miravet-Ten\'es, Mishra, Mishra, Mistry, Mitra, Mitrofanov, Mitselmakher, Mittleman, Mo, Moguel, Mogushi, Mohapatra, Mohite, Molina, Molina-Ruiz, Mondin, Montani, Moore, Moraru, Morawski, More, Moreno, Moreno, Morisaki, Mours, Mow-Lowry, Mozzon, Muciaccia, Mukherjee,
  Mukherjee, Mukherjee, Mukherjee, Mukherjee, Mukund, Mullavey, Munch, Mu\~niz, Murray, Musenich, Muusse, Nadji, Nagar, Napolano, Nardecchia, Naticchioni, Nayak, Nayak, Neil, Neilson, Nelemans, Nelson, Nery, Neubauer, Neunzert, Ng, Ng, Nguyen, Nguyen, Nguyen, Nichols, Nissanke, Nitoglia, Nocera, Norman, North, Nuttall, Oberling, O'Brien, O'Dell, Oelker, Oganesyan, Oh, Oh, Ohme, Ohta, Okada, Olivetto, Oram, O'Reilly, Ormiston, Ormsby, Ortega, O'Shaughnessy, O'Shea, Ossokine, Osthelder, Ottaway, Overmier, Pace, Pagano, Page, Pagliaroli, Pai, Pai, Palamos, Palashov, Palomba, Pan, Panda, Pang, Pankow, Pannarale, Pant, Panther, Paoletti, Paoli, Paolone, Park, Parker, Pascucci, Pasqualetti, Passaquieti, Passuello, Patel, Pathak, Patricelli, Patron, Patrone, Paul, Payne, Pedraza, Pegoraro, Pele, Penn, Perego, Pereira, Pereira, Perez, P\'erigois, Perkins, Perreca, Perri\`es, Petermann, Petterson, Pfeiffer, Pham, Phukon, Piccinni, Pichot, Piendibene, Piergiovanni, Pierini, Pierro, Pillant, Pillas, Pilo, Pinard, Pinto,
  Pinto, Piotrzkowski, Pirello, Pitkin, Placidi, Planas, Plastino, Pluchar, Poggiani, Polini, Pong, Ponrathnam, Popolizio, Porter, Poulton, Powell, Pracchia, Pradier, Prajapati, Prasai, Prasanna, Pratten, Principe, Prodi, Prokhorov, Prosposito, Prudenzi, Puecher, Punturo, Puosi, Puppo, P\"urrer, Qi, Quetschke, Quitzow-James, Raab, Raaijmakers, Radkins, Radulesco, Raffai, Rail, Raja, Rajan, Ramirez, Ramirez, Ramos-Buades, Rana, Rapagnani, Rapol, Ray, Raymond, Raza, Razzano, Read, Rees, Regimbau, Rei, Reid, Reid, Reitze, Relton, Renzini, Rettegno, Reza, Rezac, Ricci, Richards, Richardson, Richardson, Riemenschneider, Riles, Rinaldi, Rink, Rizzo, Robertson, Robie, Robinet, Rocchi, Rodriguez, Rolland, Rollins, Romanelli, Romano, Romel, Romero-Rodr\'{\i}guez, Romero-Shaw, Romie, Ronchini, Rosa, Rose, Rosell, Rosi\ifmmode~\acute{n}\else \'{n}\fi{}ska, Ross, Rowan, Rowlinson, Roy, Roy, Roy, Rozza, Ruggi, Ruiz-Rocha, Ryan, Sachdev, Sadecki, Sadiq, Sakellariadou, Salafia, Salconi, Saleem, Salemi, Samajdar, Sanchez,
  Sanchez, Sanchez, Sanchis-Gual, Sanders, Sanuy, Saravanan, Sarin, Sassolas, Satari, Sauter, Savage, Sawant, Sawant, Sayah, Schaetzl, Scheel, Scheuer, Schiworski, Schmidt, Schmidt, Schnabel, Schneewind, Schofield, Sch\"onbeck, Schulte, Schutz, Schwartz, Scott, Scott, Seglar-Arroyo, Sellers, Sengupta, Sentenac, Seo, Sequino, Sergeev, Setyawati, Shaffer, Shahriar, Shams, Sharma, Sharma, Shawhan, Shcheblanov, Shikauchi, Shoemaker, Shoemaker, ShyamSundar, Sieniawska, Sigg, Singer, Singh, Singh, Singha, Sintes, Sipala, Skliris, Slagmolen, Slaven-Blair, Smetana, Smith, Smith, Soldateschi, Somala, Son, Soni, Soni, Sordini, Sorrentino, Sorrentino, Soulard, Souradeep, Sowell, Spagnuolo, Spencer, Spera, Srinivasan, Srivastava, Srivastava, Staats, Stachie, Steer, Steinhoff, Steinlechner, Steinlechner, Stevenson, Stops, Stover, Strain, Strang, Stratta, Strunk, Sturani, Stuver, Sudhagar, Sudhir, Suh, Summerscales, Sun, Sun, Sunil, Sur, Suresh, Sutton, Swinkels, Szczepa\ifmmode~\acute{n}\else \'{n}\fi{}czyk, Szewczyk,
  Tacca, Tait, Talbot, Talbot, Tanasijczuk, Tanner, Tao, Tao, Mart\'{\i}n, Taranto, Tasson, Tenorio, Terhune, Terkowski, Thirugnanasambandam, Thomas, Thomas, Thomas, Thompson, Thondapu, Thorne, Thrane, Tiwari, Tiwari, Tiwari, Toivonen, Toland, Tolley, Tonelli, Torres-Forn\'e, Torrie, e~Melo, T\"oyr\"a, Trapananti, Travasso, Traylor, Trevor, Tringali, Tripathee, Troiano, Trovato, Trozzo, Trudeau, Tsai, Tsai, Tsang, Tse, Tso, Tsukada, Tsuna, Tsutsui, Turbang, Turconi, Ubhi, Udall, Ueno, Unnikrishnan, Urban, Utina, Vahlbruch, Vajente, Vajpeyi, Valdes, Valentini, Valsan, van Bakel, van Beuzekom, van~den Brand, Van Den~Broeck, Vander-Hyde, van~der Schaaf, van Heijningen, Vanosky, van Remortel, Vardaro, Vargas, Varma, Vas\'uth, Vecchio, Vedovato, Veitch, Veitch, Venneberg, Venugopalan, Verkindt, Verma, Verma, Veske, Vetrano, Vicer\'e, Vidyant, Viets, Vijaykumar, Villa-Ortega, Vinet, Virtuoso, Vitale, Vo, Vocca, von Reis, von Wrangel, Vorvick, Vyatchanin, Wade, Wade, Wagner, Walet, Walker, Wallace, Wallace, Walsh,
  Wang, Wang, Ward, Warner, Was, Washington, Watchi, Weaver, Webster, Weinert, Weinstein, Weiss, Weller, Weller, Wellmann, Wen, We\ss{}els, Wette, Whelan, White, Whiting, Whittle, Wilken, Williams, Williams, Williamson, Willis, Willke, Wilson, Winkler, Wipf, Wlodarczyk, Woan, Woehler, Wofford, Wong, Wu, Wysocki, Xiao, Yamamoto, Yang, Yang, Yang, Yang, Yap, Yeeles, Yelikar, Ying, Yoo, Yu, Yu, Zadro\ifmmode~\dot{z}\else \.{z}\fi{}ny, Zanolin, Zelenova, Zendri, Zevin, Zhang, Zhang, Zhang, Zhang, Zhao, Zhao, Zhao, Zhou, Zhou, Zhu, Zimmerman, Zlochower, Zucker, \& Zweizig}]{Abbott2024gwtc21}
Abbott, R., Abbott, T.~D., Acernese, F., {et~al.} 2024, Phys. Rev. D, 109, 022001

\bibitem[{{Andrews} \& {Martini}(2013)}]{Andrews2013}
{Andrews}, B.~H. \& {Martini}, P. 2013, \apj, 765, 140

\bibitem[{{Artale} {et~al.}(2020){Artale}, {Mapelli}, {Bouffanais}, {Giacobbo}, {Pasquato}, \& {Spera}}]{Artale2020}
{Artale}, M.~C., {Mapelli}, M., {Bouffanais}, Y., {et~al.} 2020, \mnras, 491, 3419

\bibitem[{{Astropy Collaboration} {et~al.}(2022){Astropy Collaboration}, {Price-Whelan}, {Lim}, {Earl}, {Starkman}, {Bradley}, {Shupe}, {Patil}, {Corrales}, {Brasseur}, {N{"o}the}, {Donath}, {Tollerud}, {Morris}, {Ginsburg}, {Vaher}, {Weaver}, {Tocknell}, {Jamieson}, {van Kerkwijk}, {Robitaille}, {Merry}, {Bachetti}, {G{"u}nther}, {Aldcroft}, {Alvarado-Montes}, {Archibald}, {B{'o}di}, {Bapat}, {Barentsen}, {Baz{'a}n}, {Biswas}, {Boquien}, {Burke}, {Cara}, {Cara}, {Conroy}, {Conseil}, {Craig}, {Cross}, {Cruz}, {D'Eugenio}, {Dencheva}, {Devillepoix}, {Dietrich}, {Eigenbrot}, {Erben}, {Ferreira}, {Foreman-Mackey}, {Fox}, {Freij}, {Garg}, {Geda}, {Glattly}, {Gondhalekar}, {Gordon}, {Grant}, {Greenfield}, {Groener}, {Guest}, {Gurovich}, {Handberg}, {Hart}, {Hatfield-Dodds}, {Homeier}, {Hosseinzadeh}, {Jenness}, {Jones}, {Joseph}, {Kalmbach}, {Karamehmetoglu}, {Ka{l}uszy{'n}ski}, {Kelley}, {Kern}, {Kerzendorf}, {Koch}, {Kulumani}, {Lee}, {Ly}, {Ma}, {MacBride}, {Maljaars}, {Muna}, {Murphy}, {Norman}, {O'Steen},
  {Oman}, {Pacifici}, {Pascual}, {Pascual-Granado}, {Patil}, {Perren}, {Pickering}, {Rastogi}, {Roulston}, {Ryan}, {Rykoff}, {Sabater}, {Sakurikar}, {Salgado}, {Sanghi}, {Saunders}, {Savchenko}, {Schwardt}, {Seifert-Eckert}, {Shih}, {Jain}, {Shukla}, {Sick}, {Simpson}, {Singanamalla}, {Singer}, {Singhal}, {Sinha}, {Sip{H{o}}cz}, {Spitler}, {Stansby}, {Streicher}, {{{S}}umak}, {Swinbank}, {Taranu}, {Tewary}, {Tremblay}, {Val-Borro}, {Van Kooten}, {Vasovi{'c}}, {Verma}, {de Miranda Cardoso}, {Williams}, {Wilson}, {Winkel}, {Wood-Vasey}, {Xue}, {Yoachim}, {Zhang}, {Zonca}, \& {Astropy Project Contributors}}]{astropy:2022}
{Astropy Collaboration}, {Price-Whelan}, A.~M., {Lim}, P.~L., {et~al.} 2022, \apj, 935, 167

\bibitem[{{Astropy Collaboration} {et~al.}(2018){Astropy Collaboration}, {Price-Whelan}, {Sip{\H{o}}cz}, {G{\"u}nther}, {Lim}, {Crawford}, {Conseil}, {Shupe}, {Craig}, {Dencheva}, {Ginsburg}, {Vand erPlas}, {Bradley}, {P{\'e}rez-Su{\'a}rez}, {de Val-Borro}, {Aldcroft}, {Cruz}, {Robitaille}, {Tollerud}, {Ardelean}, {Babej}, {Bach}, {Bachetti}, {Bakanov}, {Bamford}, {Barentsen}, {Barmby}, {Baumbach}, {Berry}, {Biscani}, {Boquien}, {Bostroem}, {Bouma}, {Brammer}, {Bray}, {Breytenbach}, {Buddelmeijer}, {Burke}, {Calderone}, {Cano Rodr{\'\i}guez}, {Cara}, {Cardoso}, {Cheedella}, {Copin}, {Corrales}, {Crichton}, {D'Avella}, {Deil}, {Depagne}, {Dietrich}, {Donath}, {Droettboom}, {Earl}, {Erben}, {Fabbro}, {Ferreira}, {Finethy}, {Fox}, {Garrison}, {Gibbons}, {Goldstein}, {Gommers}, {Greco}, {Greenfield}, {Groener}, {Grollier}, {Hagen}, {Hirst}, {Homeier}, {Horton}, {Hosseinzadeh}, {Hu}, {Hunkeler}, {Ivezi{\'c}}, {Jain}, {Jenness}, {Kanarek}, {Kendrew}, {Kern}, {Kerzendorf}, {Khvalko}, {King}, {Kirkby}, {Kulkarni},
  {Kumar}, {Lee}, {Lenz}, {Littlefair}, {Ma}, {Macleod}, {Mastropietro}, {McCully}, {Montagnac}, {Morris}, {Mueller}, {Mumford}, {Muna}, {Murphy}, {Nelson}, {Nguyen}, {Ninan}, {N{\"o}the}, {Ogaz}, {Oh}, {Parejko}, {Parley}, {Pascual}, {Patil}, {Patil}, {Plunkett}, {Prochaska}, {Rastogi}, {Reddy Janga}, {Sabater}, {Sakurikar}, {Seifert}, {Sherbert}, {Sherwood-Taylor}, {Shih}, {Sick}, {Silbiger}, {Singanamalla}, {Singer}, {Sladen}, {Sooley}, {Sornarajah}, {Streicher}, {Teuben}, {Thomas}, {Tremblay}, {Turner}, {Terr{\'o}n}, {van Kerkwijk}, {de la Vega}, {Watkins}, {Weaver}, {Whitmore}, {Woillez}, {Zabalza}, \& {Astropy Contributors}}]{astropy:2018}
{Astropy Collaboration}, {Price-Whelan}, A.~M., {Sip{\H{o}}cz}, B.~M., {et~al.} 2018, \aj, 156, 123

\bibitem[{{Astropy Collaboration} {et~al.}(2013){Astropy Collaboration}, {Robitaille}, {Tollerud}, {Greenfield}, {Droettboom}, {Bray}, {Aldcroft}, {Davis}, {Ginsburg}, {Price-Whelan}, {Kerzendorf}, {Conley}, {Crighton}, {Barbary}, {Muna}, {Ferguson}, {Grollier}, {Parikh}, {Nair}, {Unther}, {Deil}, {Woillez}, {Conseil}, {Kramer}, {Turner}, {Singer}, {Fox}, {Weaver}, {Zabalza}, {Edwards}, {Azalee Bostroem}, {Burke}, {Casey}, {Crawford}, {Dencheva}, {Ely}, {Jenness}, {Labrie}, {Lim}, {Pierfederici}, {Pontzen}, {Ptak}, {Refsdal}, {Servillat}, \& {Streicher}}]{astropy:2013}
{Astropy Collaboration}, {Robitaille}, T.~P., {Tollerud}, E.~J., {et~al.} 2013, \aap, 558, A33

\bibitem[{Baibhav {et~al.}(2019)Baibhav, Berti, Gerosa, Mapelli, Giacobbo, Bouffanais, \& Di~Carlo}]{Baibhav2019}
Baibhav, V., Berti, E., Gerosa, D., {et~al.} 2019, Phys. Rev. D, 100, 064060

\bibitem[{{Barber} \& {Antonini}(2025)}]{Barber2024}
{Barber}, J. \& {Antonini}, F. 2025, \mnras, 538, 639

\bibitem[{{Bavera} {et~al.}(2021){Bavera}, {Fragos}, {Zevin}, {Berry}, {Marchant}, {Andrews}, {Coughlin}, {Dotter}, {Kovlakas}, {Misra}, {Serra-Perez}, {Qin}, {Rocha}, {Rom{\'a}n-Garza}, {Tran}, \& {Zapartas}}]{Bavera2021}
{Bavera}, S.~S., {Fragos}, T., {Zevin}, M., {et~al.} 2021, \aap, 647, A153

\bibitem[{{Belczynski} {et~al.}(2010){Belczynski}, {Dominik}, {Bulik}, {O'Shaughnessy}, {Fryer}, \& {Holz}}]{Belczynski2010}
{Belczynski}, K., {Dominik}, M., {Bulik}, T., {et~al.} 2010, \apjl, 715, L138

\bibitem[{{Belczynski} {et~al.}(2002){Belczynski}, {Kalogera}, \& {Bulik}}]{Belczynski2002}
{Belczynski}, K., {Kalogera}, V., \& {Bulik}, T. 2002, \apj, 572, 407

\bibitem[{Belczynski {et~al.}(2022)Belczynski, Romagnolo, Olejak, Klencki, Chattopadhyay, Stevenson, Miller, Lasota, \& Crowther}]{Belczynski2022}
Belczynski, K., Romagnolo, A., Olejak, A., {et~al.} 2022, The Astrophysical Journal, 925, 69

\bibitem[{{Biscoveanu} {et~al.}(2022){Biscoveanu}, {Callister}, {Haster}, {Ng}, {Vitale}, \& {Farr}}]{Biscoveanu2022}
{Biscoveanu}, S., {Callister}, T.~A., {Haster}, C.-J., {et~al.} 2022, \apjl, 932, L19

\bibitem[{{Bisigello} {et~al.}(2018){Bisigello}, {Caputi}, {Grogin}, \& {Koekemoer}}]{Bisigello2018}
{Bisigello}, L., {Caputi}, K.~I., {Grogin}, N., \& {Koekemoer}, A. 2018, \aap, 609, A82

\bibitem[{{Boco} {et~al.}(2021){Boco}, {Lapi}, {Chruslinska}, {Donevski}, {Sicilia}, \& {Danese}}]{Boco2021}
{Boco}, L., {Lapi}, A., {Chruslinska}, M., {et~al.} 2021, \apj, 907, 110

\bibitem[{{Boco} {et~al.}(2019){Boco}, {Lapi}, {Goswami}, {Perrotta}, {Baccigalupi}, \& {Danese}}]{Boco2019}
{Boco}, L., {Lapi}, A., {Goswami}, S., {et~al.} 2019, \apj, 881, 157

\bibitem[{{Boesky} {et~al.}(2024{\natexlab{a}}){Boesky}, {Broekgaarden}, \& {Berger}}]{Boesky2024MRD}
{Boesky}, A.~P., {Broekgaarden}, F.~S., \& {Berger}, E. 2024{\natexlab{a}}, \apj, 976, 24

\bibitem[{{Boesky} {et~al.}(2024{\natexlab{b}}){Boesky}, {Broekgaarden}, \& {Berger}}]{Boesky2024Delay}
{Boesky}, A.~P., {Broekgaarden}, F.~S., \& {Berger}, E. 2024{\natexlab{b}}, \apj, 976, 23

\bibitem[{{Boogaard} {et~al.}(2018){Boogaard}, {Brinchmann}, {Bouch{\'e}}, {Paalvast}, {Bacon}, {Bouwens}, {Contini}, {Gunawardhana}, {Inami}, {Marino}, {Maseda}, {Mitchell}, {Nanayakkara}, {Richard}, {Schaye}, {Schreiber}, {Tacchella}, {Wisotzki}, \& {Zabl}}]{Boogaard2018}
{Boogaard}, L.~A., {Brinchmann}, J., {Bouch{\'e}}, N., {et~al.} 2018, \aap, 619, A27

\bibitem[{{Branchesi} {et~al.}(2023){Branchesi}, {Maggiore}, {Alonso}, {Badger}, {Banerjee}, {Beirnaert}, {Belgacem}, {Bhagwat}, {Boileau}, {Borhanian}, {Brown}, {Leong Chan}, {Cusin}, {Danilishin}, {Degallaix}, {De Luca}, {Dhani}, {Dietrich}, {Dupletsa}, {Foffa}, {Franciolini}, {Freise}, {Gemme}, {Goncharov}, {Ghosh}, {Gulminelli}, {Gupta}, {Kumar Gupta}, {Harms}, {Hazra}, {Hild}, {Hinderer}, {Siong Heng}, {Iacovelli}, {Janquart}, {Janssens}, {Jenkins}, {Kalaghatgi}, {Koroveshi}, {Li}, {Li}, {Loffredo}, {Maggio}, {Mancarella}, {Mapelli}, {Martinovic}, {Maselli}, {Meyers}, {Miller}, {Mondal}, {Muttoni}, {Narola}, {Oertel}, {Oganesyan}, {Pacilio}, {Palomba}, {Pani}, {Pasqualetti}, {Perego}, {P{\'e}rigois}, {Pieroni}, {Piccinni}, {Puecher}, {Puppo}, {Ricciardone}, {Riotto}, {Ronchini}, {Sakellariadou}, {Samajdar}, {Santoliquido}, {Sathyaprakash}, {Steinlechner}, {Steinlechner}, {Utina}, {Van Den Broeck}, \& {Zhang}}]{Branchesi2023}
{Branchesi}, M., {Maggiore}, M., {Alonso}, D., {et~al.} 2023, \jcap, 2023, 068

\bibitem[{{Bressan} {et~al.}(2012){Bressan}, {Marigo}, {Girardi}, {Salasnich}, {Dal Cero}, {Rubele}, \& {Nanni}}]{Bressan2012}
{Bressan}, A., {Marigo}, P., {Girardi}, L., {et~al.} 2012, \mnras, 427, 127

\bibitem[{Briel {et~al.}(2022)Briel, Eldridge, Stanway, Stevance, \& Chrimes}]{Briel2022}
Briel, M.~M., Eldridge, J.~J., Stanway, E.~R., Stevance, H.~F., \& Chrimes, A.~A. 2022, Monthly Notices of the Royal Astronomical Society, 514, 1315

\bibitem[{{Broekgaarden} {et~al.}(2021){Broekgaarden}, {Berger}, {Neijssel}, {Vigna-G{\'o}mez}, {Chattopadhyay}, {Stevenson}, {Chruslinska}, {Justham}, {de Mink}, \& {Mandel}}]{Broekgaarden2021}
{Broekgaarden}, F.~S., {Berger}, E., {Neijssel}, C.~J., {et~al.} 2021, \mnras, 508, 5028

\bibitem[{{Broekgaarden} {et~al.}(2022){Broekgaarden}, {Berger}, {Stevenson}, {Justham}, {Mandel}, {Chru{\'s}li{\'n}ska}, {van Son}, {Wagg}, {Vigna-G{\'o}mez}, {de Mink}, {Chattopadhyay}, \& {Neijssel}}]{Broekgaarden2022}
{Broekgaarden}, F.~S., {Berger}, E., {Stevenson}, S., {et~al.} 2022, \mnras, 516, 5737

\bibitem[{{Caffau} {et~al.}(2011){Caffau}, {Ludwig}, {Steffen}, {Freytag}, \& {Bonifacio}}]{Caffau2011}
{Caffau}, E., {Ludwig}, H.~G., {Steffen}, M., {Freytag}, B., \& {Bonifacio}, P. 2011, \solphys, 268, 255

\bibitem[{{Callister} \& {Farr}(2024)}]{Callister2024}
{Callister}, T.~A. \& {Farr}, W.~M. 2024, Physical Review X, 14, 021005

\bibitem[{{Caputi} {et~al.}(2017){Caputi}, {Deshmukh}, {Ashby}, {Cowley}, {Bisigello}, {Fazio}, {Fynbo}, {Le F{\`e}vre}, {Milvang-Jensen}, \& {Ilbert}}]{Caputi2017}
{Caputi}, K.~I., {Deshmukh}, S., {Ashby}, M.~L.~N., {et~al.} 2017, \apj, 849, 45

\bibitem[{{Casey} {et~al.}(2018){Casey}, {Zavala}, {Spilker}, {da Cunha}, {Hodge}, {Hung}, {Staguhn}, {Finkelstein}, \& {Drew}}]{Casey2018}
{Casey}, C.~M., {Zavala}, J.~A., {Spilker}, J., {et~al.} 2018, \apj, 862, 77

\bibitem[{{Chru{\'s}li{\'n}ska}(2024)}]{Chruslinska2024}
{Chru{\'s}li{\'n}ska}, M. 2024, Annalen der Physik, 536, 2200170

\bibitem[{{Chruslinska} \& {Nelemans}(2019)}]{Chruslinska2019}
{Chruslinska}, M. \& {Nelemans}, G. 2019, \mnras, 488, 5300

\bibitem[{{Chruslinska} {et~al.}(2019){Chruslinska}, {Nelemans}, \& {Belczynski}}]{Chruslinska2019MRD}
{Chruslinska}, M., {Nelemans}, G., \& {Belczynski}, K. 2019, \mnras, 482, 5012

\bibitem[{{Chru{\'s}li{\'n}ska} {et~al.}(2021){Chru{\'s}li{\'n}ska}, {Nelemans}, {Boco}, \& {Lapi}}]{Chruslinska2021}
{Chru{\'s}li{\'n}ska}, M., {Nelemans}, G., {Boco}, L., \& {Lapi}, A. 2021, \mnras, 508, 4994

\bibitem[{{Claeys} {et~al.}(2014){Claeys}, {Pols}, {Izzard}, {Vink}, \& {Verbunt}}]{Claeys2014}
{Claeys}, J.~S.~W., {Pols}, O.~R., {Izzard}, R.~G., {Vink}, J., \& {Verbunt}, F.~W.~M. 2014, \aap, 563, A83

\bibitem[{COMPAS {et~al.}(2022)COMPAS, Riley, Agrawal, Barrett, Boyett, Broekgaarden, Chattopadhyay, Gaebel, Gittins, Hirai, Howitt, Justham, Khandelwal, Kummer, Lau, Mandel, de~Mink, Neijssel, Riley, van Son, Stevenson, Vigna-Gómez, Vinciguerra, Wagg, \& Willcox}]{Riley2022}
COMPAS, T., Riley, J., Agrawal, P., {et~al.} 2022, The Astrophysical Journal Supplement Series, 258, 34

\bibitem[{{Costa} {et~al.}(2019{\natexlab{a}}){Costa}, {Girardi}, {Bressan}, {Chen}, {Goudfrooij}, {Marigo}, {Rodrigues}, \& {Lanza}}]{Costa2019b}
{Costa}, G., {Girardi}, L., {Bressan}, A., {et~al.} 2019{\natexlab{a}}, \aap, 631, A128

\bibitem[{{Costa} {et~al.}(2019{\natexlab{b}}){Costa}, {Girardi}, {Bressan}, {Marigo}, {Rodrigues}, {Chen}, {Lanza}, \& {Goudfrooij}}]{Costa2019a}
{Costa}, G., {Girardi}, L., {Bressan}, A., {et~al.} 2019{\natexlab{b}}, \mnras, 485, 4641

\bibitem[{{Curti} {et~al.}(2020){Curti}, {Mannucci}, {Cresci}, \& {Maiolino}}]{Curti2020}
{Curti}, M., {Mannucci}, F., {Cresci}, G., \& {Maiolino}, R. 2020, \mnras, 491, 944

\bibitem[{{Di Carlo} {et~al.}(2020){Di Carlo}, {Mapelli}, {Giacobbo}, {Spera}, {Bouffanais}, {Rastello}, {Santoliquido}, {Pasquato}, {Ballone}, {Trani}, {Torniamenti}, \& {Haardt}}]{Dicarlo2020}
{Di Carlo}, U.~N., {Mapelli}, M., {Giacobbo}, N., {et~al.} 2020, \mnras, 498, 495

\bibitem[{{Di Stefano} {et~al.}(2023){Di Stefano}, {Kruckow}, {Gao}, {Neunteufel}, \& {Kobayashi}}]{Distefano2023}
{Di Stefano}, R., {Kruckow}, M.~U., {Gao}, Y., {Neunteufel}, P.~G., \& {Kobayashi}, C. 2023, \apj, 944, 87

\bibitem[{{Dominik} {et~al.}(2012){Dominik}, {Belczynski}, {Fryer}, {Holz}, {Berti}, {Bulik}, {Mandel}, \& {O'Shaughnessy}}]{Dominik2012}
{Dominik}, M., {Belczynski}, K., {Fryer}, C., {et~al.} 2012, \apj, 759, 52

\bibitem[{{Dominik} {et~al.}(2013){Dominik}, {Belczynski}, {Fryer}, {Holz}, {Berti}, {Bulik}, {Mandel}, \& {O'Shaughnessy}}]{Dominik2013}
{Dominik}, M., {Belczynski}, K., {Fryer}, C., {et~al.} 2013, \apj, 779, 72

\bibitem[{{Dorozsmai} \& {Toonen}(2024)}]{Dorozsmai2024}
{Dorozsmai}, A. \& {Toonen}, S. 2024, \mnras, 530, 3706

\bibitem[{Eldridge {et~al.}(2018)Eldridge, Stanway, \& Tang}]{Eldridge2018}
Eldridge, J.~J., Stanway, E.~R., \& Tang, P.~N. 2018, Monthly Notices of the Royal Astronomical Society, 482, 870

\bibitem[{{Everson} {et~al.}(2025){Everson}, {MacLeod}, \& {Ramirez-Ruiz}}]{Everson2024}
{Everson}, R.~W., {MacLeod}, M., \& {Ramirez-Ruiz}, E. 2025, \apjl, 979, L11

\bibitem[{{Fishbach}(2025)}]{Fishbach2025}
{Fishbach}, M. 2025, Classical and Quantum Gravity, 42, 055009

\bibitem[{{Fryer} {et~al.}(2012){Fryer}, {Belczynski}, {Wiktorowicz}, {Dominik}, {Kalogera}, \& {Holz}}]{Fryer2012}
{Fryer}, C.~L., {Belczynski}, K., {Wiktorowicz}, G., {et~al.} 2012, \apj, 749, 91

\bibitem[{{Gallegos-Garcia} {et~al.}(2021){Gallegos-Garcia}, {Berry}, {Marchant}, \& {Kalogera}}]{Gallegos2021}
{Gallegos-Garcia}, M., {Berry}, C. P.~L., {Marchant}, P., \& {Kalogera}, V. 2021, \apj, 922, 110

\bibitem[{{Ge} {et~al.}(2010){Ge}, {Hjellming}, {Webbink}, {Chen}, \& {Han}}]{Ge2010}
{Ge}, H., {Hjellming}, M.~S., {Webbink}, R.~F., {Chen}, X., \& {Han}, Z. 2010, \apj, 717, 724

\bibitem[{{Ge} {et~al.}(2015){Ge}, {Webbink}, {Chen}, \& {Han}}]{Ge2015}
{Ge}, H., {Webbink}, R.~F., {Chen}, X., \& {Han}, Z. 2015, \apj, 812, 40

\bibitem[{{Ge} {et~al.}(2020){Ge}, {Webbink}, {Chen}, \& {Han}}]{Ge2020}
{Ge}, H., {Webbink}, R.~F., {Chen}, X., \& {Han}, Z. 2020, \apj, 899, 132

\bibitem[{{Giacobbo} \& {Mapelli}(2018)}]{Giacobbo2018}
{Giacobbo}, N. \& {Mapelli}, M. 2018, \mnras, 480, 2011

\bibitem[{{Giacobbo} \& {Mapelli}(2020)}]{Giacobbo2020}
{Giacobbo}, N. \& {Mapelli}, M. 2020, \apj, 891, 141

\bibitem[{{Gruppioni} {et~al.}(2020){Gruppioni}, {B{\'e}thermin}, {Loiacono}, {Le F{\`e}vre}, {Capak}, {Cassata}, {Faisst}, {Schaerer}, {Silverman}, {Yan}, {Bardelli}, {Boquien}, {Carraro}, {Cimatti}, {Dessauges-Zavadsky}, {Ginolfi}, {Fujimoto}, {Hathi}, {Jones}, {Khusanova}, {Koekemoer}, {Lagache}, {Lemaux}, {Oesch}, {Pozzi}, {Riechers}, {Rodighiero}, {Romano}, {Talia}, {Vallini}, {Vergani}, {Zamorani}, \& {Zucca}}]{Gruppioni2020}
{Gruppioni}, C., {B{\'e}thermin}, M., {Loiacono}, F., {et~al.} 2020, \aap, 643, A8

\bibitem[{{Hall} \& {Evans}(2019)}]{Hall2019}
{Hall}, E.~D. \& {Evans}, M. 2019, Classical and Quantum Gravity, 36, 225002

\bibitem[{Harris {et~al.}(2020)Harris, Millman, van~der Walt, Gommers, Virtanen, Cournapeau, Wieser, Taylor, Berg, Smith, Kern, Picus, Hoyer, van Kerkwijk, Brett, Haldane, del R{'{\i}}o, Wiebe, Peterson, G{'{e}}rard-Marchant, Sheppard, Reddy, Weckesser, Abbasi, Gohlke, \& Oliphant}]{Harris20}
Harris, C.~R., Millman, K.~J., van~der Walt, S.~J., {et~al.} 2020, Nature, 585, 357

\bibitem[{{Hirai} \& {Mandel}(2022)}]{Hirai2022}
{Hirai}, R. \& {Mandel}, I. 2022, \apjl, 937, L42

\bibitem[{{Hobbs} {et~al.}(2005){Hobbs}, {Lorimer}, {Lyne}, \& {Kramer}}]{Hobbs2005}
{Hobbs}, G., {Lorimer}, D.~R., {Lyne}, A.~G., \& {Kramer}, M. 2005, \mnras, 360, 974

\bibitem[{Hunter(2007)}]{Hunter2007}
Hunter, J.~D. 2007, Computing in Science \& Engineering, 9, 90

\bibitem[{{Hurley} {et~al.}(2002){Hurley}, {Tout}, \& {Pols}}]{Hurley2002}
{Hurley}, J.~R., {Tout}, C.~A., \& {Pols}, O.~R. 2002, \mnras, 329, 897

\bibitem[{{Iorio} {et~al.}(2023){Iorio}, {Mapelli}, {Costa}, {Spera}, {Escobar}, {Sgalletta}, {Trani}, {Korb}, {Santoliquido}, {Dall'Amico}, {Gaspari}, \& {Bressan}}]{Iorio2023}
{Iorio}, G., {Mapelli}, M., {Costa}, G., {et~al.} 2023, \mnras, 524, 426

\bibitem[{{Ivanova} {et~al.}(2013){Ivanova}, {Justham}, {Chen}, {De Marco}, {Fryer}, {Gaburov}, {Ge}, {Glebbeek}, {Han}, {Li}, {Lu}, {Marsh}, {Podsiadlowski}, {Potter}, {Soker}, {Taam}, {Tauris}, {van den Heuvel}, \& {Webbink}}]{Ivanova2013}
{Ivanova}, N., {Justham}, S., {Chen}, X., {et~al.} 2013, \aapr, 21, 59

\bibitem[{{Kalogera} {et~al.}(2021){Kalogera}, {Sathyaprakash}, {Bailes}, {Bizouard}, {Buonanno}, {Burrows}, {Colpi}, {Evans}, {Fairhurst}, {Hild}, {Kasliwal}, {Lehner}, {Mandel}, {Mandic}, {Nissanke}, {Alessandra Papa}, {Reddy}, {Rosswog}, {Van Den Broeck}, {Ajith}, {Anand}, {Andreoni}, {Arun}, {Barausse}, {Baryakhtar}, {Belgacem}, {Berry}, {Bertacca}, {Brito}, {Caprini}, {Chatziioannou}, {Coughlin}, {Cusin}, {Dietrich}, {Dirian}, {East}, {Fan}, {Figueroa}, {Foffa}, {Ghosh}, {Hall}, {Harms}, {Harry}, {Hinderer}, {Janka}, {Justham}, {Kasen}, {Kotake}, {Lovelace}, {Maggiore}, {Mangiagli}, {Mapelli}, {Maselli}, {Matas}, {McIver}, {Messer}, {Mezzacappa}, {Mills}, {Mueller}, {M{\"u}ller}, {P{\"u}rrer}, {Pani}, {Pratten}, {Regimbau}, {Sakellariadou}, {Schneider}, {Sesana}, {Shao}, {Sotiriou}, {Tamanini}, {Tauris}, {Thrane}, {Valiante}, {van de Meent}, {Varma}, {Vines}, {Vitale}, {Yang}, {Yunes}, {Zumalacarregui}, {Punturo}, {Reitze}, {Couvares}, {Katsanevas}, {Kajita}, {Lueck}, {McClelland}, {Rowan}, {Sanders},
  {Shoemaker}, \& {van den Brand}}]{Kalogera2021}
{Kalogera}, V., {Sathyaprakash}, B.~S., {Bailes}, M., {et~al.} 2021, arXiv e-prints, arXiv:2111.06990

\bibitem[{{Klencki} {et~al.}(2018){Klencki}, {Moe}, {Gladysz}, {Chruslinska}, {Holz}, \& {Belczynski}}]{Klencki2018}
{Klencki}, J., {Moe}, M., {Gladysz}, W., {et~al.} 2018, \aap, 619, A77

\bibitem[{{Klencki} {et~al.}(2021){Klencki}, {Nelemans}, {Istrate}, \& {Chruslinska}}]{Klencki2021}
{Klencki}, J., {Nelemans}, G., {Istrate}, A.~G., \& {Chruslinska}, M. 2021, \aap, 645, A54

\bibitem[{{Klencki} {et~al.}(2020){Klencki}, {Nelemans}, {Istrate}, \& {Pols}}]{Klencki2020}
{Klencki}, J., {Nelemans}, G., {Istrate}, A.~G., \& {Pols}, O. 2020, \aap, 638, A55

\bibitem[{{Kroupa}(2001)}]{kroupa2001}
{Kroupa}, P. 2001, \mnras, 322, 231

\bibitem[{{Lamberts} {et~al.}(2018){Lamberts}, {Garrison-Kimmel}, {Hopkins}, {Quataert}, {Bullock}, {Faucher-Gigu{\`e}re}, {Wetzel}, {Kere{\v{s}}}, {Drango}, \& {Sanderson}}]{Lamberts2018}
{Lamberts}, A., {Garrison-Kimmel}, S., {Hopkins}, P.~F., {et~al.} 2018, \mnras, 480, 2704

\bibitem[{Lau {et~al.}(2022)Lau, Hirai, González-Bolívar, Price, De Marco, \& Mandel}]{Lau2022}
Lau, M. Y.~M., Hirai, R., González-Bolívar, M., {et~al.} 2022, Monthly Notices of the Royal Astronomical Society, 512, 5462

\bibitem[{{Madau} \& {Dickinson}(2014)}]{Madau2014}
{Madau}, P. \& {Dickinson}, M. 2014, \araa, 52, 415

\bibitem[{{Madau} \& {Fragos}(2017)}]{Madau2017}
{Madau}, P. \& {Fragos}, T. 2017, \apj, 840, 39

\bibitem[{{Maiolino} \& {Mannucci}(2019)}]{Maiolino2019}
{Maiolino}, R. \& {Mannucci}, F. 2019, \aapr, 27, 3

\bibitem[{{Mandel} \& {Broekgaarden}(2022)}]{Mandel2022}
{Mandel}, I. \& {Broekgaarden}, F.~S. 2022, Living Reviews in Relativity, 25, 1

\bibitem[{{Mannucci} {et~al.}(2010){Mannucci}, {Cresci}, {Maiolino}, {Marconi}, \& {Gnerucci}}]{Mannucci2010}
{Mannucci}, F., {Cresci}, G., {Maiolino}, R., {Marconi}, A., \& {Gnerucci}, A. 2010, \mnras, 408, 2115

\bibitem[{{Mannucci} {et~al.}(2011){Mannucci}, {Salvaterra}, \& {Campisi}}]{Mannucci2011}
{Mannucci}, F., {Salvaterra}, R., \& {Campisi}, M.~A. 2011, \mnras, 414, 1263

\bibitem[{{Mapelli}(2020)}]{Mapelli2020}
{Mapelli}, M. 2020, Frontiers in Astronomy and Space Sciences, 7, 38

\bibitem[{{Mapelli}(2021)}]{Mapelli2021}
{Mapelli}, M. 2021, in Handbook of Gravitational Wave Astronomy, ed. C.~{Bambi}, S.~{Katsanevas}, \& K.~D. {Kokkotas}, 16

\bibitem[{{Mapelli} {et~al.}(2022){Mapelli}, {Bouffanais}, {Santoliquido}, {Arca Sedda}, \& {Artale}}]{Mapelli2022}
{Mapelli}, M., {Bouffanais}, Y., {Santoliquido}, F., {Arca Sedda}, M., \& {Artale}, M.~C. 2022, \mnras, 511, 5797

\bibitem[{{Mapelli} \& {Giacobbo}(2018)}]{Mapelli2018}
{Mapelli}, M. \& {Giacobbo}, N. 2018, \mnras, 479, 4391

\bibitem[{{Mapelli} {et~al.}(2017){Mapelli}, {Giacobbo}, {Ripamonti}, \& {Spera}}]{Mapelli2017}
{Mapelli}, M., {Giacobbo}, N., {Ripamonti}, E., \& {Spera}, M. 2017, \mnras, 472, 2422

\bibitem[{{Mapelli} {et~al.}(2019){Mapelli}, {Giacobbo}, {Santoliquido}, \& {Artale}}]{Mapelli2019}
{Mapelli}, M., {Giacobbo}, N., {Santoliquido}, F., \& {Artale}, M.~C. 2019, \mnras, 487, 2

\bibitem[{{Marchant} {et~al.}(2021){Marchant}, {Pappas}, {Gallegos-Garcia}, {Berry}, {Taam}, {Kalogera}, \& {Podsiadlowski}}]{Marchant2021}
{Marchant}, P., {Pappas}, K. M.~W., {Gallegos-Garcia}, M., {et~al.} 2021, \aap, 650, A107

\bibitem[{{Misra} {et~al.}(2020){Misra}, {Fragos}, {Tauris}, {Zapartas}, \& {Aguilera-Dena}}]{Misra2020}
{Misra}, D., {Fragos}, T., {Tauris}, T.~M., {Zapartas}, E., \& {Aguilera-Dena}, D.~R. 2020, \aap, 642, A174

\bibitem[{{Moe} \& {Di Stefano}(2017)}]{Moe2017}
{Moe}, M. \& {Di Stefano}, R. 2017, \apjs, 230, 15

\bibitem[{{Nagarajan} \& {El-Badry}(2025)}]{Nagarajan2024}
{Nagarajan}, P. \& {El-Badry}, K. 2025, \pasp, 137, 034203

\bibitem[{{Nakajima} {et~al.}(2023){Nakajima}, {Ouchi}, {Isobe}, {Harikane}, {Zhang}, {Ono}, {Umeda}, \& {Oguri}}]{Nakajima2023}
{Nakajima}, K., {Ouchi}, M., {Isobe}, Y., {et~al.} 2023, \apjs, 269, 33

\bibitem[{{Neijssel} {et~al.}(2019){Neijssel}, {Vigna-G{\'o}mez}, {Stevenson}, {Barrett}, {Gaebel}, {Broekgaarden}, {de Mink}, {Sz{\'e}csi}, {Vinciguerra}, \& {Mandel}}]{Neijssel2019}
{Neijssel}, C.~J., {Vigna-G{\'o}mez}, A., {Stevenson}, S., {et~al.} 2019, \mnras, 490, 3740

\bibitem[{{Nguyen} {et~al.}(2022){Nguyen}, {Costa}, {Girardi}, {Volpato}, {Bressan}, {Chen}, {Marigo}, {Fu}, \& {Goudfrooij}}]{nguyen2022}
{Nguyen}, C.~T., {Costa}, G., {Girardi}, L., {et~al.} 2022, \aap, 665, A126

\bibitem[{{Nitz} {et~al.}(2023){Nitz}, {Kumar}, {Wang}, {Kastha}, {Wu}, {Sch{\"a}fer}, {Dhurkunde}, \& {Capano}}]{Nitz2023}
{Nitz}, A.~H., {Kumar}, S., {Wang}, Y.-F., {et~al.} 2023, \apj, 946, 59

\bibitem[{{Olejak} {et~al.}(2021){Olejak}, {Belczynski}, \& {Ivanova}}]{Olejak2021}
{Olejak}, A., {Belczynski}, K., \& {Ivanova}, N. 2021, \aap, 651, A100

\bibitem[{{Olejak} {et~al.}(2022){Olejak}, {Fryer}, {Belczynski}, \& {Baibhav}}]{Olejak2022}
{Olejak}, A., {Fryer}, C.~L., {Belczynski}, K., \& {Baibhav}, V. 2022, \mnras, 516, 2252

\bibitem[{{Pavlovskii} {et~al.}(2017){Pavlovskii}, {Ivanova}, {Belczynski}, \& {Van}}]{Pavlovskii2017}
{Pavlovskii}, K., {Ivanova}, N., {Belczynski}, K., \& {Van}, K.~X. 2017, \mnras, 465, 2092

\bibitem[{{Popesso} {et~al.}(2023){Popesso}, {Concas}, {Cresci}, {Belli}, {Rodighiero}, {Inami}, {Dickinson}, {Ilbert}, {Pannella}, \& {Elbaz}}]{Popesso2023}
{Popesso}, P., {Concas}, A., {Cresci}, G., {et~al.} 2023, \mnras, 519, 1526

\bibitem[{{Rauf} {et~al.}(2023){Rauf}, {Howlett}, {Davis}, \& {Lagos}}]{Rauf2023}
{Rauf}, L., {Howlett}, C., {Davis}, T.~M., \& {Lagos}, C. D.~P. 2023, \mnras, 523, 5719

\bibitem[{Ray {et~al.}(2023)Ray, Hernandez, Mohite, Creighton, \& Kapadia}]{Ray2023}
Ray, A., Hernandez, I.~M., Mohite, S., Creighton, J., \& Kapadia, S. 2023, The Astrophysical Journal, 957, 37

\bibitem[{{Riley} {et~al.}(2021){Riley}, {Mandel}, {Marchant}, {Butler}, {Nathaniel}, {Neijssel}, {Shortt}, \& {Vigna-G{\'o}mez}}]{Riley2021}
{Riley}, J., {Mandel}, I., {Marchant}, P., {et~al.} 2021, \mnras, 505, 663

\bibitem[{{Rinaldi} {et~al.}(2022){Rinaldi}, {Caputi}, {van Mierlo}, {Ashby}, {Caminha}, \& {Iani}}]{Rinaldi2022}
{Rinaldi}, P., {Caputi}, K.~I., {van Mierlo}, S.~E., {et~al.} 2022, \apj, 930, 128

\bibitem[{{Rinaldi} {et~al.}(2024){Rinaldi}, {Del Pozzo}, {Mapelli}, {Lorenzo-Medina}, \& {Dent}}]{Rinaldi2024}
{Rinaldi}, S., {Del Pozzo}, W., {Mapelli}, M., {Lorenzo-Medina}, A., \& {Dent}, T. 2024, \aap, 684, A204

\bibitem[{Rodighiero {et~al.}(2015)Rodighiero, Brusa, Daddi, Negrello, Mullaney, Delvecchio, Lutz, Renzini, Franceschini, Baronchelli, Pozzi, Gruppioni, Strazzullo, Cimatti, \& Silverman}]{Rodighiero2015}
Rodighiero, G., Brusa, M., Daddi, E., {et~al.} 2015, The Astrophysical Journal Letters, 800, L10

\bibitem[{{Rodriguez} \& {Loeb}(2018)}]{Rodriguez2018}
{Rodriguez}, C.~L. \& {Loeb}, A. 2018, \apjl, 866, L5

\bibitem[{{Romagnolo} {et~al.}(2023){Romagnolo}, {Belczynski}, {Klencki}, {Agrawal}, {Shenar}, \& {Sz{\'e}csi}}]{Romagnolo2023}
{Romagnolo}, A., {Belczynski}, K., {Klencki}, J., {et~al.} 2023, \mnras, 525, 706

\bibitem[{{Rom{\'a}n-Garza} {et~al.}(2021){Rom{\'a}n-Garza}, {Bavera}, {Fragos}, {Zapartas}, {Misra}, {Andrews}, {Coughlin}, {Dotter}, {Kovlakas}, {Serra}, {Qin}, {Rocha}, \& {Tran}}]{Romangarza2021}
{Rom{\'a}n-Garza}, J., {Bavera}, S.~S., {Fragos}, T., {et~al.} 2021, \apjl, 912, L23

\bibitem[{{Sana} {et~al.}(2012){Sana}, {de Mink}, {de Koter}, {Langer}, {Evans}, {Gieles}, {Gosset}, {Izzard}, {Le Bouquin}, \& {Schneider}}]{Sana2012}
{Sana}, H., {de Mink}, S.~E., {de Koter}, A., {et~al.} 2012, Science, 337, 444

\bibitem[{{Santoliquido} {et~al.}(2022){Santoliquido}, {Mapelli}, {Artale}, \& {Boco}}]{Santoliquido2022}
{Santoliquido}, F., {Mapelli}, M., {Artale}, M.~C., \& {Boco}, L. 2022, \mnras, 516, 3297

\bibitem[{{Santoliquido} {et~al.}(2020){Santoliquido}, {Mapelli}, {Bouffanais}, {Giacobbo}, {Di Carlo}, {Rastello}, {Artale}, \& {Ballone}}]{Santoliquido2020}
{Santoliquido}, F., {Mapelli}, M., {Bouffanais}, Y., {et~al.} 2020, \apj, 898, 152

\bibitem[{{Santoliquido} {et~al.}(2021){Santoliquido}, {Mapelli}, {Giacobbo}, {Bouffanais}, \& {Artale}}]{Santoliquido2021}
{Santoliquido}, F., {Mapelli}, M., {Giacobbo}, N., {Bouffanais}, Y., \& {Artale}, M.~C. 2021, \mnras, 502, 4877

\bibitem[{{Santoliquido} {et~al.}(2023){Santoliquido}, {Mapelli}, {Iorio}, {Costa}, {Glover}, {Hartwig}, {Klessen}, \& {Merli}}]{Santoliquido2023}
{Santoliquido}, F., {Mapelli}, M., {Iorio}, G., {et~al.} 2023, \mnras, 524, 307

\bibitem[{{Sargent} {et~al.}(2012){Sargent}, {B{\'e}thermin}, {Daddi}, \& {Elbaz}}]{Sargent2012}
{Sargent}, M.~T., {B{\'e}thermin}, M., {Daddi}, E., \& {Elbaz}, D. 2012, \apjl, 747, L31

\bibitem[{{Schneider} {et~al.}(2021){Schneider}, {Podsiadlowski}, \& {M{\"u}ller}}]{Schneider2021}
{Schneider}, F.~R.~N., {Podsiadlowski}, P., \& {M{\"u}ller}, B. 2021, \aap, 645, A5

\bibitem[{{Schreiber, C.} {et~al.}(2015){Schreiber, C.}, {Pannella, M.}, {Elbaz, D.}, {Béthermin, M.}, {Inami, H.}, {Dickinson, M.}, {Magnelli, B.}, {Wang, T.}, {Aussel, H.}, {Daddi, E.}, {Juneau, S.}, {Shu, X.}, {Sargent, M. T.}, {Buat, V.}, {Faber, S. M.}, {Ferguson, H. C.}, {Giavalisco, M.}, {Koekemoer, A. M.}, {Magdis, G.}, {Morrison, G. E.}, {Papovich, C.}, {Santini, P.}, \& {Scott, D.}}]{Schreiber2015}
{Schreiber, C.}, {Pannella, M.}, {Elbaz, D.}, {et~al.} 2015, \aap, 575, A74

\bibitem[{{Sgalletta} {et~al.}(2023){Sgalletta}, {Iorio}, {Mapelli}, {Artale}, {Boco}, {Chattopadhyay}, {Lapi}, {Possenti}, {Rinaldi}, \& {Spera}}]{Sgalletta2023}
{Sgalletta}, C., {Iorio}, G., {Mapelli}, M., {et~al.} 2023, \mnras, 526, 2210

\bibitem[{{Shao} \& {Li}(2021)}]{Shao2021}
{Shao}, Y. \& {Li}, X.-D. 2021, \apj, 920, 81

\bibitem[{{Speagle} {et~al.}(2014){Speagle}, {Steinhardt}, {Capak}, \& {Silverman}}]{Speagle2014}
{Speagle}, J.~S., {Steinhardt}, C.~L., {Capak}, P.~L., \& {Silverman}, J.~D. 2014, \apjs, 214, 15

\bibitem[{{Spera} \& {Mapelli}(2017)}]{Spera2017}
{Spera}, M. \& {Mapelli}, M. 2017, \mnras, 470, 4739

\bibitem[{{Spera} {et~al.}(2019){Spera}, {Mapelli}, {Giacobbo}, {Trani}, {Bressan}, \& {Costa}}]{Spera2019}
{Spera}, M., {Mapelli}, M., {Giacobbo}, N., {et~al.} 2019, \mnras, 485, 889

\bibitem[{{Srinivasan} {et~al.}(2023){Srinivasan}, {Lamberts}, {Bizouard}, {Bruel}, \& {Mastrogiovanni}}]{Srinivasan2023}
{Srinivasan}, R., {Lamberts}, A., {Bizouard}, M.~A., {Bruel}, T., \& {Mastrogiovanni}, S. 2023, \mnras, 524, 60

\bibitem[{{Stevenson} \& {Clarke}(2022)}]{Stevenson2022}
{Stevenson}, S. \& {Clarke}, T.~A. 2022, \mnras, 517, 4034

\bibitem[{{Tang} {et~al.}(2020){Tang}, {Eldridge}, {Stanway}, \& {Bray}}]{Tang2020}
{Tang}, P.~N., {Eldridge}, J.~J., {Stanway}, E.~R., \& {Bray}, J.~C. 2020, \mnras, 493, L6

\bibitem[{{The pandas development Team}(2024)}]{Pandas2024}
{The pandas development Team}. 2024, {pandas-dev/pandas: Pandas}

\bibitem[{{Trani} {et~al.}(2022){Trani}, {Rieder}, {Tanikawa}, {Iorio}, {Martini}, {Karelin}, {Glanz}, \& {Portegies Zwart}}]{Trani2022}
{Trani}, A.~A., {Rieder}, S., {Tanikawa}, A., {et~al.} 2022, \prd, 106, 043014

\bibitem[{{van Son} {et~al.}(2022){van Son}, {de Mink}, {Callister}, {Justham}, {Renzo}, {Wagg}, {Broekgaarden}, {Kummer}, {Pakmor}, \& {Mandel}}]{VanSon2022}
{van Son}, L.~A.~C., {de Mink}, S.~E., {Callister}, T., {et~al.} 2022, \apj, 931, 17

\bibitem[{{van Son} {et~al.}(2023){van Son}, {de Mink}, {Chru{\'s}li{\'n}ska}, {Conroy}, {Pakmor}, \& {Hernquist}}]{VanSon2023}
{van Son}, L.~A.~C., {de Mink}, S.~E., {Chru{\'s}li{\'n}ska}, M., {et~al.} 2023, \apj, 948, 105

\bibitem[{{van Son} {et~al.}(2025){van Son}, {Roy}, {Mandel}, {Farr}, {Lam}, {Merritt}, {Broekgaarden}, {Sander}, \& {Andrews}}]{Vanson2025}
{van Son}, L.~A.~C., {Roy}, S.~K., {Mandel}, I., {et~al.} 2025, \apj, 979, 209

\bibitem[{{Verbunt} {et~al.}(2017){Verbunt}, {Igoshev}, \& {Cator}}]{Verbunt2017}
{Verbunt}, F., {Igoshev}, A., \& {Cator}, E. 2017, \aap, 608, A57

\bibitem[{Virtanen {et~al.}(2020)Virtanen, Gommers, Oliphant, Haberland, Reddy, Cournapeau, Burovski, Peterson, Weckesser, Bright, {van der Walt}, Brett, Wilson, Millman, Mayorov, Nelson, Jones, Kern, Larson, Carey, Polat, Feng, Moore, {VanderPlas}, Laxalde, Perktold, Cimrman, Henriksen, Quintero, Harris, Archibald, Ribeiro, Pedregosa, {van Mulbregt}, \& {SciPy 1.0 Contributors}}]{SciPy2020}
Virtanen, P., Gommers, R., Oliphant, T.~E., {et~al.} 2020, Nature Methods, 17, 261

\bibitem[{{Webbink}(1984)}]{Webbink1984}
{Webbink}, R.~F. 1984, \apj, 277, 355

\bibitem[{{Zevin} {et~al.}(2021){Zevin}, {Bavera}, {Berry}, {Kalogera}, {Fragos}, {Marchant}, {Rodriguez}, {Antonini}, {Holz}, \& {Pankow}}]{Zevin2021}
{Zevin}, M., {Bavera}, S.~S., {Berry}, C. P.~L., {et~al.} 2021, \apj, 910, 152

\bibitem[{{Zevin} {et~al.}(2020){Zevin}, {Spera}, {Berry}, \& {Kalogera}}]{Zevin2020}
{Zevin}, M., {Spera}, M., {Berry}, C. P.~L., \& {Kalogera}, V. 2020, \apjl, 899, L1

\end{thebibliography}

\begin{appendix}

\onecolumn
\section{Observational scaling relations} \label{sec:osr_appendix}

Below, we provide the details of the observational scaling relations adopted in this work. We follow the approach presented by \cite{Santoliquido2022}. 

\subsection{Galaxy stellar mass function} \label{sec:gsmf}

Our adopted galaxy stellar mass function takes the form of a Schechter function:
\begin{equation}
    \phi(M_\ast, z) \,{}dM_\ast = \phi_{\rm N}(z) \,{} e^{-M_\ast / M_{\rm cut}(z)} \left( \frac{M_\ast}{M_{\rm cut}(z)}\right)^{-\alpha_{\rm GSMF}} dM_\ast,
\end{equation}

where $\phi_{\rm N}(z)$ is a normalization factor and $M_{\rm cut}(z)$ is the galaxy stellar mass at which the function transitions from a power law at low masses to an  exponential decay at high masses. In our models, we use the fit derived by \cite{Chruslinska2019} and based on a wide catalog of galaxy stellar mass function relations over separate redshift bins.

\subsection{Star formation rate (SFR)} \label{sec:mainsequence}

We draw the SFR of galaxies from the following double lognormal distribution \citep{Sargent2012, Rodighiero2015, Schreiber2015}:
\begin{equation}
    \mathcal{P} (\log{\rm SFR} | M_\ast, z) = A_{\rm MS} \exp\left[  - \frac{\left( \log{\rm SFR} - \langle \log{\rm SFR} \rangle_{\rm MS} \right)^{2}}{ 2 \sigma_{\rm MS}^{2}}\right] + A_{rm SB} \exp\left[  - \frac{\left( \log{\rm SFR} - \langle \log{\rm SFR} \rangle_{\rm SB} \right)^{2}}{ 2 \sigma_{\rm SB}^{2}}\right],
\end{equation}

where $A_{\rm MS}=0.97$ and $A_{\rm SB} = 0.243$ constants \citep{Sargent2012}. $ \langle \log{\rm SFR} \rangle_{\rm MS} $ and $\sigma_{\rm MS}$ are the average SFR of the main sequence and its dispersion. We adopt several definitions of the galaxy main sequence, as described in section \ref{sec:osr}. For all of our models we use the fiducial value $\sigma_{\rm MS}=0.188$ dex \citep{Santoliquido2022}. 
In a similar way, $ \langle \log{\rm SFR} \rangle_{\rm SB} $ and $\sigma_{\rm SB}$ are the average and the standard deviation of the galaxy starburst sequence.  The latter is defined as in \cite{Sargent2012},
\begin{equation}
    \langle \log{\rm SFR} \rangle_{\rm SB} = \langle \log{\rm SFR} \rangle_{\rm MS} + 0.59 
\end{equation}
Moreover, we assume $\sigma_{SB} = 0.243$ dex. We couple the galaxy stellar mass function with the distribution of SFR as described above.

\subsection{Fundamental metallicity relations}

\begin{figure*}[h!]
    \centering
    \includegraphics[width=.9\textwidth]{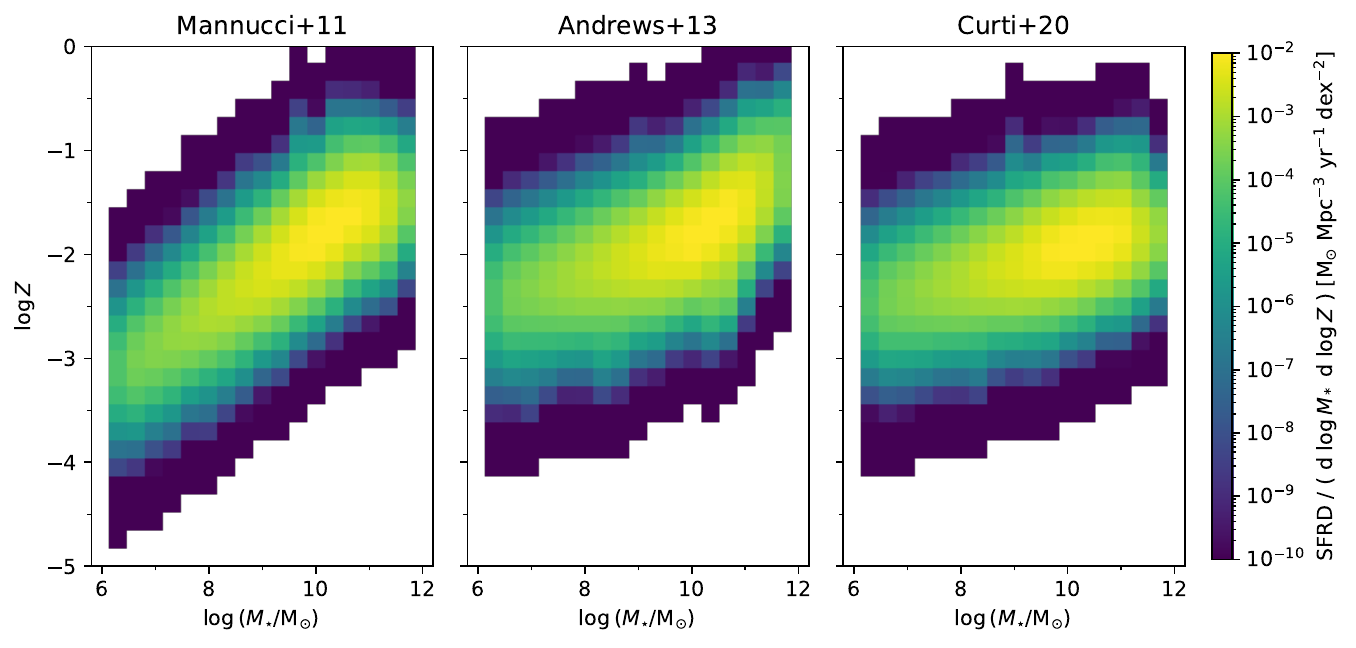}
    \caption{SFR density per galaxy mass (x axes) and metallicity bins (y axes) at $z=0$. Each panel shows the result for a different FMR relation, from left to right: \cite{Mannucci2011}, \cite{Andrews2013} and \cite{Curti2020}. }
    \label{fig:SFRD_by_M}
\end{figure*}

We consider three different fundamental metallicity relations. We adopt the fit from \cite{Mannucci2011}:
\begin{equation}
    \begin{split}
        12 + \log \left(\mathrm{O/H} \right) &=  8.90+0.37 m - 0.14 s - 0.19 m^2 + 0.12 m s - 0.054 s^2 \quad \; \mathrm{for} \; \mu_{0.32} \geq 9.5 \\
    &= 8.93 + 0.51(\mu_{0.32} - 10) \quad \; \mathrm{for} \; \mu_{0.32} < 9.5
\end{split}
\end{equation}

where $m=\log M_\ast - 10$ and $s=\log{\rm SFR}$ and $\mu_{0.32} = \log M_\ast - 0.32 \log{\rm SFR}$, all the quantities are in solar units.

We derive an additional fit  for the fundamental metallicity relation based on figure 12 by \citet{Andrews2013} 
\begin{equation}
    12 + \log \left(\mathrm{O/H} \right) = 0.43 ( \log M_\ast - 0.66 \log{\rm SFR} ) + 4.59
\end{equation}

Lastly, we consider the metallicity relation calculated in \cite{Curti2020}:
\begin{equation}
    12 + \log \left(\mathrm{O/H} \right) = Z_{0} - \gamma / \beta \log \left( 1 + (M_\ast / M_0(\mathrm{SFR}))^{-\beta}\right)
\end{equation}

where $M_0(\mathrm{SFR}) = m_0 + m_1 \log{\rm SFR}$ and $Z_0 = 8.779 \pm 0.005$, $m_0 = 10.11 \pm 0.03$, $m_1 = 0.56 \pm 0.01$, $\gamma = 0.31 \pm 0.01$ and $\beta = 2.1 \pm 0.4$. For the purposes of our simulations, we  convert these relations into absolute metallicities. We adopt the solar metallicity values from \cite{Caffau2011}: $Z_{\odot} = 0.0153$ and $12 + \log (\mathrm{O/H})_{\odot} = 8.76$.

Figure \ref{fig:SFRD_by_M} compares the distributions of SFR density as a function of galaxy stellar mass and metallicity, for different fundamental metallicity relations at $z=0$. The differences are particularly noticeable for low--mass galaxies, where the model by \citet{Mannucci2011} clearly predicts lower metallicities compared to the other two prescriptions. \cite{Curti2020} predicts the flattest relation, with less than an order of magnitude difference in metallicty between the low and high--mass galaxies.

\section{Merger rate densities} \label{sec:mrd_allfmr}

\begin{figure*}[h!]
    \centering
    \begin{subfigure}[t]{0.49\textwidth}
        \centering
        \includegraphics[width=\linewidth]{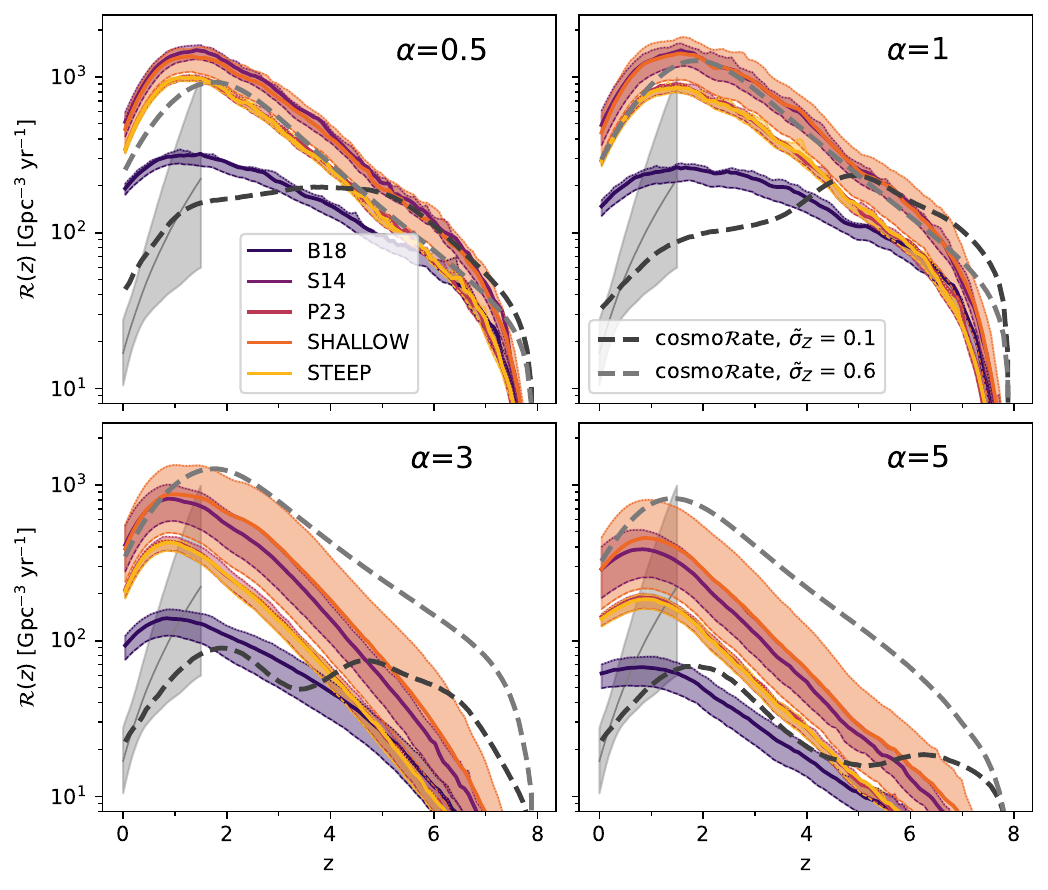}
        \caption{Same as Fig.~\ref{fig:MRD_all} but for the FMR by \citet{Curti2020}.}
        \label{fig:MRD_all_curti}
    \end{subfigure}
    \hfill
    \begin{subfigure}[t]{0.49\textwidth}
        \centering
        \includegraphics[width=\linewidth]{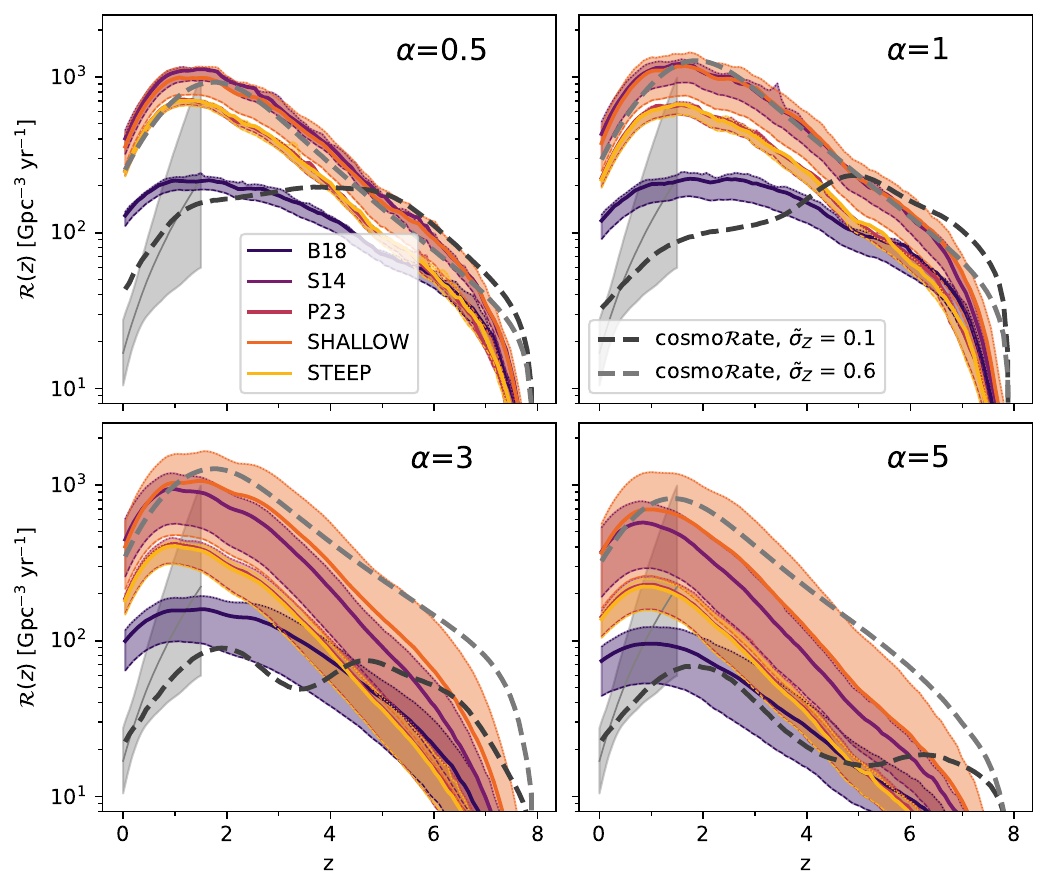}
        \caption{Same as Fig.~\ref{fig:MRD_all} but for the FMR by \citet{Mannucci2011}.}
        \label{fig:MRD_all_mannucci}
    \end{subfigure}
    \caption{BBH cosmic merger rate densities for different SFR--$M_*$ relations, using two different fundamental metallicity relations.}
    \label{fig:MRD_all_side_by_side}
\end{figure*}

Figures \ref{fig:MRD_all_curti} and \ref{fig:MRD_all_mannucci} show the BBH cosmic merger rate densities obtained for different SFR--$M_*$ relations, adopting the fundamental metallicity relations by \citet{Curti2020} and \citet{Mannucci2011}, respectively. 

\end{appendix}

\end{document}